\documentclass[aps,showpacs,superscriptaddress,preprint]{revtex4-2}

\usepackage{titlesec}
\usepackage{amssymb,amsmath,amsfonts,latexsym,graphicx,epsfig,bm}
\usepackage{epstopdf}

\usepackage{tabularx} 
\usepackage{multirow}

\usepackage{enumitem} 

\titleformat{\section}{\large\bfseries}{\thesection}{1em}{}

\newcommand{\bea}{\begin{eqnarray}}
\newcommand{\ena}{\end{eqnarray}}
\newcommand{\nn}{\nonumber\\}
\newcommand{\be}{\begin{equation}}
\newcommand{\en}{\end{equation}}

\newcommand{\ed}{\end{document}}

\newcommand{\slp}{p\kern-5pt/}
\newcommand{\Tr}{\mbox{\rm{tr}}}

\makeatletter
\def\switch@array{}  
\makeatother

\begin{document}

\title{Detailed analysis of possible new-physics effects in the semileptonic decay 
\boldmath{$B_s \to D_s^{(*)}\tau\bar{\nu}$}}

\author{M. A. Ivanov}
\email{ivanovm@theor.jinr.ru}
\affiliation{Bogoliubov Laboratory of Theoretical Physics, 
	Joint Institute for Nuclear Research, 141980 Dubna, Russia}

\author{J.~N.~Pandya}
\email{jnpandya-phy@spuvvn.edu}
\affiliation{Department of Physics, Sardar Patel University, \\ Vallabh Vidyanagar 388120, Gujarat, India}

\author{P.~Santorelli}
\email{Pietro.Santorelli@na.infn.it}
\affiliation{
	Dipartimento di Fisica ``E.~Pancini'', Universit\`a di Napoli Federico II,
	\hbox{ Complesso Universitario di Monte S.~Angelo,
		Via Cintia, Edificio 6, 80126 Napoli, Italy}}
\affiliation{
	\hbox{Istituto Nazionale di Fisica Nucleare, 
		Sezione di Napoli, 80126 Napoli, Italy}}
	
\author{N.~R.~Soni}
\email{nakul.soni-phy@msubaroda.ac.in}
\affiliation{Department of Physics, Faculty of Science, \\ The Maharaja Sayajirao University of Baroda, Vadodara 390002, Gujarat, India}

\author{C.~T.~Tran}
\email{thangtc@hcmute.edu.vn (corresponding author)}
\affiliation{
	\hbox{Department of Engineering Physics, Ho Chi Minh City University of Technology and Engineering, }\\
	Vo Van Ngan 1, 700000 Ho Chi Minh City, Vietnam}

\author{H.~C.~Tran}
\email{catth@hcmute.edu.vn}
\affiliation{
	\hbox{Department of Engineering Physics, Ho Chi Minh City University of Technology and Engineering, }\\
	Vo Van Ngan 1, 700000 Ho Chi Minh City, Vietnam}

\author{Vo Quoc Phong}
\email{vqphong@hcmus.edu.vn}
\affiliation{Department of Theoretical Physics, Faculty of Physics and Engineering Physics, University of Science, Ho Chi Minh City 700000, Vietnam}
\affiliation{
	\hbox{Vietnam National University, Ho Chi Minh City 700000, Vietnam}}


\begin{abstract}
We study the semileptonic decays $B_s \to D_s^{(*)}\tau\bar{\nu}$ as a promising probe for new physics (NP) beyond the standard model (SM). The extension of the SM is done through the introduction of four-fermion operators beyond the $V-A$ structure with the corresponding Wilson coefficients characterizing their contribution. The constraints on these coefficients are obtained from recent experimental data. Form factors describing hadron transitions are calculated in our covariant quark model with infrared confinement. Theoretical predictions for the full set of observables in these channels are provided. We analyze possible NP effects to be tested in future experiments.
\end{abstract}



\maketitle
\newpage

\section{INTRODUCTION}
\label{sec:intro} 
For more than a decade, the tension between the standard model (SM) prediction and experimental data for the ratios of branching fractions $R(D^{(*)})\equiv \mathcal{B}(B\to D^{(*)}\tau^-\bar{\nu}_\tau)/\mathcal{B}(B\to D^{(*)}\ell^-\bar{\nu}_\ell)$ $(\ell = e, \mu)$ has never disappeared. It has become well known among particle physicists and is referred to as ``the $R(D^{(*)})$ puzzle''. The puzzle implies possible violation of lepton flavor universality (LFU) and has fueled the search for new physics (NP) beyond the SM, on both theoretical and experimental fronts. Ever since the first observation by the \emph{BABAR} Collaboration~\cite{BaBar:2012obs}, the tension has been confirmed by many other experiments at all $B$ factories including the \emph{BABAR}~\cite{BaBar:2013mob}, Belle~\cite{Belle:2015qfa, Belle:2016ure, Belle:2016dyj, Belle:2017ilt, Belle:2019rba}, and, recently, LHCb~\cite{LHCb:2023zxo, LHCb:2023uiv, LHCb:2024jll} and Belle~II~\cite{Belle-II:2024ami, Belle-II:2025yjp} Collaborations. The most up-to-date average values of the mentioned experimental data are given by the Heavy Flavor Averaging Group (HFLAV) and read~\cite{HFLAV:2022esi}
\begin{equation}
R(D)|_{\rm expt} = 0.358 \pm 0.024 ,
\qquad
R(D^\ast)|_{\rm expt} = 0.281 \pm 0.011 ,
\label{eq:RD-expt}
\end{equation}
which exceed the average SM predictions~\cite{Bigi:2016mdz, Bordone:2019vic, Martinelli:2021onb, Bernlochner:2022ywh, Ray:2023xjn, FlavourLatticeAveragingGroupFLAG:2024oxs, BaBar:2019vpl, Gambino:2019sif, Martinelli:2023fwm}
\begin{equation}
R(D)|_{\rm SM} = 0.296 \pm 0.004 ,
\qquad
R(D^\ast)|_{\rm SM} = 0.254 \pm 0.005 ,
\label{eq:RD-SM}
\end{equation}
by 2.5$\sigma$ and 2.3$\sigma$, respectively [see also~\cite{MILC:2015uhg, Na:2015kha, Fajfer:2012vx}]. Taking into account the correlations between the two ratios, the difference with the SM expectations corresponds to about 3.8$\sigma$~\cite{HFLAV:2022esi}.

In addition to the $R(D^{(*)})$ puzzle, many other hints of NP in the flavor-changing charged current  $b\to c\tau\nu$ has been accumulated. For instance, the ratio of branching fractions $R(J/\psi)\equiv \mathcal{B}(B_c \to J/\psi \tau\nu)/\mathcal{B}(B_c \to J/\psi \mu\nu)$ has been measured by LHCb and CMS at CERN. The two collaborations reported $R(J/\psi)=0.71\pm 0.17(\rm{stat})\pm 0.18(\rm{syst})$ (LHCb)~\cite{LHCb:2017vlu} and $R(J/\psi)=0.49\pm 0.26$ (CMS)~\cite{CMS:2025jfx}. The combined result $R(J/\psi)=0.61\pm 0.18$~\cite{Iguro:2024hyk} shows an excess of about $1.9\sigma$ over the SM prediction $R(J/\psi)=0.2582(38)$~\cite{Harrison:2020nrv}. Besides, as pointed out in Ref.~\cite{Colangelo:2016ymy}, the flavor anomalies in $b\to c\tau\nu$ can be linked to the current tension between the inclusive and exclusive determinations of the Cabibbo--Kobayashi--Maskawa (CKM) matrix element $|V_{cb}|$. It is therefore quite natural to explore other semileptonic decays induced by the quark-level current $b\to c\tau\nu$ to search for NP effects which can be tested by experiment. A recent review on the prospects of using semileptonic $b$-hadron decays as a LFU laboratory is given in Ref.~\cite{Bernlochner:2021vlv}.

Inspired by the recent measurement of $|V_{cb}|$ with the $B_s\to D_s^{(*)}\mu\nu$ channels by the LHCb Collaboration~\cite{LHCb:2020cyw}, we consider the semileptonic decays $B_s \to D_s^{(*)}\tau\nu$ as a probe for NP in the $b\to c\tau\nu$ transition. We use the standard model effective field theory as a model-independent approach, in which the $b\to c\ell\nu$ transition is described by a general effective Hamiltonian of the form~\cite{Goldberger:1999yh,Buchmuller:1985jz,Grzadkowski:2010es}
\begin{eqnarray}
\label{eq:Heff}
\mathcal{H}_{eff} &=&\frac{4G_F V_{cb}}{\sqrt{2}} \Big(\mathcal{O}_{V_L}+
\sum\limits_{X=S_{L(R)},V_{L(R)},T_L} \delta_{\tau\ell}X\mathcal{O}_{X}\Big),
\end{eqnarray}
where $X$'s are complex Wilson coefficients governing the NP contributions, and $\mathcal{O}_{X}$'s are four-fermion operators of dimension-six which read 
\begin{eqnarray}
\mathcal{O}_{V_{L(R)}} &=&
\left(\bar{c}\gamma^{\mu}P_{L(R)} b\right)
\left(\bar{\ell}\gamma_{\mu}P_L\nu_{\ell}\right),\\
\mathcal{O}_{S_{L(R)}} &=& \left(\bar{c}P_{L(R)} b\right)\left(\bar{\ell}P_L\nu_{\ell}\right),
\\
\mathcal{O}_{T_L} &=& \left(\bar{c}\sigma^{\mu\nu}P_Lb\right)
\left(\bar{\ell}\sigma_{\mu\nu}P_L\nu_{\ell}\right).
\end{eqnarray}
Here, $\sigma_{\mu\nu}=i\left[\gamma_{\mu},\gamma_{\nu}\right]/2$ and
$P_{L,R}=(1\mp\gamma_5)/2$. The tensor operator with a right-handed quark current is identically equal to zero and is therefore omitted. In the SM, one simply finds $V_{L,R}=S_{L,R}=T_L=0$. The delta function $\delta_{\tau\ell}$ in Eq.~(\ref{eq:Heff}) implies that NP effects are supposed to appear in the tau mode only, as suggested by current experimental data.  We have assumed also that neutrinos are left-handed. Using experimental data regarding the $b\to c\ell\nu$ transition, we will obtain constraints on the Wilson coefficients. After that, we will provide predictions for the branching fractions and other polarization observables under the effects of these NP Wilson coefficients. The study therefore provides many predictions to probe NP in future experiments at $B$ factories such as Belle~II and LHCb Collaborations.

The idea of using the semileptonic decays $B_s \to D_s^{(*)}\tau\nu$ to search for NP beyond the SM is quite natural. Therefore, several studies on this topic have been done recently. Using the Ward identity for hadronic matrix elements, Schacht and Soni established relations between differential decay rates of various semileptonic decays induced by the quark-level transition $b\to c\tau\nu_\tau$ that hold in the scalar scenarios~\cite{Schacht:2020qot}. These relations efficiently help in probing the contribution from other operators. In Ref.~\cite{Dutta:2018jxz}, the authors considered the signature of LFU violation in the decay $B_s \to D_s\tau\nu$ using a similar effective Hamiltonian approach. The hadronic form factors (FFs) used in this analysis were calculated in lattice QCD (LQCD) by the HPQCD Collaboration~\cite{Monahan:2017uby}. However, the tensor operator was omitted due to the lack of the corresponding tensor-current FFs. One should also mention the work of the HPQCD Collaboration~\cite{Harrison:2021tol} in which the vector and axial-vector FFs of the transition $B_s \to D_s^*$ were calculated in LQCD. A discussion of NP effects was also provided. However, they did not include scalar and tensor four-fermion operators. Regarding physical observables, they only considered the branching fractions, the forward-backward asymmetry, the longitudinal polarization fraction of the final $D_s^*$ meson, and the lepton polarization asymmetry. Using these HPQCD FFs~\cite{Harrison:2021tol}, the authors of Refs.~\cite{Das:2019cpt, Das:2021lws} analyzed NP effects in the decay $B_s \to D_s^*\tau\nu$ taking into account also the scalar and tensor operators. However, the relevant tensor FFs in this study were obtained by using equations of motion, rather than direct evaluation. Recently, the HPQCD Collaboration finally provided their LQCD calculation of the tensor FFs~\cite{Harrison:2023dzh}. Using SM FFs and the new tensor FF provided by the HPQCD Collaboration~\cite{McLean:2019qcx, Harrison:2023dzh}, the authors of Ref.~\cite{Penalva:2023snz} applied the heavy quark effective theory (HQET) to obtain other (pseudo)scalar and tensor FFs. They then analyzed NP effects in the decays $B_s \to D_s^{(*)}\tau\nu$ with a particular focus on the final tau polarization. Recently, a binwise study of NP vector couplings in the decays $B_s \to D_s^{(*)}\tau\nu$ was conducted~\cite{Yadav:2024zrn}. This study also makes use of the form factors calculated in Refs.~\cite{McLean:2019qcx, Harrison:2023dzh}. Apart from the model-independent approach mentioned above, there is also the model-dependent approach to the investigation of NP in the $B_s \to D_s^{(*)}\tau\nu$ decay. In the latter, one employs a specific model for the implementation of NP such as two-Higgs doublet models and leptoquark models~\cite{Crivellin:2012ye, Wang:2021zfp, Sahoo:2020wnk, Wang:2020kov, Crivellin:2023sig}. 

There are three key elements in the NP analyses mentioned above: (i) the hadronic FFs, (ii) the experimental constraints, and (iii) the set of physical observables to be affected by NP operators. Despite the fact that several similar analyses have been carried out, our study  contributes new insights into the topics since we calculate all the FFs (including NP ones) directly using our model (the covariant confined quark model, CCQM), we use the most up-to-date experimental data, and we consider the full set of observables in a systematic manner.

Within the SM, hadronic FFs for the $B_s \to D_s^{(*)}\ell\nu$ transitions [corresponding to the operator $\mathcal{O}_{V_L}$ in the effective Hamiltonian in Eq.~(\ref{eq:Heff})] have been evaluated in various approaches including the framework of LQCD~\cite{Bailey:2012rr, Monahan:2017uby, McLean:2019qcx, Harrison:2017fmw, Harrison:2021tol}, QCD sum rules (QCDSR)~\cite{Blasi:1993fi, Azizi:2008tt, Zhang:2021wnv}, light-cone sum rules~\cite{Li:2009wq, Bordone:2019guc, Cui:2023jiw, Zhang:2025tlr}, constituent quark model~\cite{Zhao:2006at}, covariant light-front quark model~\cite{Cheng:2003sm, Verma:2011yw, S:2025uej}, Bethe-Salpeter equation~\cite{Chen:2011ut}, relativistic quark models~\cite{Faustov:2012mt, Patnaik:2025fry}, perturbative QCD~\cite{Hu:2019bdf},  and the CCQM~\cite{Soni:2021fky, Pandya:2023ldv, Dubnicka:2025feg}. For a full analysis of NP operators given in Eq.~(\ref{eq:Heff}), one needs also the scalar and tensor FFs, which are rare in literature. To our knowledge, there is only one direct calculation of such FFs given recently by the HPQCD Collaboration~\cite{Harrison:2023dzh}. Therefore, the calculation of scalar and tensor FFs for the $B_s \to D_s^{(*)}\ell\nu$ decays using other approaches is necessary.

In this study, all the FFs (in and beyond the SM) are directly calculated in the framework of the covariant confined quark model with embedded infrared confinement, which has been developed in some earlier papers by us [see Refs.~\cite{Efimov:1988yd, Efimov:zg, Branz:2009cd,Ivanov:2011aa} and references therein]. In our model, all form factors are calculated in the full physical range of momentum transfer squared $q^2$ with no extrapolations, making the predictions for physical observables more reliable. Moreover, to obtain the allowed regions for the NP Wilson coefficients, one has to calculate other decay processes and compare with experimental data. These processes usually include the semileptonic decays $B \to D^{(*)}\ell\nu$ and $B_c \to J/\psi\ell\nu$, and the leptonic decay $B_c \to \ell\nu$. Therefore, to investigate NP in $B_s \to D_s^{(*)}\tau\nu$, one has to calculate also FFs for $B \to D^{(*)}$ and $B_c \to J/\psi$ semileptonic transitions, as well as the leptonic decay constant for the $B_c$ meson~\cite{Gershtein:1976mv, Khlopov:1978id}. Most of the NP analyses use FFs obtained in LQCD as well as other approaches, and rely on the HQET~\cite{Neubert:1993mb, Grozin:2004yc} to evaluate the required hadronic FFs, which are expressed through a few universal functions in the heavy quark limit. Regarding NP analyses in $B_s \to D_s^{(*)}\tau\nu$, our study is unique in the sense that we calculate all FFs, including the ones for $B \to D^{(*)}$ and $B_c \to J/\psi$, and the ones for $B_s \to D_s^{(*)}$ in one framework without relying on HQET or other calculations of FFs. This allows us to investigate NP operators in a self-consistent manner, and independently of the HQET.     

Finally, regarding the physical observables in the semileptonic decays $B_s \to D_s^{(*)}\tau\nu$, existing NP studies usually consider only several main quantities. To provide the full set of observables to be tested in future experiments, we consider the fourfold full angular distribution of the cascade decay $B_s \to D_s^*(\to D_s\pi)\tau\nu$ and define relevant observables based on the coefficient functions of the distribution. We also consider the longitudinal, transverse, and normal polarizations of the final tau lepton as promising NP probes. Inspired by the recent measurement of coefficient functions of the full angular distribution in the decay $B \to D^*(\to D\pi)\ell\nu$ by the Belle Collaboration~\cite{Belle:2023xgj}, we also consider NP effects on these functions. 


The rest of the paper is organized as follows. In Sec.~\ref{sec:model}, we briefly introduce the main features of the CCQM. In Sec.~\ref{sec:FF}, we parametrize the hadronic matrix elements in terms of the invariant form factors, and present our result for the form factors in the whole $q^2$ range. In Sec.~\ref{sec:constraint}, we obtain constraints on the NP Wilson coefficients from available experimental data. Theoretical predictions for the physical observables in the SM and beyond are presented in Sec.~\ref{sec:prediction}. Finally, we briefly conclude in Sec.~\ref{sec:summary}.

\section{THE COVARIANT CONFINED QUARK MODEL IN A NUTSHELL}
\label{sec:model}
The FFs required in this study are calculated in the framework of our model, the CCQM~\cite{Efimov:1988yd, Efimov:zg, Branz:2009cd,Ivanov:2011aa}. The introduction of the model as well as its applications have been given in many previous papers by our group. For instance, a detailed description of the model calculation for similar semileptonic transitions $\bar{B}^0 \to D^{(\ast)} \tau \nu$ and $B_c\to J/\psi \tau\nu$ can be found in Refs.~\cite{Ivanov:2015tru, Ivanov:2016qtw, Tran:2018kuv}. To keep the main text short and focus more on the new results, we list here several key features of the CCQM and refer the reader to the references above for more details.

The CCQM is an effective quantum field theory approach to hadron physics. The starting point of the model  is a Lagrangian describing the quark-hadron interaction with a nonlocal characteristic, which, in the case of a meson $M$, has the form
\begin{equation}
	\mathcal{L}_{\mathrm{int}}(x) =g_M M(x)J(x)+\mathrm{H.c.},\qquad
	J(x) = \int dx_1\int dx_2 F_M(x;x_1,x_2)[\bar{q}_2(x_2)\Gamma_M q_1(x_1)].
\end{equation}  
Here, $g_M$ is the meson-quark coupling constant, $\Gamma_M$ is the relevant Dirac matrix, and $F_M(x;x_1,x_2)$ is the vertex function whose form is chosen as
\begin{equation}
	F_M(x;x_1,x_2) = \delta^{(4)}(x-\omega_1 x_1-\omega_2 x_2)\Phi_M[(x_1-x_2)^2],
\end{equation} 
where $\omega_{i}=m_{q_i}/(m_{q_1}+m_{q_2})$ and $m_{q_i}$ is mass of the constituent quark $q_i$. The chosen form is invariant under translational transformations, i.e., $F_M(x+a;x_1+a,x_2+a)=F_M(x;x_1,x_2)$, for any given 4-vector $a$. 

The function $\Phi_M[(x_1-x_2)^2]$ can be calculated, in principle, using the Bethe-Salpeter equation for meson bound states. However, we have observed that hadronic observables are insensitive to the concrete functional form of the quark-hadron vertex function~\cite{Anikin:1995cf, Ivanov:1997ug, Faessler:2003yf}. The essential requirement for this function is that its Fourier transform, denoted by $\widetilde{\Phi}_M(-k^2)$, must fall off sufficiently fast in the Euclidean space to render ultraviolet finites in Feynman diagrams. Therefore, we adopt the following Gaussian form for calculational convenience:
\begin{equation}
	\widetilde{\Phi}_M(-p^2)=\exp(p^2/\Lambda^2_M),
	\label{eq:vertexf}
\end{equation}
where $\Lambda_M$ is a parameter of the model. 

\begin{figure}[ht]
	\includegraphics[width=0.40\textwidth]{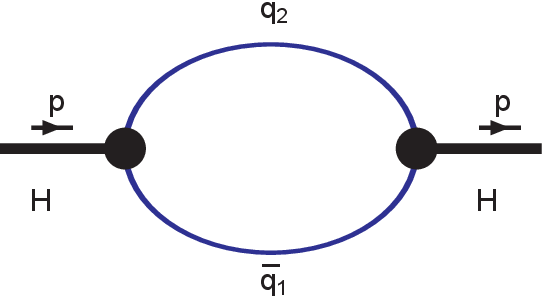}
	\caption{One-loop self-energy diagram for a meson.}
	\label{fig:mass}
\end{figure}

The coupling $g_H$ is determined by using the so-called compositeness condition~\cite{Salam:1962ap, Weinberg:1962hj, Hayashi:1967bjx}, which imposes that the wave function renormalization constant of the hadron is equal to zero $Z_H=0$. A more detailed description of how the compositeness condition works can be found in our recent study~\cite{Tran:2023hrn}. For mesons, the condition has the form
$Z_H=1-\Pi'_H (m^2_H)=0$, where $\Pi'_H (m^2_H)$ is the derivative of the hadron mass operator, which corresponds to the self-energy diagram in Fig.~\ref{fig:mass}
and has the following form:
\bea
\Pi_P(p^2) &=& 3g_H^2 \int\!\! \frac{dk}{(2\pi)^4i}
\widetilde\Phi^2_P \left(-k^2\right)
\Tr\left[ S_1(k+w_1p)\gamma^5 S_2(k-w_2p)\gamma^5 \right],
\nn
\Pi_V(p^2) &=& g_H^2 \Big(g^{\mu\nu} - \frac{p^{\mu}p^{\nu}}{p^2}\Big) 
\int\!\! \frac{dk}{(2\pi)^4i}\widetilde\Phi^2_V \left(-k^2\right)
\Tr\left[ S_1(k+w_1p)\gamma_{\mu} S_2(k-w_2p)\gamma_{\nu} \right],
\ena
for pseudoscalar and vector mesons, respectively. Here, $S_{1,2}$ are quark propagators, for which we use the Fock-Schwinger representation
\be
S_i(k) = (m_{q_i}+\not\! k)
\int\limits_0^\infty\! d\alpha_i \exp[-\alpha_i(m_{q_i}^2-k^2)].
\en
It should be noted that all loop integrations  are carried out in Euclidean space.

Similarly to the hadron mass operator, matrix elements of hadronic transitions are represented by quark-loop diagrams, which are described as convolutions of the corresponding quark propagators and vertex functions. Using various techniques described in our previous papers, any hadronic matrix element $\Pi$ can be finally written in the form
$\Pi =  \int\limits_0^\infty\! d^n \alpha\, F(\alpha_1,\ldots,\alpha_n)$, where $F$
is the resulting integrand corresponding to a given diagram. It is more convenient to turn the set of Fock-Schwinger parameters into a simplex by adding the integral
$1=\int\limits_0^\infty\!dt\,\delta\big(t-\sum\limits_{i = 1}^2 {{\alpha _i}}\big)$
as follows:
\be
\Pi   = \int\limits_0^\infty\! dt t^{n-1} \int\limits_0^1\! d^n \alpha  
\delta\Big(1 -  \sum\limits_{i=1}^n \alpha_i \Big)  
F(t\alpha_1,\ldots,t\alpha_n). 
\label{eq:Pi}
\en
The integral in Eq.~(\ref{eq:Pi}) begins to diverge when $t\to\infty$, if the kinematic variables allow the appearance of branching point corresponding to the creation of free quarks. However, these possible threshold singularities disappear if one cuts off the integral at the upper limit:
\be
\Pi^{\rm c}   = \int\limits_0^{1/\lambda^2}\! dt t^{n-1} 
\int\limits_0^1\! d^n \alpha  
\delta\Big(1 -  \sum\limits_{i=1}^n \alpha_i \Big) 
F(t\alpha_1,\ldots,t\alpha_n). 
\en
The parameter $\lambda$ effectively guarantees the confinement of quarks inside a hadron and is called the infrared cutoff parameter.

The CCQM has several free parameters including the constituent quark masses $m_q$, the hadron size parameters $\Lambda_H$, and a universal cutoff parameter $\lambda$ which guarantees the confinement of constituent quarks inside hadrons. These parameters are obtained by fitting to available experimental data and/or LQCD. Once they are fixed, the CCQM can be used to calculate hadronic quantities in a straightforward manner. The parameters relevant to this study are collected in Tables~\ref{tab:para1} and~\ref{tab:para2}.
\begin{table}[ht]
	\caption{Quark masses and infrared cutoff parameter (all in GeV).}\label{tab:para1}
	\renewcommand{\arraystretch}{0.8}
	\begin{ruledtabular}
		\begin{tabular}{ccccc}
			$m_{u/d}$ & $m_s$ &  $m_c$ &  $m_b$ &  $\lambda$ \\
			\hline
			0.241 & 0.428 & 1.672 & 5.046 & 0.181
		\end{tabular}
	\end{ruledtabular}
\end{table}
\begin{table}[ht]
	\caption{Meson size parameters (all in GeV).}\label{tab:para2}
	\renewcommand{\arraystretch}{0.9}
	\begin{ruledtabular}
		\begin{tabular}{cccccccccc}
			$\Lambda_D$ & $\Lambda_{D_s}$ &  $\Lambda_B$ & $\Lambda_{B_s}$ &	$\Lambda_{B_c}$ &  $\Lambda_{D^*}$ & $\Lambda_{D_s^*}$ &  $\Lambda_{B^*}$ & $\Lambda_{B_s^*}$ & $\Lambda_{J/\psi}$  \\
			\hline
			1.600 & 1.748 & 1.963 & 2.050 & 2.728 & 1.529 & 1.558 & 1.805 & 1.794 & 1.738
		\end{tabular}
	\end{ruledtabular}
\end{table}

Finally, we briefly discuss some estimates of the theoretical errors within our model.
The CCQM consists of several free parameters: the constituent quark masses $m_q$, the hadron size parameters $\Lambda_H$, and the universal infrared cutoff parameter $\lambda$. These parameters are determined by
minimizing the functional
$\chi^2 = \sum\limits_i\frac{(y_i^{\rm expt}-y_i^{\rm theor})^2}{\sigma^2_i}$,
where $\sigma_i$ is the experimental  uncertainty. We use the weak leptonic decay constants of mesons and their radiative decay widths to fit the parameters [see Refs.~\cite{Branz:2009cd,Ivanov:2011aa}].
The fitted values of the decay constants and widths deviate from the experimental data by less than 5$\%$. Furthermore, we have observed that our predictions for the branching fraction of various decay channels most often agree with experimental data within the range of about 5--10$\%$. We therefore estimate the theoretical error of the CCQM to be of the order of 10$\%$. It is worth mentioning that the search for possible NP requires highly precise theoretical predictions, such as those obtained in LQCD. In this regard, our predictions suffer from larger uncertainty compared to those available within LQCD. However, the purpose of this study is to analyze all possible effects of NP operators on the full set of physical observables and to provide a roadmap toward the disentangling of different NP scenarios. Therefore, albeit the weakness of the model in treating the theoretical error, the study can well serve as a ``first-skimming" of possible NP effects in the decays $B_s \to D_s^{(*)}\tau\nu$ and can provide useful insights. 


\section{HADRONIC MATRIX ELEMENTS AND FORM FACTORS}
\label{sec:FF}
Starting with the general effective Hamiltonian given in Eq.~(\ref{eq:Heff}), the matrix element of the semileptonic decays $B_s \to D_s^{(*)} \tau\bar{\nu}_\tau$ is written as
\begin{eqnarray}
\mathcal{M}&=&
\frac{G_FV_{cb}}{\sqrt{2}}\Big[
(1+V_R+V_L)\langle D_s^{(\ast)}|\bar{c}\gamma^\mu b|B_s\rangle 
\bar{\tau}\gamma_\mu(1-\gamma^5)\nu_\tau\nn
&&
+(V_R-V_L)\langle D_s^{(\ast)}|\bar{c}\gamma^\mu\gamma^5b|B_s\rangle
\bar{\tau}\gamma_\mu(1-\gamma^5)\nu_\tau\nn
&&+(S_R+S_L)\langle D_s^{(\ast)}|\bar{c}b|B_s\rangle
\bar{\tau}(1-\gamma^5)\nu_\tau\nn
&&+(S_R-S_L)\langle D_s^{(\ast)}|\bar{c}\gamma^5 b|B_s\rangle 
\bar{\tau}(1-\gamma^5)\nu_\tau\nn
&&+T_L\langle D_s^{(\ast)}|\bar{c}\sigma^{\mu\nu}(1-\gamma^5)b|B_s\rangle
\bar{\tau}\sigma_{\mu\nu}(1-\gamma^5)\nu_\tau\Big].
\label{eq:amplitude-full}
\end{eqnarray}
Note that the axial and pseudoscalar hadronic currents do not contribute to 
$P\to P^\prime$ transitions, while the scalar hadronic current does not 
contribute to $P\to V$ ones. As a result, NP scenarios corresponding to pure $V_R-V_L$ or $S_R\pm S_L$ couplings are ruled out, since we assume that NP appears simultaneously in $P\to P^\prime$ and $P\to V$ cases.

The hadronic matrix elements in Eq.~(\ref{eq:amplitude-full}) are parametrized by a set of invariant FFs depending on the momentum transfer squared $q^2$ between the initial and final mesons as follows:
\begin{eqnarray}
\langle D_s(p_2)
|\bar{c} \gamma^\mu b
| B_s(p_1) \rangle
&=& F_+(q^2) P^\mu + F_-(q^2) q^\mu \equiv T_1^\mu,\nn
\langle D_s(p_2)
|\bar{c}b
| B_s(p_1) \rangle &=& (m_1+m_2)F^S(q^2) \equiv T_2,\nn
\langle D_s(p_2)|\bar{c}\sigma^{\mu\nu}(1-\gamma^5)b|B_s(p_1)\rangle 
&=&\frac{iF^T(q^2)}{m_1+m_2}\left(P^\mu q^\nu - P^\nu q^\mu 
+i \varepsilon^{\mu\nu Pq}\right) \equiv T_3^{\mu\nu},\nn
\langle D_s^*(p_2)
|\bar{c} \gamma^\mu(1\mp\gamma^5)b
| B_s(p_1) \rangle
&=& \frac{\epsilon^{\dagger}_{2\alpha}}{m_1+m_2}
\Big[ \mp g^{\mu\alpha}PqA_0(q^2) \pm P^{\mu}P^{\alpha}A_+(q^2)\nn
&&\pm q^{\mu}P^\alpha A_-(q^2) 
+ i\varepsilon^{\mu\alpha P q}V(q^2)\Big] \equiv \epsilon^\dagger_{2\alpha} \mathcal{T}_{1L(R)}^{\mu\alpha},
\nn
\langle D_s^*(p_2)
|\bar{c}\gamma^5 b
| B_s(p_1) \rangle &=& \epsilon^\dagger_{2\alpha}P^\alpha G^P(q^2) \equiv \epsilon^\dagger_{2\alpha}\mathcal{T}_2^\alpha,
\nn
\langle D_s^*(p_2)|\bar{c}\sigma^{\mu\nu}(1-\gamma^5)b|B_s(p_1)\rangle
&=&-i\epsilon^\dagger_{2\alpha}\Big[
\left(P^\mu g^{\nu\alpha} - P^\nu g^{\mu\alpha} 
+i \varepsilon^{P\mu\nu\alpha}\right)G_1^T(q^2)\nn
&&+\left(q^\mu g^{\nu\alpha} - q^\nu g^{\mu\alpha}
+i \varepsilon^{q\mu\nu\alpha}\right)G_2^T(q^2)\nn
&&+\left(P^\mu q^\nu - P^\nu q^\mu 
+ i\varepsilon^{Pq\mu\nu}\right)P^\alpha\frac{G_0^T(q^2)}{(m_1+m_2)^2}
\Big] \equiv \epsilon^\dagger_{2\alpha}\mathcal{T}_3^{\mu\nu\alpha},\nn
\label{eq:ff}
\end{eqnarray}
where $P=p_1+p_2$, $q=p_1-p_2$, and $\epsilon_2$ is the $D_s^\ast$ polarization vector which satisfies the condition $\epsilon_2^\dagger\cdot p_2=0$. The mesons are on-shell: $p_1^2=m_1^2=m_{B_s}^2$ and
$p_2^2=m_2^2=m_{D_s^{(\ast)}}^2$. We have adopted the shorthand notation $\varepsilon^{pqrs}=\varepsilon^{\mu\nu\alpha\beta}p_\mu q_\nu r_\alpha s_\beta$ for any 4-vectors $p$, $q$, $r$, and $s$. 

\begin{figure}[htbp]
	\begin{tabular}{lr}
		\includegraphics[scale=0.8]{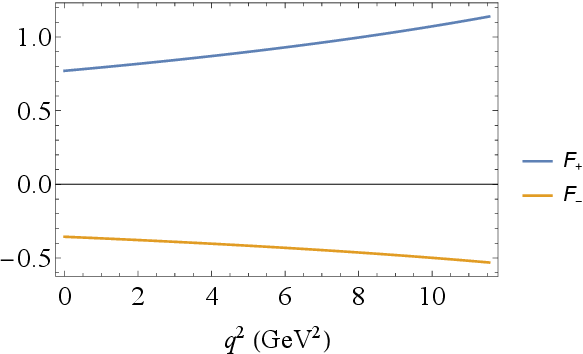}
		& 
		\includegraphics[scale=0.8]{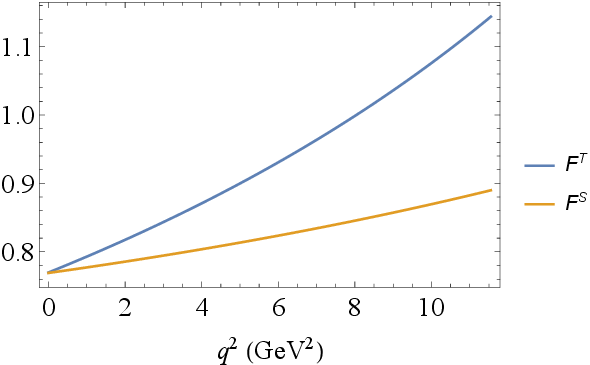}\\
		\includegraphics[scale=0.8]{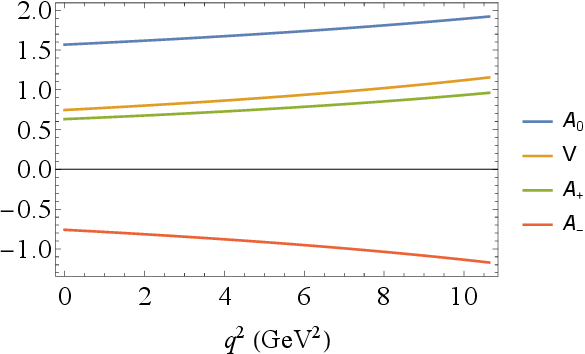}
		& 
		\includegraphics[scale=0.8]{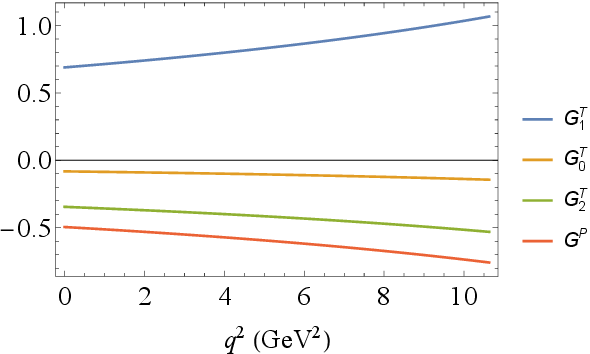}
	\end{tabular}
	\caption{Form factors of the transitions $B_s\to D_s$ (upper panels)
		and $B_s \to D_s^*$ (lower panels). Note that $F^T(0)\approx F^S(0)$ is only a coincidence.}
	\label{fig:formfactor}
\end{figure}

The $B_s\to D_s^{(*)}$ hadronic transitions are calculated from their one-loop quark diagrams. For a more detailed description of the calculation techniques, we refer to Ref.~\cite{Ivanov:2016qtw} where we computed similar form factors for the $\bar{B}^0\to D^{(\ast)}$ transitions. In the framework of the CCQM, the interested form factors  are represented by threefold integrals
which are calculated by using \textsc{fortran} codes in the full kinematical momentum 
transfer region $0\le q^2 \le q^2_{max}=(m_{B_s}-m_{D_s^{(*)}})^2$. The numerical results for the form factors are well approximated by a double-pole parametrization:
\be
F(q^2)=\frac{F(0)}{1 - a s + b s^2}, \quad s=\frac{q^2}{m_{B_s}^2}. 
\label{eq:ff-para}
\en
The parameters of the $B_s \to D_s^{(*)}$ form factors  are listed in Table~\ref{tab:ff-param}. Their $q^2$ dependence in the full momentum transfer range $0\le q^2 \le q^2_{max}=(m_{B_s}-m_{D_s^{(*)}})^2$ is shown in Fig.~\ref{fig:formfactor}.
\begin{table}[htbp]
	\caption{Parameters of the dipole approximation in Eq.~(\ref{eq:ff-para}) for  $B_s \to D_s^{(*)}$ form factors. Zero-recoil (or $q^2_{\rm max}$) values of the form factors are also listed.}
\centering
		\begin{tabular}{lccccccccccccc}
			\hline\hline
			\multicolumn{1}{c}{} &\multicolumn{8}{c}{$B_s \to D_s^*$} &\multicolumn{1}{c}{} 
			&\multicolumn{4}{c}{$B_s \to D_s$} \\
			\cline{2-9}\cline{11-14}
			& $ A_0 $ & $  A_+  $ & $  A_-  $ & $  V  $ 
			& $ G^P $ & $G_0^T$ & $G_1^T$ & $  G_2^T$ & {} & $F_+$ & $F_-$ & $F^S$ & $F^T$ 
			\\
			\hline
			$F(0)$ &  1.57 & 0.63  & $-0.76$ & 0.75 & $-0.49$ & $-0.08$ & 0.69 & $-0.35$ & {} &  0.77   & $-0.36$ &  0.77 & 0.77  
			\\
			$a$    &  0.43 & 0.96  &  0.99 & 1.00 & 0.98 & 1.31 & 1.00 & 0.99 & {} &  0.84   &  0.85 &  0.30 & 0.85  
			\\
			$b$    & $-0.18$ & 0.09 & 0.11 & 0.12 & 0.11 & 0.38 & 0.12 & 0.11 & {} &  0.08  & 0.08 & $-0.11$ & 0.08 
			\\ 
			$F(q^2_{\rm max})$ &  1.92 & 0.96  & $-1.17$ & 1.16 & $-0.76$ & $-0.14$ & 1.07 & $-0.53$ & {} &  1.14  & $-0.53$ & 0.89 & 1.14  
			\\
			\hline\hline
		\end{tabular}
		\label{tab:ff-param}
\end{table}

We compare our form factor results with the LQCD predictions provided in Refs.~\cite{Harrison:2023dzh, McLean:2019qcx}. In Ref.~\cite{McLean:2019qcx}, the authors calculate the form factors $F_+(q^2)$ and $F_0(q^2)$, the latter of which is related to our FFs via the relation $F_0(q^2)=F_+(q^2)+q^2 F_-(q^2)/(m_1^2-m_2^2)$. In Ref.~\cite{Harrison:2023dzh}, the HQET-basis FFs $h_Y(w)$ are used, where $w=(m_1^2+m_2^2-q^2)/(2m_1m_2)$. These FFs are translated to our FF definitions using the relations
\begin{eqnarray}
	A_0 &=& \frac{(m_1+m_2)^2-q^2}{2\sqrt{m_1 m_2}(m_1-m_2)} h_{A_1},\qquad
	V = \frac{m_1+m_2}{2\sqrt{m_1m_2}}h_V,\nn
	A_{\pm} &=& \frac{m_1+m_2}{2}\sqrt{\frac{m_2}{m_1}}\left(\frac{h_{A_2}}{m_1}\pm \frac{h_{A_3}}{m_2}\right),\nn
	G^T_{1/2} &=& \sqrt{m_1m_2}\left( \frac{h_{T_1}+h_{T_2}}{m_1} \pm \frac{h_{T_1}-h_{T_2}}{m_2} \right),\nn
	G^T_0 &=& \frac{(m_1+m_2)^2}{2m_1\sqrt{m_1m_2}}h_{T_3}.
\end{eqnarray}
\begin{figure}[htbp]
	\begin{tabular}{ccc}
		\includegraphics[width=0.33\textwidth]{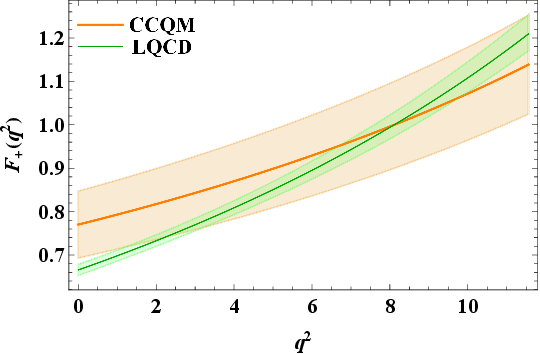}
		& 
		\includegraphics[width=0.33\textwidth]{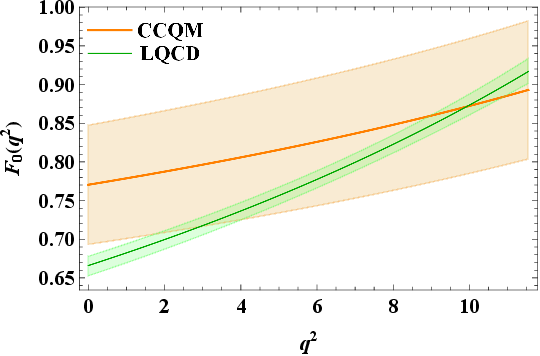}
		&\includegraphics[width=0.33\textwidth]{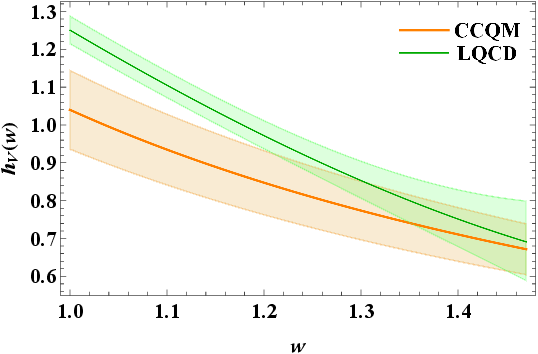}\\
		\includegraphics[width=0.33\textwidth]{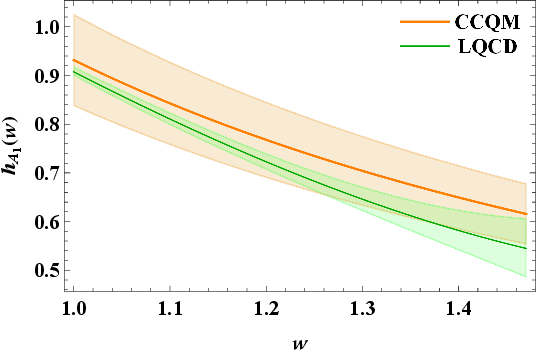}
		& 
		\includegraphics[width=0.33\textwidth]{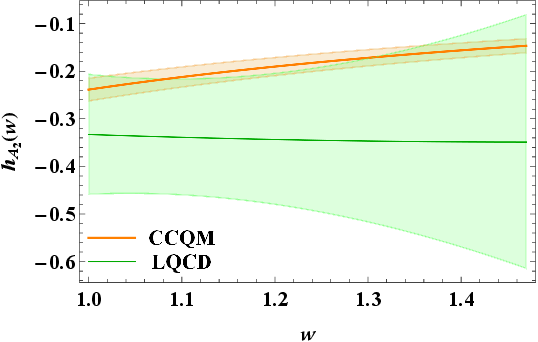}
		& 
		\includegraphics[width=0.33\textwidth]{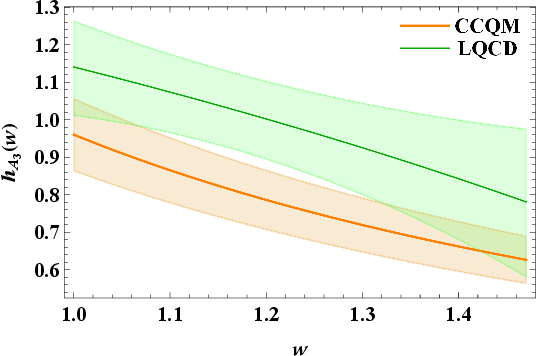}\\
		\includegraphics[width=0.33\textwidth]{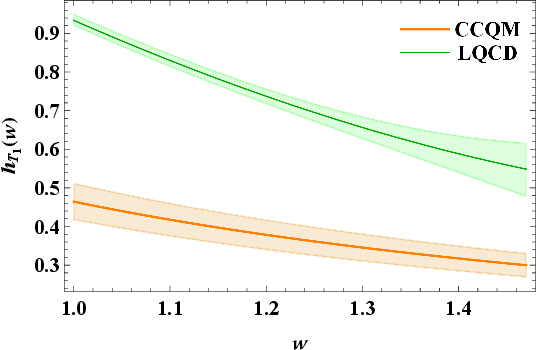}
		& 
		\includegraphics[width=0.33\textwidth]{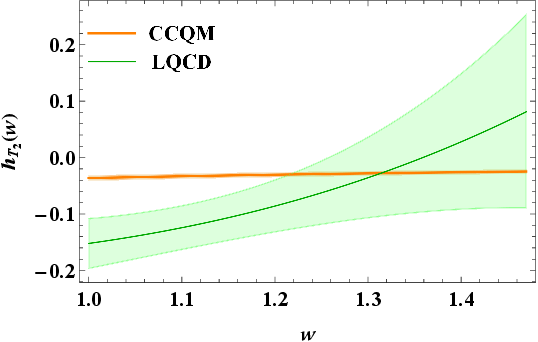}
		& 
		\includegraphics[width=0.33\textwidth]{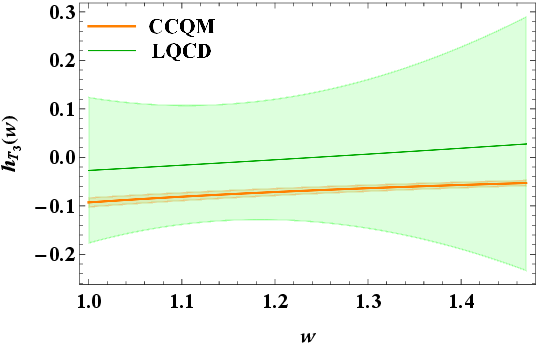}
	\end{tabular}
	\caption{Comparison of form factors in the CCQM and in LQCD~\cite{Harrison:2023dzh, McLean:2019qcx}.}
	\label{fig:CCQM-vs-LQCD}
\end{figure}

As shown in Fig.~\ref{fig:CCQM-vs-LQCD}, our FFs exhibit a generally less steep slope than the LQCD results. Within the SM, our FFs are consistent with LQCD results within the current uncertainty. This consistency is also reflected in the predicted physical observables. Specifically, our predictions for $R(D_s)=0.27(3)$, $R(D^*_s)=0.24(3)$, and $F^\tau_L(D^*_s)=0.46(5)$ are in good agreement with the LQCD values: $R(D_s)=0.2987(46)$~\cite{McLean:2019qcx}, $R(D^*_s)=0.266(9)$, and $F^\tau_L(D^*_s)=0.420(12)$~\cite{Harrison:2023dzh}. Notably, for $R(D^*_s)$, when LQCD FFs are combined with experimental data fits, the ratio yields $R(D^*_s)=0.2459(34)$~\cite{Harrison:2023dzh}, which aligns closely with our central value. Regarding the tensor FFs, our $h_{T_3}$ shows strong agreement with LQCD, while $h_{T_2}$ matches at large $w$ but diverges at small $w$. Finally, $h_{T_1}$ exhibits a significant discrepancy between the two approaches: they differ by approximately a factor of two in the whole $w$ region.

In this paper, we also impose the constraint from the leptonic decay channel of $B_c$ on the Wilson coefficients. Therefore, we present here the leptonic branching in the presence of NP operators. In the SM, the purely leptonic decays $B_c\to \ell \nu$ 
proceed via the annihilation of the quark pair into an off-shell $W$ boson. Assuming the effective Hamiltonian Eq.~(\ref{eq:Heff}), the tau mode of these decays receives NP contributions from all operators except $\mathcal{O}_{T_L}$. The branching fraction of the leptonic decay in the presence of NP is given by~\cite{Ivanov:2017hun}
\be
 \mathcal{B}(B_c \to \tau \nu)=
\frac{G_F^2}{8\pi}|V_{cb}|^2\tau_{B_c}m_{B_c}m_{\tau}^2\left(1-\frac{m_{\tau}^2}{m_{B_c}^2}\right)^2f_{B_c}^2
 \times
\left|
1-g_A+\frac{m_{B_c}}{m_\tau} \frac{f_{B_c}^P}{f_{B_c}}g_P
\right|^2,
\en
where $g_A\equiv V_R-V_L$, $g_P\equiv S_R-S_L$, $\tau_{B_c}$ is the $B_c$ lifetime, $f_{B_c}$ is the leptonic decay constant of $B_c$, and $f_{B_c}^P$ is a new  constant corresponding to the new quark current structure. One has
\be
\langle 0
|\bar{q} \gamma^\mu \gamma_5 b
| B_c(p) \rangle
= -f_{B_c} p^\mu,\qquad
\langle 0
|\bar{q}\gamma_5b
| B_c(p) \rangle = m_{B_c} f_{B_c}^P.
\en
In the CCQM, we obtain the following values for these constants (all in MeV):
\be
f_{B_c}=489\pm 49,\quad f_{B_c}^P=646\pm 65.
\en
\begin{figure}[htbp]
	\centering
	\begin{tabular}{cc}
		\includegraphics[width=0.4\textwidth]{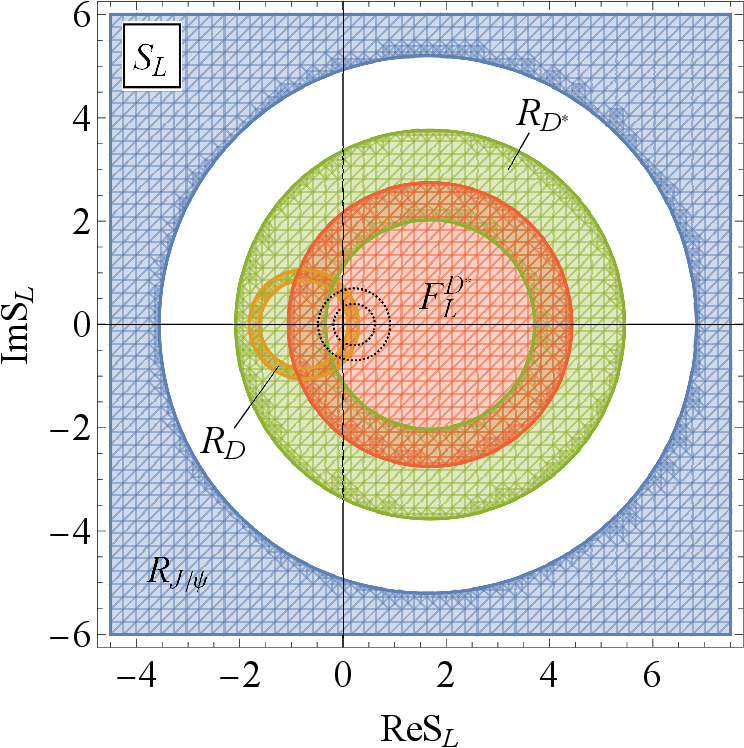}&
		\includegraphics[width=0.4\textwidth]{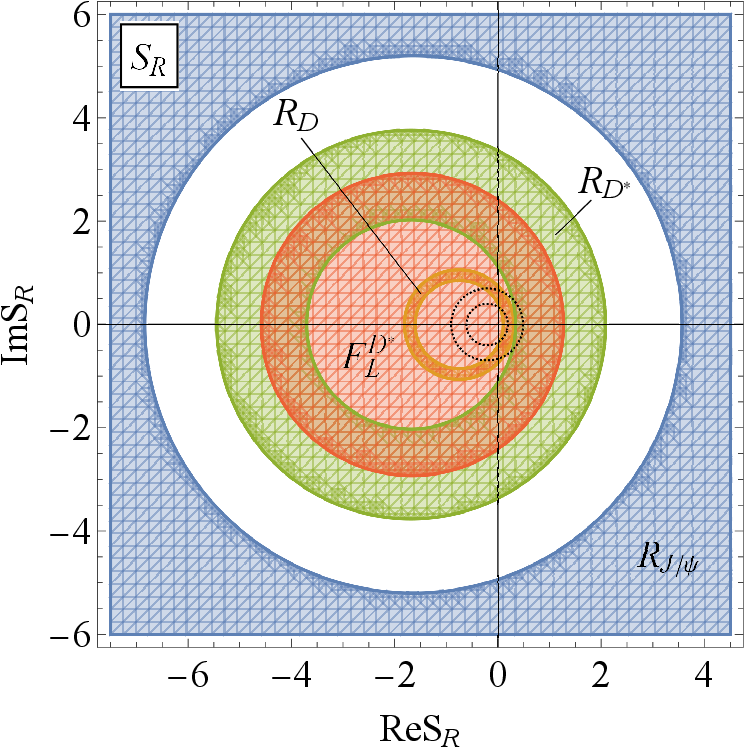}\\
		\includegraphics[width=0.4\textwidth]{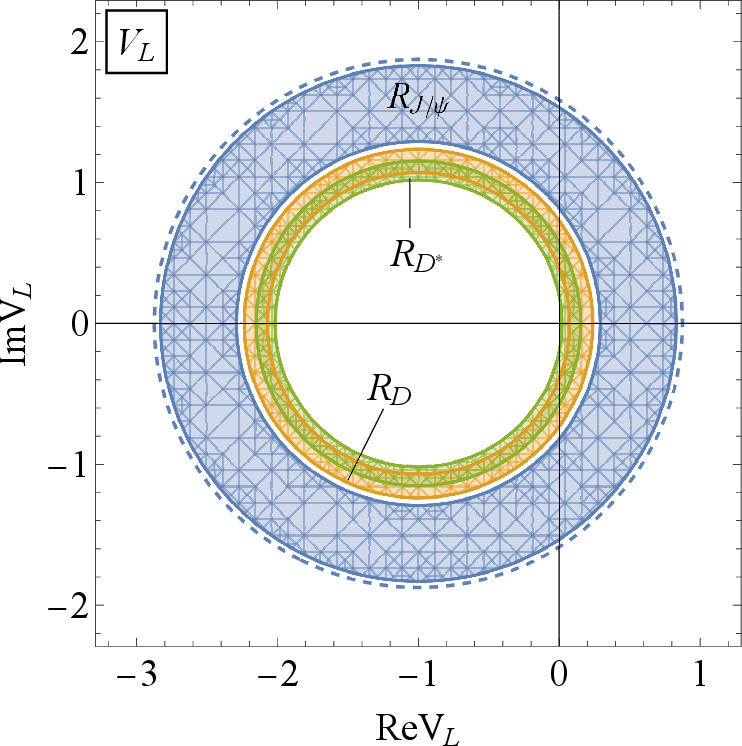}&
		\includegraphics[width=0.4\textwidth]{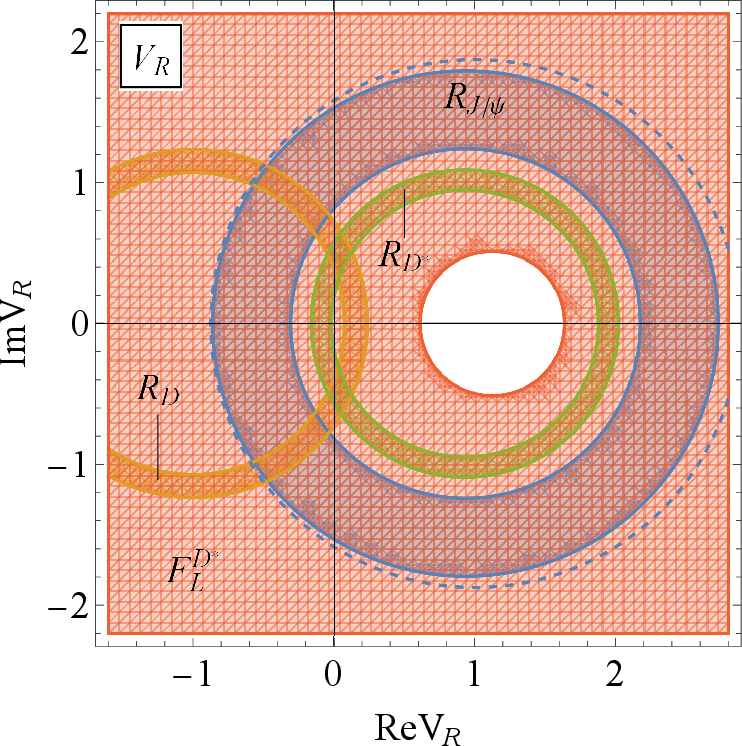}\\
		\includegraphics[width=0.4\textwidth]{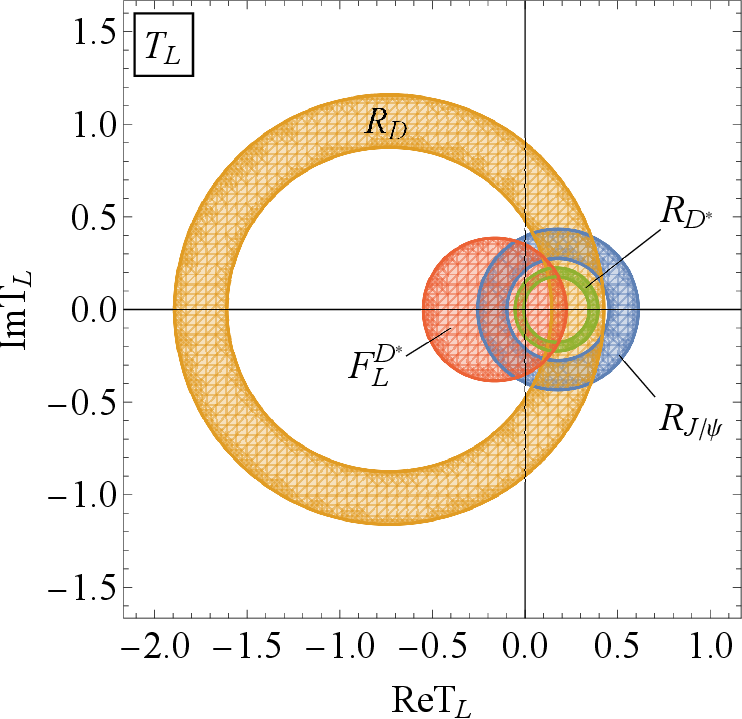}
	\end{tabular}
	\caption{$1\sigma$ constraints on the complex Wilson coefficients from experimental data of $R_D$, $R_{D^*}$, $R_{J/\psi}$, and $F_L^{D^*}$. The dashed curves represent the upper limits $\mathcal{B}(B_c\to\tau\bar{\nu}_\tau)\leq 10\%$ (smaller) and $\mathcal{B}(B_c\to\tau\bar{\nu}_\tau)\leq 30\%$ (larger). For $V_L$ and $V_R$, we show only the limit $\mathcal{B}(B_c\to\tau\bar{\nu}_\tau)\leq 10\%$. For $T_L$, $\mathcal{B}(B_c\to\tau\bar{\nu}_\tau)$ has no effect.}
	\label{fig:1sigma}
\end{figure}
\section{EXPERIMENTAL CONSTRAINTS}
\label{sec:constraint}
Constraints on the Wilson coefficients appearing in the effective Hamiltonian Eq.~(\ref{eq:Heff}) are obtained by using experimental data for the following observables:
\begin{enumerate}[label=\roman*.]
	\item The ratios of branching fractions: $R_D=0.358 \pm 0.024$, $R_{D^\ast}= 0.281 \pm 0.011$ (HFLAV most updated values for CKM 2025~\cite{HFLAV:2022esi}), and  $R_{J/\psi}=0.61\pm 0.18$~\cite{LHCb:2017vlu,CMS:2025jfx}.
	\item The longitudinal polarization fraction of the $D^*$ meson $F_L^{D^*}({B} \to D^{\ast} \tau\bar{\nu}_\tau)=0.41\pm 0.06\pm 0.03$~\cite{LHCb:2023ssl}.
	\item The constraint $\mathcal{B}(B_c\to \tau\nu)\leq 30\%$ derived from the $B_c$ lifetime~\cite{Alonso:2016oyd}.	
\end{enumerate}

A more stringent constraint is the requirement $\mathcal{B}(B_c\to \tau\nu)\leq 10\%$, as discussed in Ref.~\cite{Akeroyd:2017mhr} using LEP1 data. However, to be more conservative, in this paper, we use the constraint from the $B_c$ lifetime ($\mathcal{B}(B_c\to \tau\nu)\leq 30\%$) to obtain the allowed region of the Wilson coefficients. We refer the reader to Ref.~\cite{Blanke:2018yud} for a more detailed discussion on the implications of various choices for the $\mathcal{B}(B_c\to \tau\nu)$ constraint. 

Note that, along with the $q^2$-averaged value $F_L^{D^*}(B \to D^{\ast} \tau\bar{\nu}_\tau)=0.41\pm 0.06\pm 0.03$, the LHCb Collaboration also provided the values of $F_L^{D^*}$ measured in two distinct $q^2$ regions, below and above $7~\text{GeV}^2$, which read $0.52\pm 0.07\pm 0.04$ and $0.34\pm 0.08\pm 0.02$, respectively~\cite{LHCb:2023ssl}. Including these two binned values in the global fit instead of the single averaged value results in minor alterations to the final best-fit points, as well as small changes to the size and shape of the allowed regions for the Wilson coefficients. However, these shifts are minimal and do not modify the main conclusions of this section, nor do they affect the subsequent NP analysis, which forms the primary objective of our study. We refer the reader to Ref.~\cite{Martinelli:2024bov}, where the authors explicitly discriminate the effects of using the $F_L^{D^*}$ values over the full $q^2$ range versus using the two bins ($q^2 < 7\text{ GeV}^2$ and $q^2 > 7\text{ GeV}^2$) when fitting hadronic form factors. While such a bin-by-bin breakdown is highly valuable for scrutinizing the consistency between computed form factors and experimental data, using the $q^2$-averaged value is sufficient for the scope of our present work.

For reference, within the SM, our calculation yields $R_D=0.266 \pm 0.027$, $R_{D^\ast}=0.237\pm 0.024$, $R_{J/\psi}=0.248 \pm 0.025$, and $F_L^{D^*}=0.462\pm 0.046$. We take into account a theoretical error of $10\%$ for our predictions. Furthermore, we assume the dominance of only one NP operator besides the SM contribution, meaning that only one NP Wilson coefficient is turned on at a time.

Figure~\ref{fig:1sigma} shows the $1\sigma$ constraints on the complex Wilson coefficients derived from experimental data of $R_D$, $R_{D^*}$, $R_{J/\psi}$, and $F_L^{D^*}$, as well as from the upper limits of the branching fraction $\mathcal{B}(B_c\to\tau\bar{\nu}_\tau)$. We made the following observations:
\begin{enumerate}
	\item There is no available space for any NP operator within $1\sigma$ when all constraints mentioned above are taken into account. 
	\item The right-handed scalar operator $\mathcal{O}_{S_R}$ is the most strictly constrained. Any combination of two ratios from $R_D$, $R_{D^*}$, and $R_{J/\psi}$ can rule out $\mathcal{O}_{S_R}$. The upper limit of $\mathcal{B}(B_c\to \tau\nu)$ puts a severe additional constraint on this NP scenario. 
	\item The left-handed scalar operator $\mathcal{O}_{S_L}$ can survive the three constraints from $R_D$, $R_{D^*}$, and $F_L^{D^*}$. However, when one of the two constraints, $R_{J/\psi}$ or $\mathcal{B}(B_c\to \tau\nu)\leq 30\%$, is added, $\mathcal{O}_{S_L}$ is excluded. 
	\item The vector operators $\mathcal{O}_{V_{L(R)}}$ and the tensor operator $\mathcal{O}_{T_L}$ are slightly disfavored within $1\sigma$, mainly due to the additional constraint from $R_{J/\psi}$ rather than from $\mathcal{B}(B_c \to \tau\nu)$. This holds exactly in the case of $\mathcal{O}_{T_L}$ since the operator $\mathcal{O}_{T_L}$ has no effect on $\mathcal{B}(B_c \to \tau\nu)$. Note that the current experimental data for $R_{J/\psi}$ still suffers from a large error (about 30\%) compared to the cases of $R_D$ (about 7\%) and $R_{D^*}$ (about 4\%); it is possible that future updates for $R_{J/\psi}$ will allow for these operators within $1\sigma$.   
\end{enumerate}

It is worth discussing the implications of the recent measurement of the longitudinal polarization fraction $F_L^{D^*}$ by the LHCb Collaboration~\cite{LHCb:2023ssl}. Note that the precision of this new measurement is better than the first one provided by the Belle Collaboration, which reported $F_L^{D^*}=0.60\pm 0.08\pm 0.04$~\cite{Belle:2019ewo}. This observable is of great interest due to its sensitivity to all NP operators (except for $\mathcal{O}_{V_L}$)~\cite{Ivanov:2020iad}. In other words, a precise measurement of $F_L^{D^*}$ provides a stringent constraint on these NP operators. Especially, the recently measured $F_L^{D^*}$ by the LHCb Collaboration puts a very strong constraint on the tensor operator. Besides, it is important to recognize that the upper bound on the branching fraction $\mathcal{B}(B_c\to \tau\nu)$ severely restricts models featuring scalar NP operators. Consequently, several specific NP models, including those introducing charged Higgs bosons or leptoquarks, are strongly affected by this limitation and often require significant modification to remain consistent with current experimental observations [see, e.g., Refs.~\cite{Crivellin:2017zlb, Lee:2017kbi, Iguro:2017ysu}]. 

\begin{figure}[htbp]
	\centering
	\begin{tabular}{cc}
		\includegraphics[width=0.4\textwidth]{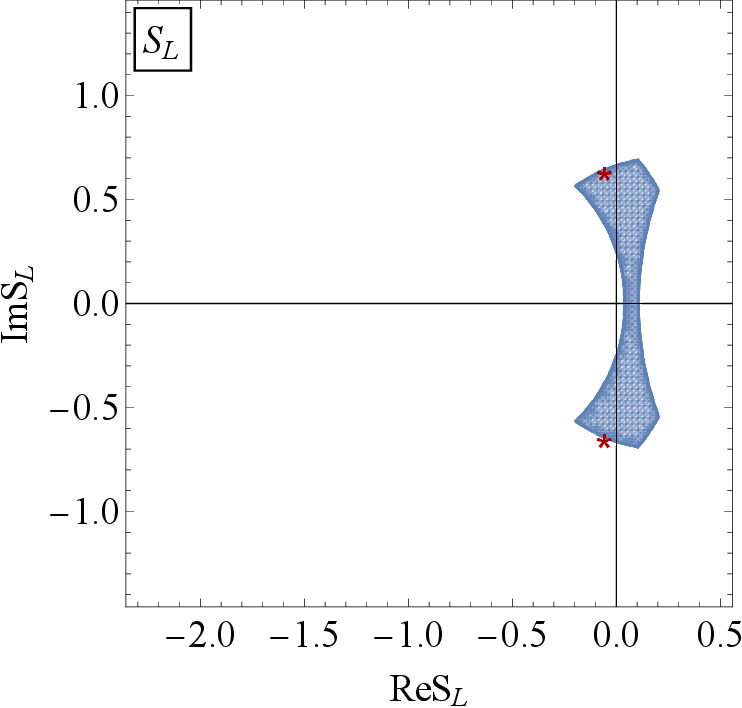}&
		\includegraphics[width=0.4\textwidth]{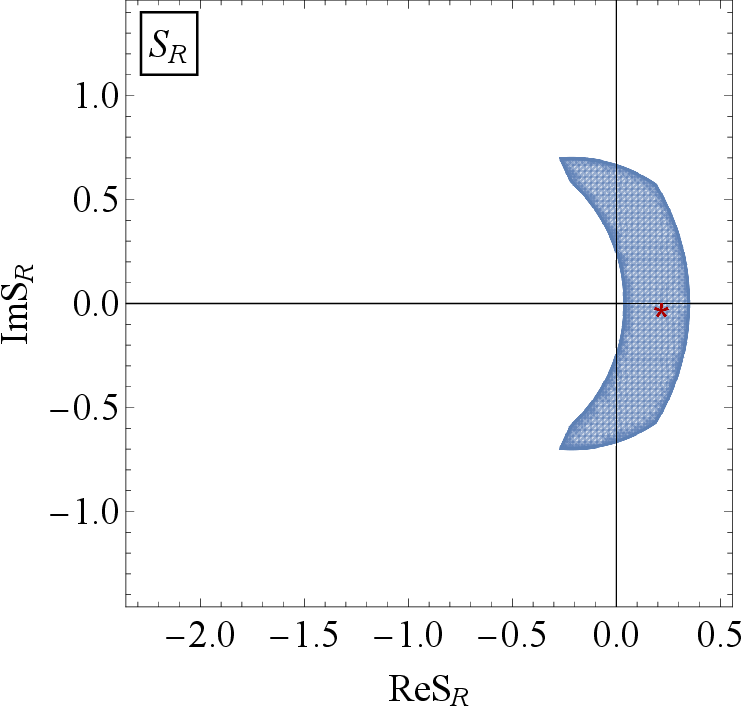}\\
		\includegraphics[width=0.4\textwidth]{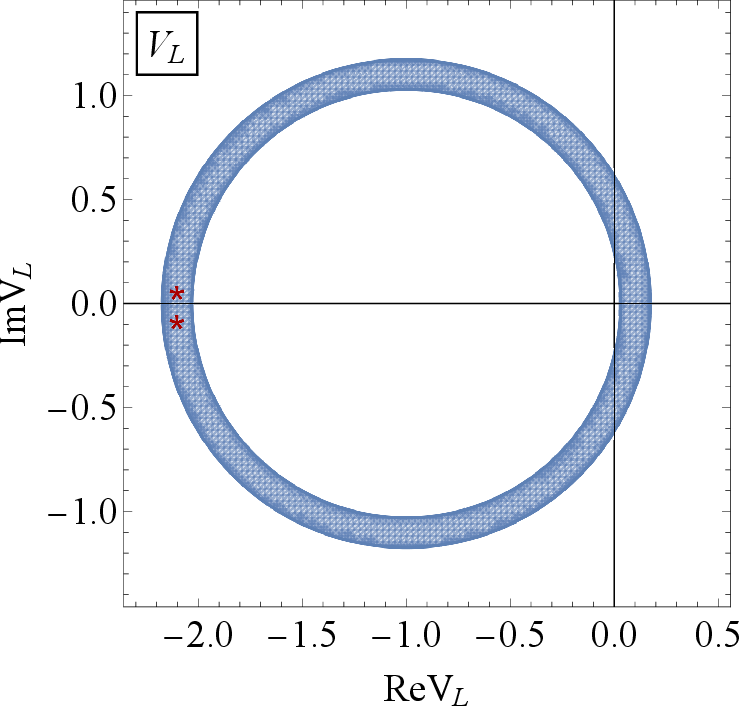}&
		\includegraphics[width=0.4\textwidth]{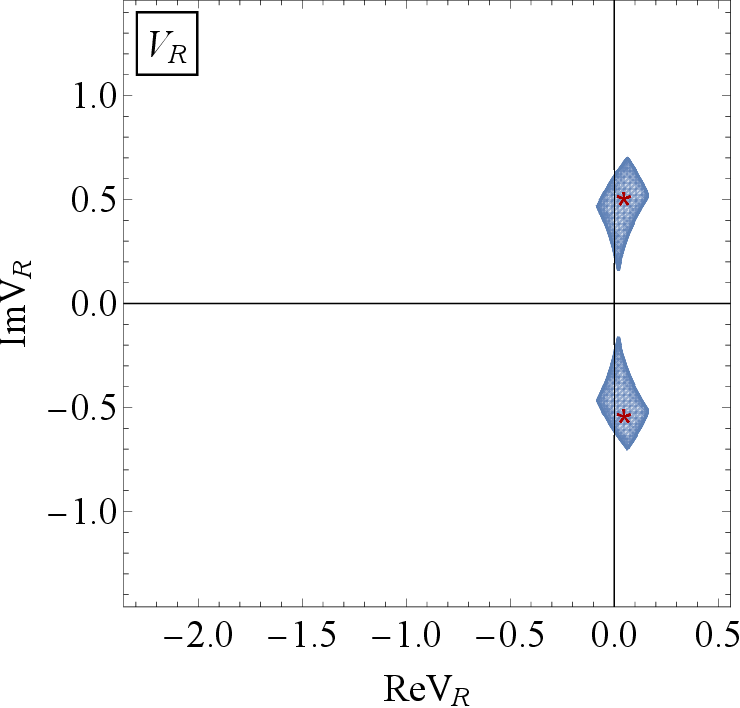}\\
		\includegraphics[width=0.4\textwidth]{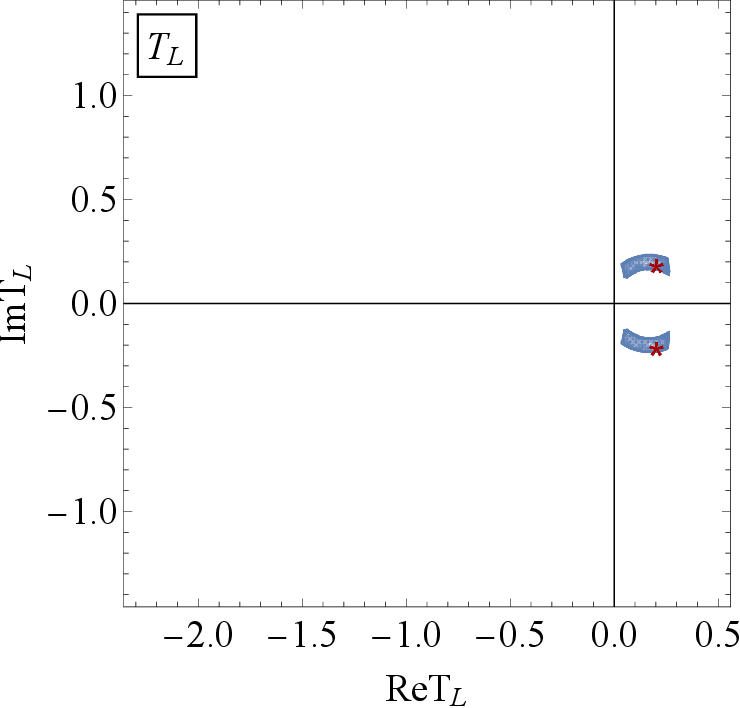}		
	\end{tabular}
	\caption{Allowed regions for the complex Wilson coefficients obtained from experimental data of $R_D$, $R_{D^*}$, $R_{J/\psi}$, and $F_L^{D^*}$ within $2\sigma$, and from the constraint $\mathcal{B}(B_c\to\tau\bar{\nu}_\tau)\leq 30\%$. Best-fit values are indicated by asterisks.}
	\label{fig:2sigma}
\end{figure}

Even though all the NP operators are excluded at $1\sigma$, it is important to recall that we have assumed the dominance of only one operator at a time. Scenarios involving several operators simultaneously, including interference terms, could potentially relax these constraints. Therefore, it is useful to analyze the effects of the NP operators on physical observables in the $B_s\to D_s^{(*)}\tau\nu$ decay, even within the single-operator scenarios. This analysis will provide further insights toward a more definite answer in the future, especially when more precise experimental data become available. To do this, we consider the $2\sigma$ allowed regions of the Wilson coefficients and present predictions for the observables within these regions. In Fig.~\ref{fig:2sigma}, we show the allowed regions for $S_{L,R}$, $V_{L,R}$, and $T_L$ within $2\sigma$. In each region, we identify best-fit values for each NP coupling, which are marked with asterisks. The best-fit couplings read
\begin{eqnarray}
S_L &=&-0.057 \pm i 0.642,\qquad	S_R = 0.217 -i 0.014,\nonumber\\
V_L &=& -2.102\pm  i 0.071,\qquad V_R = 0.047 \pm i 0.523, \qquad T_L = 0.203 \pm i 0.198.
\end{eqnarray}
In the next section, we will use the $2\sigma$ allowed regions and the best-fit values for $S_{L,R}$, $V_{L,R}$, and $T_L$ to analyze the effects of NP operators on the full set of physical observables in the concerned decays.

\section{OBSERVABLE DEFINITION AND THEORETICAL PREDICTIONS}
\label{sec:prediction}

\subsection{The fourfold decay distribution}
\label{subsec:4fold}

\begin{figure}[htbp]
	\begin{tabular}{c}
		\includegraphics[scale=0.5]{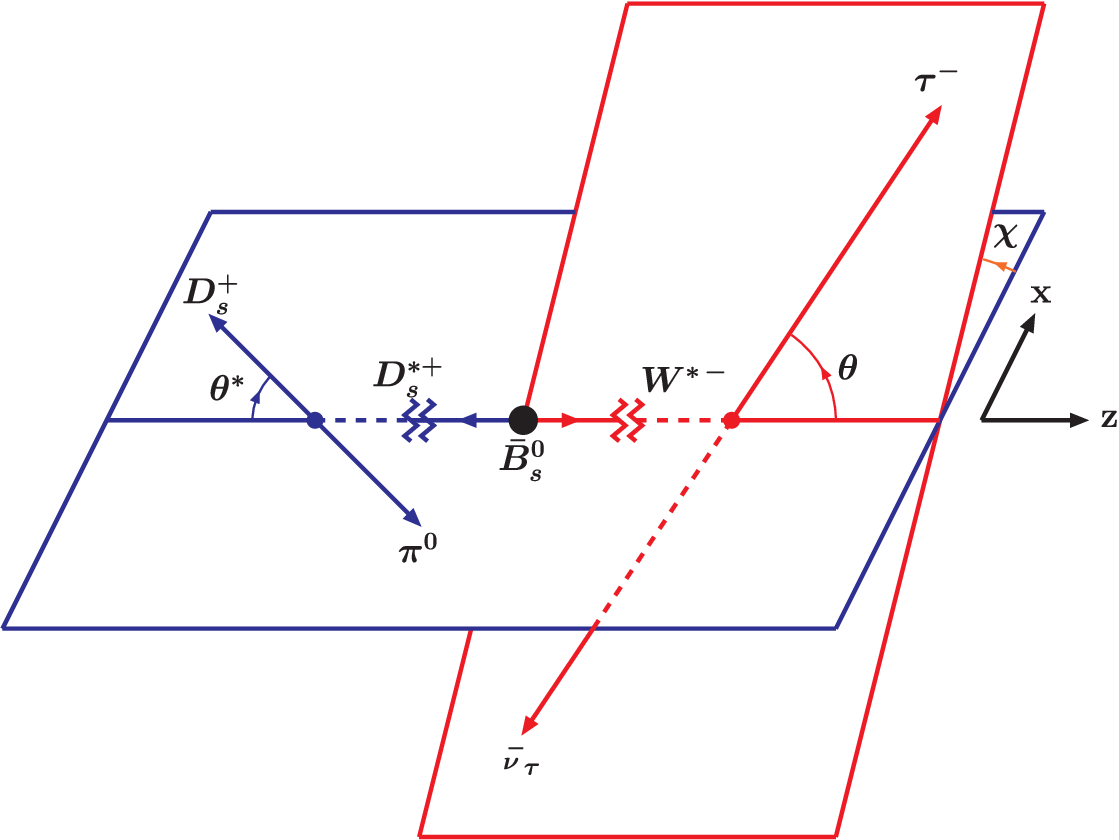}
	\end{tabular}
	\caption{Definition of the angles $\theta$, $\theta^\ast$, and $\chi$ in
		the cascade decay $\bar{B}^0_s\to D^{\ast+}_s(\to D^+_s\pi^0)\tau^-\bar{\nu}_\tau$.}
	\label{fig:angle}
\end{figure}

To obtain the full angular distribution, we consider the cascade decay $\bar{B}^0_s\to D^{\ast+}_s(\to D^+_s\pi^0)\tau^-\bar{\nu}_\tau$. The decay angles are defined as follows: $\theta$ is the angle between the direction of the $\tau^-$ in the $W^{*-}$ rest frame and the direction of $W^{*-}$ in the $B_s$ frame, $\theta^*$ is the angle between the direction of $D_s^+$ in the $D^*_s$ rest frame and the $D^*_s$ in the $B_s$ frame, and $\chi$ is the angle between the two decay planes (see Fig.~\ref{fig:angle}).

In the present work, we consider the cascade decay
$\bar B_s^0 \to D_s^{*+}(\to D_s^+\pi^0)\tau^-\bar\nu_\tau$
in order to formulate the angular observables in close correspondence with the standard angular analysis of
$B \to D^*(\to D\pi)\ell\nu$,
including the recent Belle Collaboration measurement of angular coefficients~\cite{Belle:2023xgj}. Using the pion decay mode allows us to employ the same angular basis and observable definitions for both the $B\to D^*$ and $B_s\to D_s^*$ transitions, thereby facilitating a direct comparison between the two systems.
We note that experimentally the radiative decay mode
$D_s^{*+}\to D_s^+\gamma$
has a significantly larger branching fraction. However, the radiative channel involves a different helicity structure and consequently leads to a different angular distribution and modified angular observables. As discussed in Ref.~\cite{Das:2021lws}, the $D_s\gamma$ and $D_s\pi$ channels must in general be treated as distinct angular systems, with several angular observables receiving significantly different values in the two cases. Therefore, a dedicated treatment of the radiative mode requires a separate derivation of the angular distribution and is beyond the scope of the present analysis.

The fourfold distribution has the form
\begin{eqnarray}
	\frac{d^4\Gamma(\bar{B}^0_s\to D^{\ast+}_s(\to D^+_s\pi^0) \tau^-\bar\nu_\tau)}
	{dq^2 d\cos\theta d\chi d\cos\theta^\ast}
	=\frac{9}{8\pi}|N|^2J(\theta,\theta^\ast,\chi),
	\label{eq:4fold}\\
	|N|^2=
	\frac{G_F^2 |V_{cb}|^2 |{\bf p_2}| q^2}{(2\pi)^3 12 m_1^2}\left(1-\frac{m_\ell^2}{q^2}\right)^2 \mathcal{B}(D^{\ast+}_s\to D_{s}^+\pi^0),
\end{eqnarray}
where  $|{\bf p_2}|=\lambda^{1/2}(m_1^2,m_2^2,q^2)/2m_1 $
is the momentum of the daughter meson in the $B_s$ rest frame and $\lambda(x,y,z) \equiv x^2+y^2+z^2-2(xy+yz+zx)$ is the K{\"a}ll{\'e}n function.  

The full angular distribution $J(\theta,\theta^\ast,\chi)$ in Eq.~(\ref{eq:4fold}) is expanded in a trigonometric basis and is written in terms of 12 coefficient functions $J_i$ as
\begin{eqnarray}
	J(\theta,\theta^\ast,\chi)
	&=& J_{1s}\sin^2\!\theta^\ast + J_{1c}\cos^2\!\theta^\ast
	+(J_{2s}\sin^2\!\theta^\ast + J_{2c}\cos^2\!\theta^\ast)\cos2\theta\nn
	&&+J_3\sin^2\!\theta^\ast \sin^2\!\theta \cos2\chi
	+J_4\sin2\theta^\ast \sin2\theta \cos\chi\nn
	&&+J_5\sin2\theta^\ast \sin\theta \cos\chi
	+(J_{6s}\sin^2\!\theta^\ast+J_{6c}\cos^2\!\theta^\ast)\cos\theta\nn
	&&+J_7\sin2\theta^\ast \sin\theta \sin\chi
	+J_8\sin2\theta^\ast \sin2\theta \sin\chi
	+J_9\sin^2\!\theta^\ast \sin^2\!\theta \sin2\chi .
	\label{eq:J}
\end{eqnarray}
The coefficient functions $J_i$ depend only on $q^2$ and are expressed in terms of helicity amplitudes and Wilson coefficients. The explicit expressions for these functions in the presence of NP operators can be found in our paper~\cite{Ivanov:2016qtw}.

\subsection{The $q^2$ distribution and ratios of branching fractions $R(D_s^{(*)})$}
\label{subsec:q2dist}
The differential decay width for the process $\bar{B}^0_s\to D^{\ast}_s \tau^-\bar\nu_\tau$ is obtained by performing an integration of the fourfold angular distribution over all angles. The result is expressed in terms of the total angular coefficient $J_{\rm tot}$ as follows:
\begin{equation}
	\frac{d\Gamma(\bar{B}^0_s\to D^{\ast}_s \tau^-\bar\nu_\tau)}{dq^2} =
	|N|^2 J_{\rm tot} = |N|^2 (J_L+J_T).
\end{equation}
Here, the longitudinal and transverse polarization amplitudes of the $D^\ast_s$ meson are denoted by $J_L$ and $J_T$, respectively. They are defined by the relations $J_L=3J_{1c}-J_{2c}$ and $J_T=2(3J_{1s}-J_{2s})$.

To study the decay's geometric properties independent of the total rate, we introduce the normalized full angular distribution $\hat{J}(\theta,\theta^*,\chi)$. This distribution is defined such that its integral over all the angles equals unity:
\begin{equation}
	\hat{J}(\theta,\theta^*,\chi) = \frac{9}{8\pi}\frac{J(\theta,\theta^*,\chi)}{J_{\rm tot}}.
\end{equation}
The same normalization is applied to the individual coefficient functions $J_i$, leading to $\hat{J_i} = (9/8\pi)J_i/J_{\rm tot}$. 
\begin{figure}[htbp]
\centering
\begin{tabular}{cc}
\includegraphics[width=0.33\textwidth]{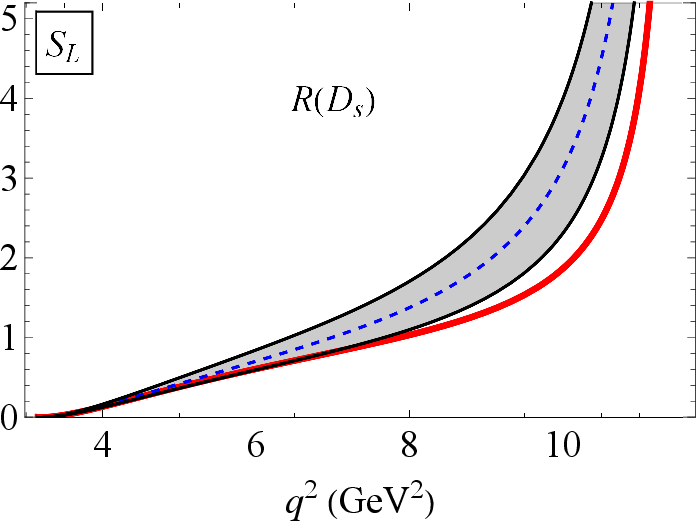}&
\includegraphics[width=0.33\textwidth]{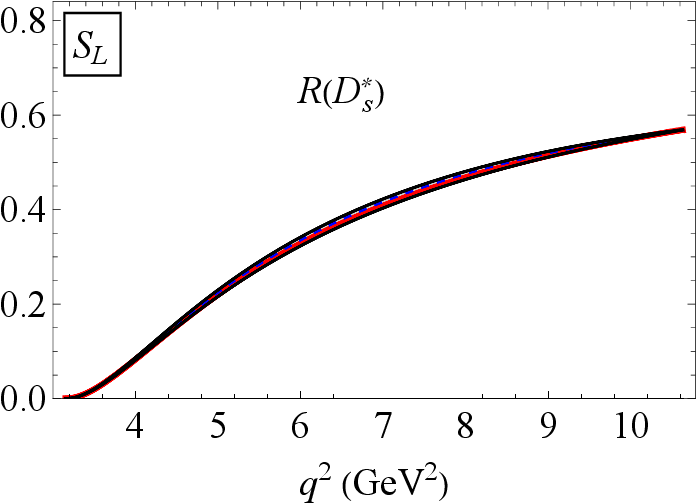}\\
\includegraphics[width=0.33\textwidth]{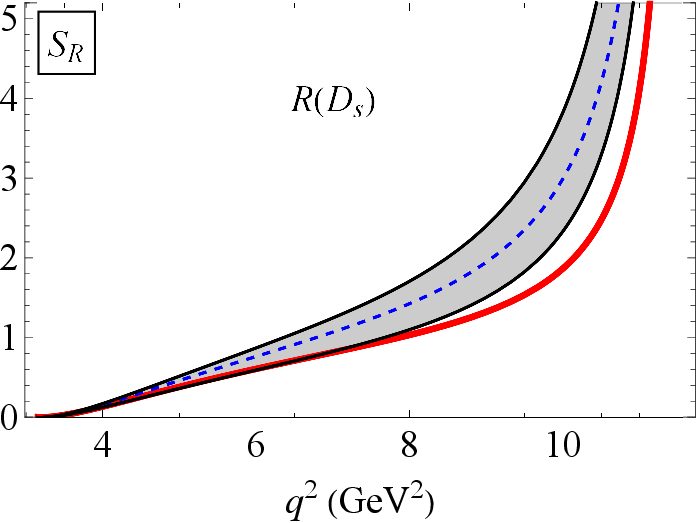}&
\includegraphics[width=0.33\textwidth]{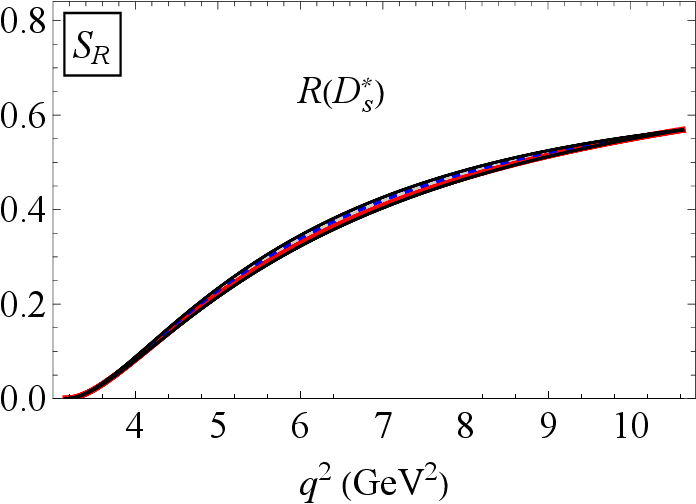}\\
\includegraphics[width=0.33\textwidth]{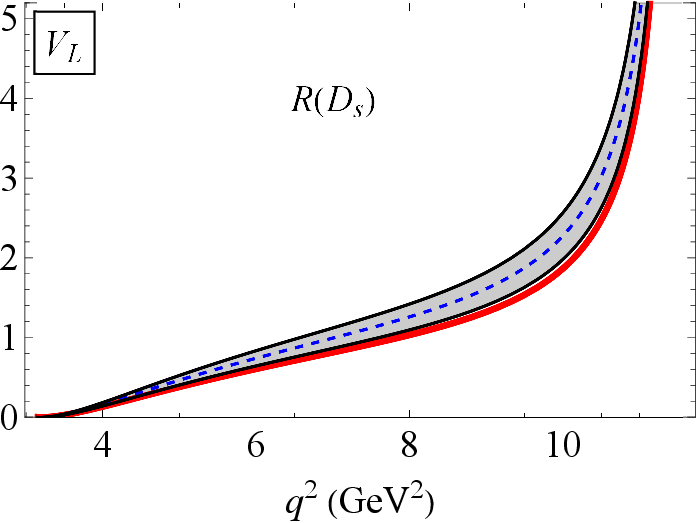}&
\includegraphics[width=0.33\textwidth]{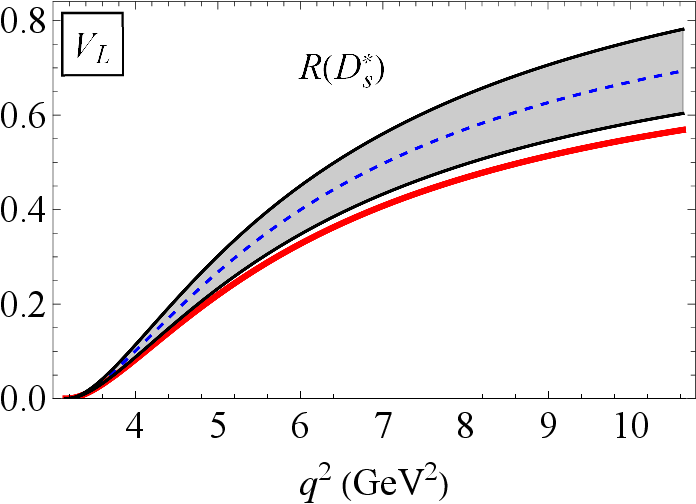}\\
\includegraphics[width=0.33\textwidth]{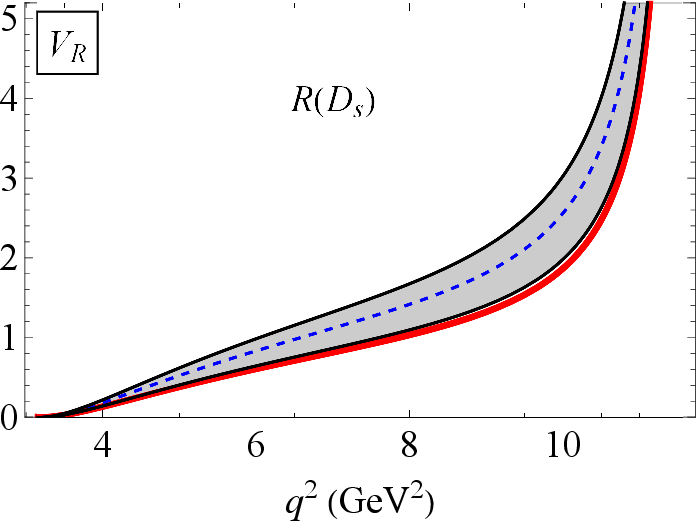}&
\includegraphics[width=0.33\textwidth]{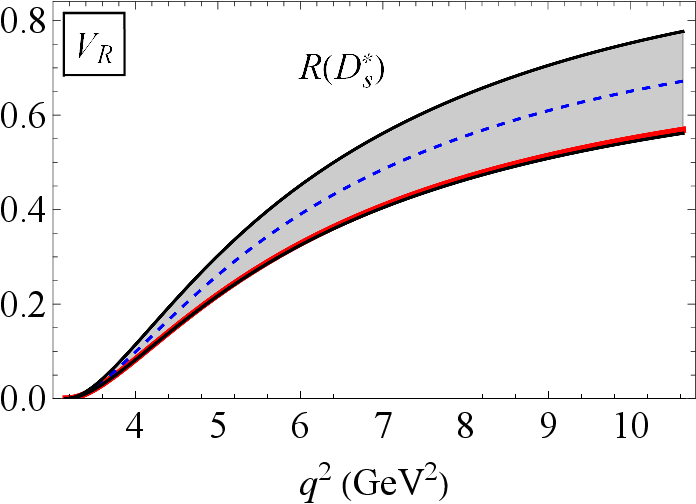}\\
\includegraphics[width=0.33\textwidth]{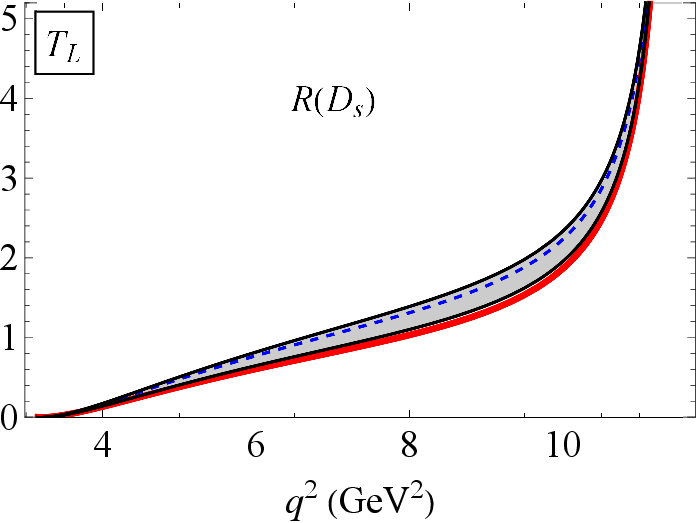}&
\includegraphics[width=0.33\textwidth]{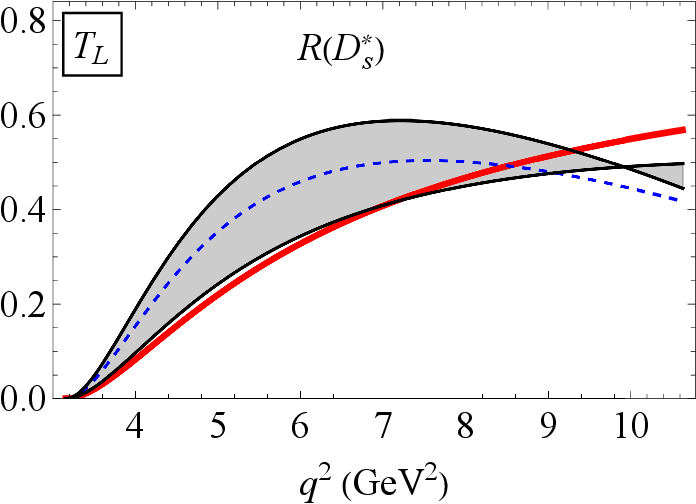}
\end{tabular}
\caption{$q^2$-Dependence of the ratios $R_{D_s}$ (left panels) and $R_{D_s^*}$ (right panels). The thick red lines represent the SM prediction; the gray bands represent NP effects corresponding to the $2\sigma$ allowed regions in Fig.~\ref{fig:2sigma}; the blue dashed lines represent the best-fit values of the NP couplings.}
\label{fig:R}
\end{figure}
 
The  $q^2$ dependence of the observable ratios $R_{D^*_s}$ and $R_{D_s}$ under various NP scenarios is illustrated in Fig.~\ref{fig:R}. Several key observations can be made. First, the inclusion of NP operators consistently leads to an increase in the predicted ratios compared to the SM. Second, the ratio $R_{D^*_s}$ is rather insensitive to the scalar NP operators $\mathcal{O}_{S_L}$ and $\mathcal{O}_{S_R}$, while the ratio $R_{D_s}$ is maximally sensitive to these two operators. Finally, a particularly notable feature arises from the tensor operator $\mathcal{O}_{T_L}$. Unlike the scalar and vector operators, $\mathcal{O}_{T_L}$ significantly modifies the functional form of $R_{D^*_s}(q^2)$, potentially introducing a characteristic peak in the distribution. This distinct shift in the $q^2$ dependence serves as a unique experimental signature. By analyzing the differential distribution of $B_s\to D_s^* \tau\nu$, one can effectively distinguish the tensor contribution from other NP sources.  

Note that the gray bands depicting the NP sensitivity limits in Fig.~\ref{fig:R} (and in other figures below) are constructed using a representative fixed-point envelope approach. For each observable, the parameter space within the $2\sigma$ allowed regions of the Wilson coefficients is sampled at a central kinematic reference point (typically $q^2 \approx 7.0\text{ GeV}^2$) to isolate the specific parameter configurations that produce the maximal and minimal values of the observable. These benchmark configurations are then evaluated across the entire phase space to define the upper and lower boundaries. Consequently, the localized ``zero-width'' behaviors of the gray band observed in some distributions (such as $R(D_s^*)$ under $T_L$ in Fig.~\ref{fig:R}) are simply geometric crossing points where these two boundary trajectories intersect. This simplified approach provides a straightforward and clean visualization of the overall magnitude and behavior of the various NP operator effects across the $q^2$ range [see also Ref.~\cite{Huang:2018nnq}].

\begin{table}[htbp] 
\centering
		\begin{tabular}{ccc}
			\hline\hline
			&\quad  $\left\langle R_{D_s}\right\rangle$ \qquad 
			&\quad  $\left\langle R_{D_s^*}\right\rangle$ \qquad   
			\\
			\hline
			SM &\quad $0.27(3)$\quad &\quad $0.24(3)$\quad \\
			$S_L$
			&\quad $(0.29,0.44)$\quad
			&\quad $(0.24,0.25)$\quad
			\\
			$S_R$
			&\quad $(0.29,0.44)$\quad
			&\quad $(0.24,0.25)$\quad
			\\
			$V_L$
			&\quad $(0.29,0.37)$\quad
			&\quad $(0.26,0.33)$\quad
			\\
			$V_R$
			&\quad $(0.29,0.44)$\quad
			&\quad $(0.24,0.33)$\quad
			\\
			$T_L$
			&\quad $(0.29,0.36)$\quad
			&\quad $(0.24,0.33)$\quad
			\\
			\hline\hline
		\end{tabular}
		\caption{The $q^2$ average of the ratios $R_{D_s}$ and $R_{D^*_s}$ in the SM and in the presence of NP.}
		\label{tab:R}
\end{table}
Table \ref{tab:R} presents the integrated averages for the $R_{D_s}$ and $R_{D^*_s}$ ratios across the entire kinematic range. The SM results (first row) are obtained using the form factors calculated in our quark model. For the NP scenarios, the predicted ranges for these ratios are derived based on the $2\sigma$ parameter space illustrated in Fig.~\ref{fig:2sigma}.
Here, the most pronounced effects come from the operators $\mathcal{O}_{S_L}$, $\mathcal{O}_{S_R}$, and $\mathcal{O}_{V_R}$, which can boost the integrated ratio $\left\langle R_{D_s}\right\rangle$ by approximately 60\% relative to the SM prediction.
 
\subsection{The \boldmath{$\cos\theta$} distribution, forward-backward asymmetry, and lepton-side convexity parameter}
\label{subsec:theta-dist}
By performing a partial integration of the complete fourfold angular distribution over the variables $\cos\theta^\ast$ and $\chi$, we isolate the differential distribution with respect to $\cos\theta$. This distribution is characterized by a tilted parabolic functional form. Its normalized representation is given by
\be
\hat{J}(\theta)=a+b\cos\theta+c\cos^{2}\theta.
\en

The linear coefficient $b$ of the distribution is isolated by defining a forward-backward asymmetry $\mathcal{A}_{FB}(q^2)$. This observable is calculated as the normalized difference between the decay rates in the forward and backward hemispheres:
\bea
\mathcal{A}_{FB}(q^2) = 
\frac{ \int_{0}^{1} d\cos\theta\, d\Gamma/d\cos\theta
	-\int_{-1}^{0} d\cos\theta\, d\Gamma/d\cos\theta }
{ \int_{0}^{1} d\cos\theta\, d\Gamma/d\cos\theta
	+\int_{-1}^{0} d\cos\theta\, d\Gamma/d\cos\theta} 
= \frac32 \frac{J_{6c}+2J_{6s}}{J_{\rm tot}}.
\label{fbAsym}
\ena

The quadratic coefficient $c$ is captured by the convexity parameter $C_F^\tau(q^2)$. This observable is derived from the second derivative of the normalized distribution: 
\be
C_F^\tau(q^2) = \frac{d^{2}\hat{J}(\theta)}{d(\cos\theta)^{2}}
= \frac{6(J_{2c}+2J_{2s})}{J_{\rm tot}}.
\label{eq:convex_lep}
\en 
\begin{figure}[htbp]
	\begin{tabular}{cc}
		&\includegraphics[width=0.33\textwidth]{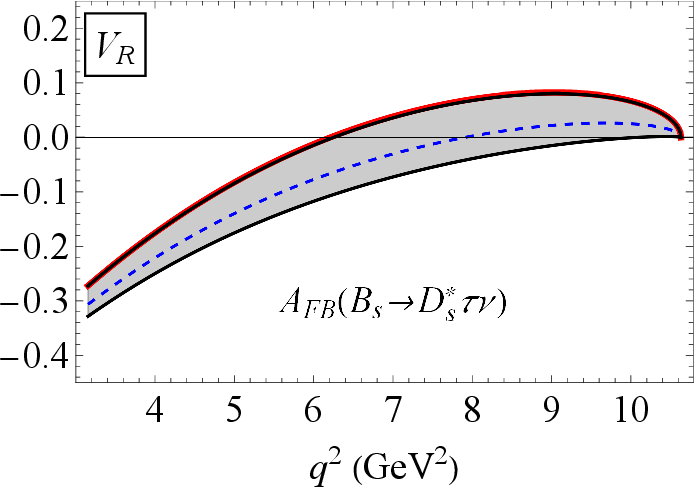}\\
		\includegraphics[width=0.33\textwidth]{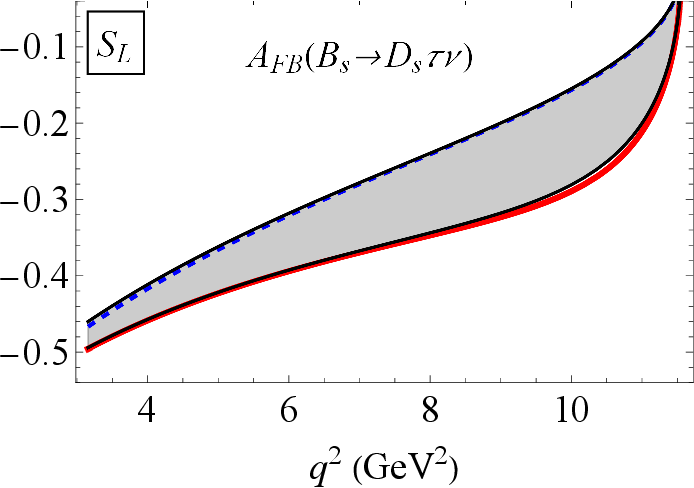}&
		\includegraphics[width=0.33\textwidth]{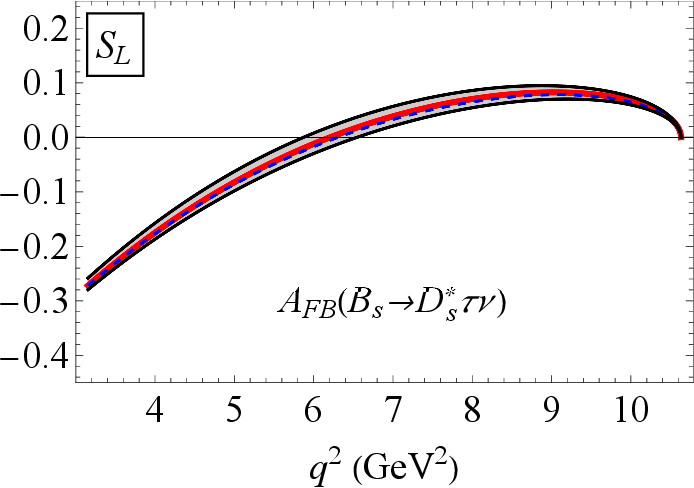}\\
 \includegraphics[width=0.33\textwidth]{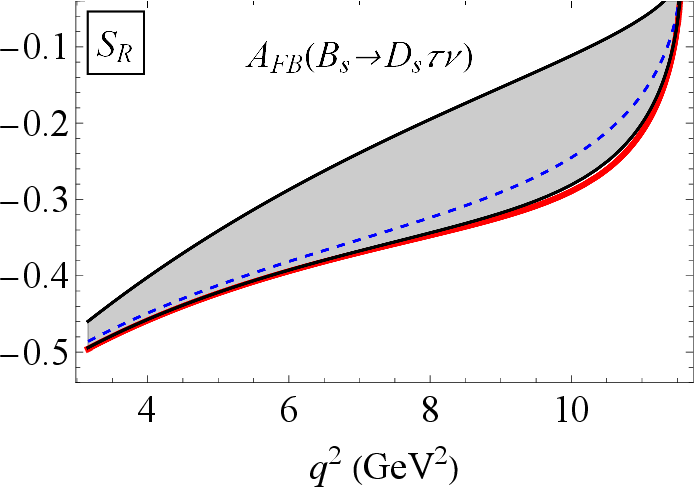}&\includegraphics[width=0.33\textwidth]{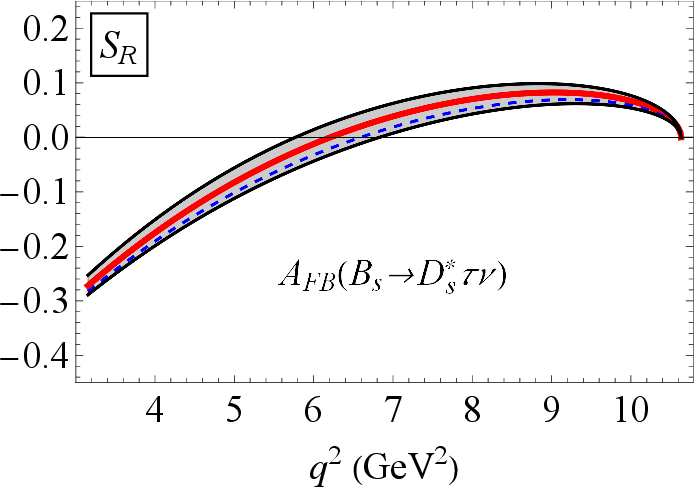}\\
		\includegraphics[width=0.33\textwidth]{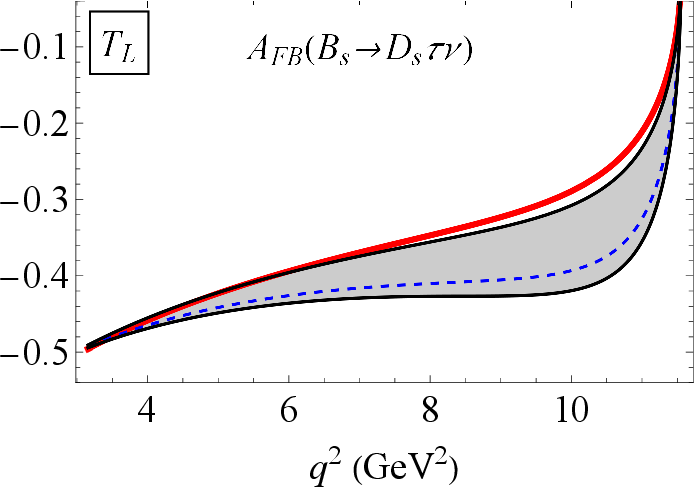}&
		\includegraphics[width=0.33\textwidth]{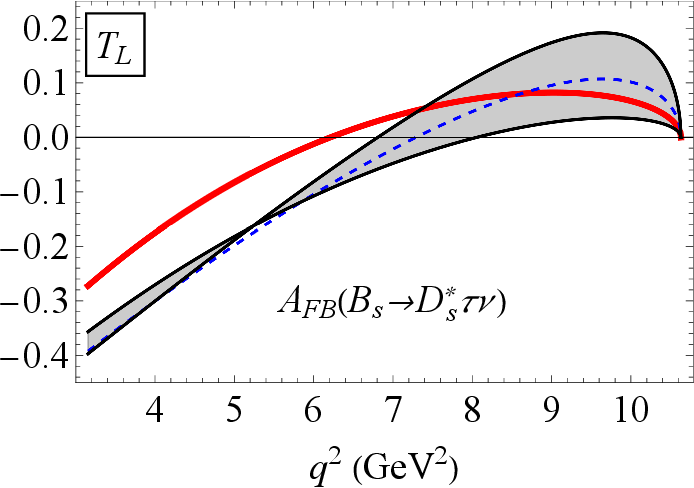}	
	\end{tabular}
	\caption{Forward-backward asymmetry $\mathcal{A}_{FB}(q^2)$ for $B_s \to D_s\tau\nu$ (left panels) and $B_s \to D_s^\ast\tau\nu$ (right panels). Notations are the same as in Fig.~\ref{fig:R}.}
	\label{fig:AFB}
\end{figure}

The sensitivity of the forward-backward asymmetry to various NP scenarios is illustrated in Fig.~\ref{fig:AFB}. While the $V_L$ coupling leaves $\mathcal{A}_{FB}$ unchanged due to its SM-like structure, other NP operators introduce distinct shifts. In $B_s\to D_s^*$ transition, $\mathcal{O}_{S_L}$ and $\mathcal{O}_{S_R}$ only slightly alter the asymmetry, while  $\mathcal{O}_{V_R}$ and $\mathcal{O}_{T_L}$ primarily suppress the observable and delay the zero-crossing point. Notably, $\mathcal{O}_{T_L}$ can enhance $\mathcal{A}_{FB}$ at high $q^2$. In $B_s\to D_s$ transition, $\mathcal{O}_{V_R}$ has no effect, while $\mathcal{O}_{S_L}$, $\mathcal{O}_{S_R}$, and  $\mathcal{O}_{T_L}$ significantly alter the asymmetry. Note that these operators affect $\mathcal{A}_{FB}$ in opposing ways; the scalar operators act to increase the observable, while the tensor one tends to lower it.
\begin{figure}[htbp]
\centering
	\begin{tabular}{cc}	
	&\includegraphics[width=0.33\textwidth]{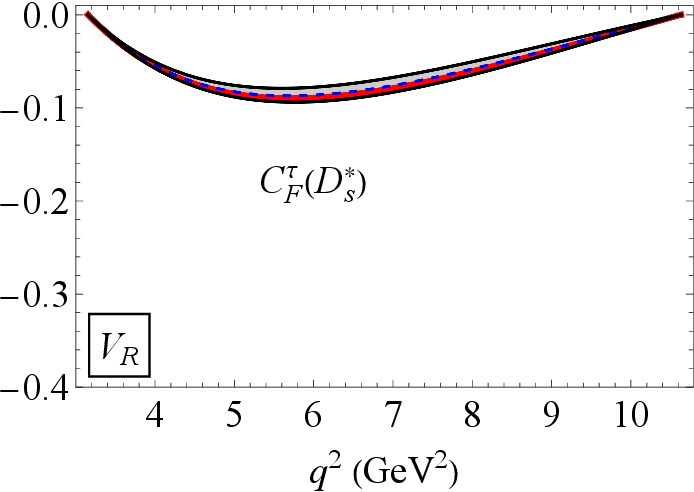}\\
	\includegraphics[width=0.33\textwidth]{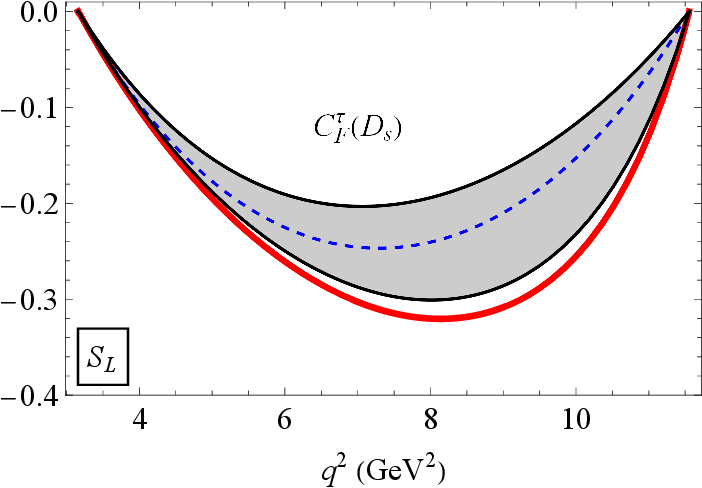}&
	\includegraphics[width=0.33\textwidth]{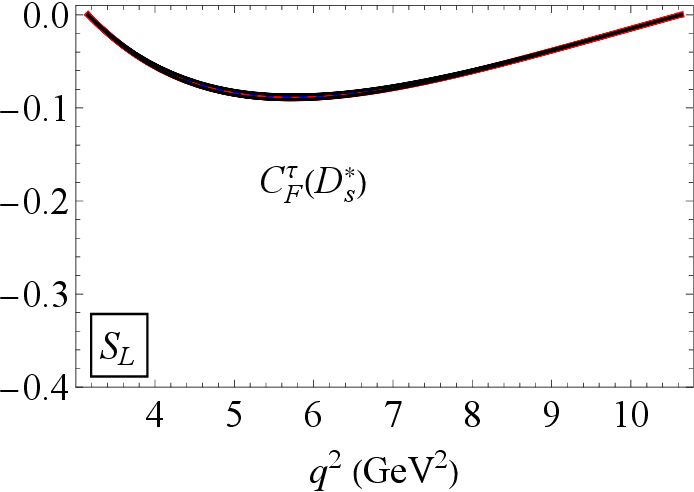}\\
	\includegraphics[width=0.33\textwidth]{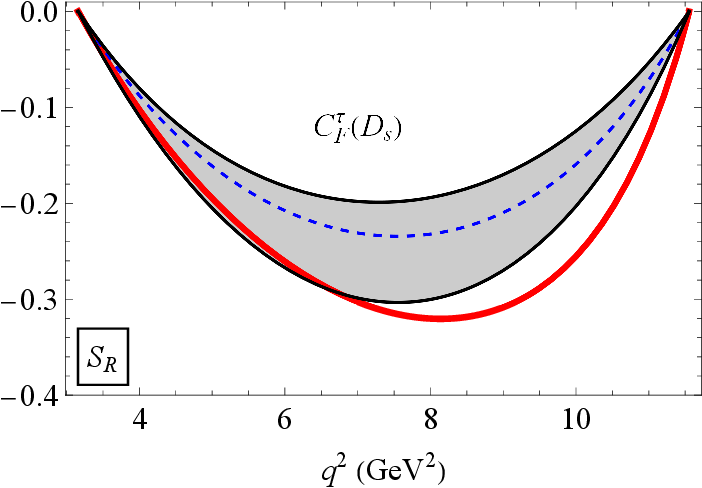}&
	\includegraphics[width=0.33\textwidth]{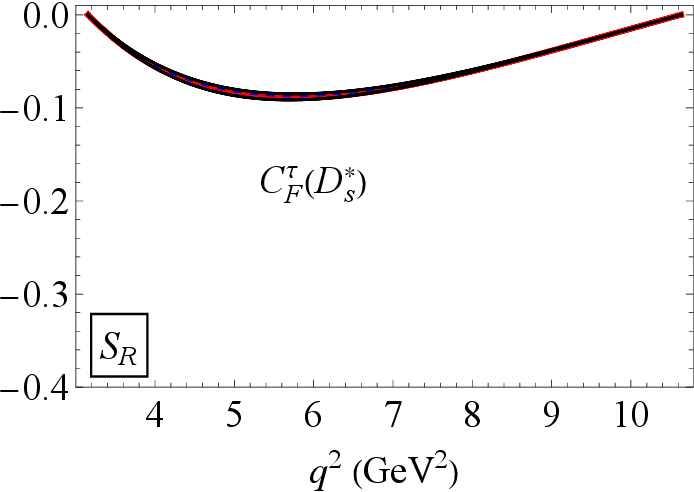}\\	
	\includegraphics[width=0.33\textwidth]{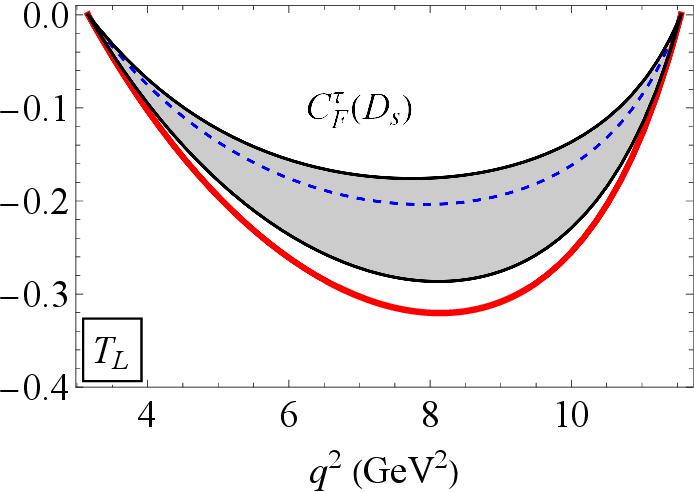}&
	\includegraphics[width=0.33\textwidth]{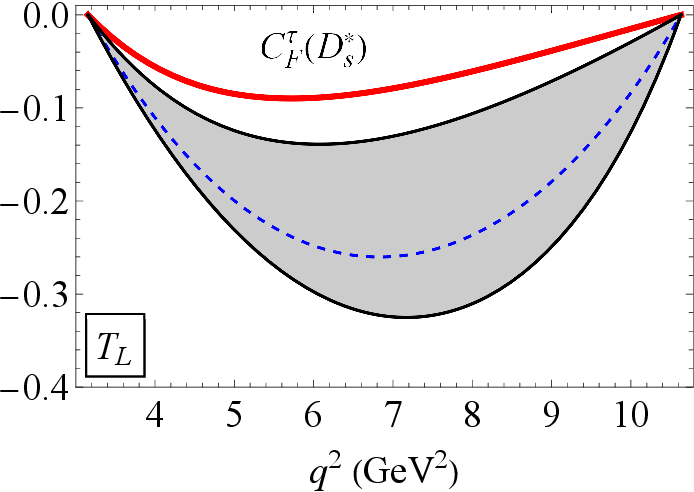}
	\end{tabular}
	\caption{Lepton-side convexity parameter $C_F^\tau(q^2)$ for $B_s \to D_s\tau\nu$ (left panels) and $B_s \to D_s^\ast\tau\nu$ (right panels). Notations are the same as in Fig.~\ref{fig:R}.}
	\label{fig:CFL}
\end{figure}

Figure~\ref{fig:CFL} displays the lepton-side convexity parameter $C_F^\tau(q^2)$. $C_F^\tau(D_s)$
shows sensitivity to a broad range of NP, including $\mathcal{O}_{S_L}$, $\mathcal{O}_{S_R}$, and $\mathcal{O}_{T_L}$. All of these operators tend to increase $C_F^\tau(D_s)$. In contrast, $C_F^\tau(D_s^{\ast})$ is almost exclusively sensitive to the tensor operator $\mathcal{O}_{T_L}$. Moreover, $\mathcal{O}_{T_L}$ consistently lowers $C_F^\tau(D_s^{\ast})$. Near $q^2 \approx 7 \text{ GeV}^2$, the tensor operator's influence is especially pronounced, leading to a fourfold change in the parameter compared to SM expectations. Therefore, $C_F^\tau(D_s^{\ast})$ can be used as a powerful probe for the tensor operator.

Next, we calculate the $q^{2}$ averages of the forward-backward asymmetry and the lepton-side convexity parameter. Note that one 
has to multiply the numerator and denominator of Eqs.~(\ref{fbAsym}) and~(\ref{eq:convex_lep}) by the 
$q^{2}$-dependent piece of the phase-space factor 
$ C(q^2) = |\mathbf{p_2}| q^2 (1-m^2_\ell/q^2)^2 $ before integrating. For example, the average forward-backward asymmetry is calculated as
\be
\langle \mathcal{A}_{FB}\rangle = \frac32 
\frac{\int dq^{2} C(q^{2})\big(J_{6c}+2J_{6s}\big)}
{\int dq^{2} C(q^{2})J_{\rm tot}}.
\label{eq:FBint}
\en
The results are summarized in Table~\ref{tab:lepton-side-avg}. As a byproduct, we also list the corresponding predictions for the muon and electron channels within the SM.
\begin{table}[htbp] 
	\caption{The $q^{2}$ averages of the forward-backward asymmetry and the lepton-side convexity parameter.}
\centering
\begin{tabular}{lcccccc}
		\hline\hline
		\multicolumn{7}{c}{ $B_s\to D_s$ }\\
		\hline
			&  $\left\langle A^\tau_{FB}\right\rangle$ 			  
			&  $\left\langle A^\mu_{FB}\right\rangle$
			&  $\left\langle A^e_{FB}\right\rangle$ 
			&  $\left\langle C_{F}^\tau\right\rangle$
			&  $\left\langle C_{F}^\mu\right\rangle$ 			 
			&  $\left\langle C_{F}^e\right\rangle$
			\\		\hline
		SM \quad & $-0.36(4)$  & $-0.015(2)$ & $-1.2(1)\times 10^{-6}$ & $-0.26(3)$ & $-1.45(15)$ &$-1.50(15)$\\ 
		$\mathcal{O}_{S_L}$\quad & $(-0.35,-0.20)$ &&& $(-0.25,-0.16)$ & &\\
		$\mathcal{O}_{S_R}$\quad & $(-0.35,-0.16)$ &&& $(-0.25,-0.16)$ & &\\
		$\mathcal{O}_{T_L}$\quad & $(-0.43,-0.37)$ &&& $(-0.23,-0.15)$ & &\\
		\hline\hline
		\multicolumn{7}{c}{ $B_s\to D_s^*$ }\\ 	\hline
			&  $\left\langle A^\tau_{FB}\right\rangle$ 			  
			&  $\left\langle A^\mu_{FB}\right\rangle$
			&  $\left\langle A^e_{FB}\right\rangle$ 
			&  $\left\langle C_{F}^\tau\right\rangle$
			&  $\left\langle C_{F}^\mu\right\rangle$ 			 
			&  $\left\langle C_{F}^e\right\rangle$ 
			\\			\hline
		SM \quad & $0.029(3)$ & $0.19(2)$ & $0.20(2$) & $-0.060(6)$ & $-0.44(5)$ & $-0.45(5)$	\\
		$\mathcal{O}_{S_L}$\quad & $(0.013,0.045)$ &&& $(-0.061,-0.059)$ & &\\
		$\mathcal{O}_{S_R}$\quad & $(0.002,0.050)$ &&& $(-0.061,-0.058)$ & &\\
		$\mathcal{O}_{V_R}$\quad & $(-0.069,0.027)$ &&& $(-0.064,-0.052)$ & &\\
		$\mathcal{O}_{T_L}$\quad & $(-0.072,0.023)$ &&& $(-0.26,-0.10)$ & &\\
			\hline\hline
		\end{tabular}
		\label{tab:lepton-side-avg}
\end{table}

\subsection{The \boldmath{$\cos\theta^*$} distribution, hadron-side convexity parameter, and polarization of the final vector meson}
\label{subsec:thetastar-dist}
By integrating the fourfold decay distribution over the lepton-side angle $\cos\theta$ and the azimuthal angle $\chi$, we isolate the distribution relative to the hadron-side angle $\theta^*$. Unlike the lepton side, this distribution follows an untilted parabolic form (lacking a linear term) and is normalized as 
\begin{equation}
	\hat{J} (\theta^\ast)=a'+c'\cos^{2}\theta^\ast.
\end{equation}
The curvature of the $\cos\theta^*$ distribution is quantified by the hadron-side convexity parameter $C_F^h(q^2)$. It is derived from the second derivative of the normalized distribution:
\be
C_F^h(q^2) = \frac{d^{2}\hat{J}(\theta^\ast)}{d(\cos\theta^{\ast})^{2}}
=\frac{c'}{2}=
\frac{3J_{1c}-J_{2c}-3J_{1s}+J_{2s}}{J_{\rm tot}/3}.
\label{eq:convex_had}
\en

The $\cos\theta^\ast$ distribution can also be written as
\begin{equation}
	\hat{J}(\theta^*) = \frac{3}{4}\left(2F_L(q^2)\cos^2\!\theta^* + F_T(q^2)\sin^2\!\theta^*\right),
\end{equation}
where $F_L(q^2)$ and $F_T(q^2)$ represent the longitudinal and transverse polarization fractions of the $D^*_s$ meson, respectively. These fractions are defined in terms of the angular coefficient functions:
\begin{equation}
	F_L(q^2) = \frac{J_L}{J_L + J_T}, \quad F_T(q^2) = \frac{J_T}{J_L + J_T}, \quad
	F_L(q^2) + F_T(q^2) = 1,
\end{equation}
where $J_L = 3J_{1c} - J_{2c}$ and $J_T = 2(3J_{1s} - J_{2s})$.

There is a direct geometric relationship between the convexity parameter and the longitudinal polarization fraction:
\be 
C_F^h(q^2)=\frac32 \left( 2F_L(q^2)-F_T(q^2) \right)=\frac32 \left(3F_L(q^2)-1\right).
\en

\begin{figure}[htbp]
	\begin{tabular}{cc}
		\includegraphics[width=0.33\textwidth]{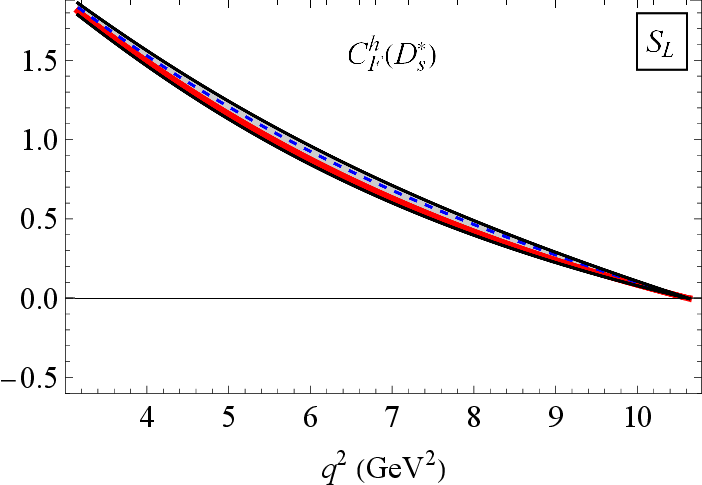}&
		\includegraphics[width=0.33\textwidth]{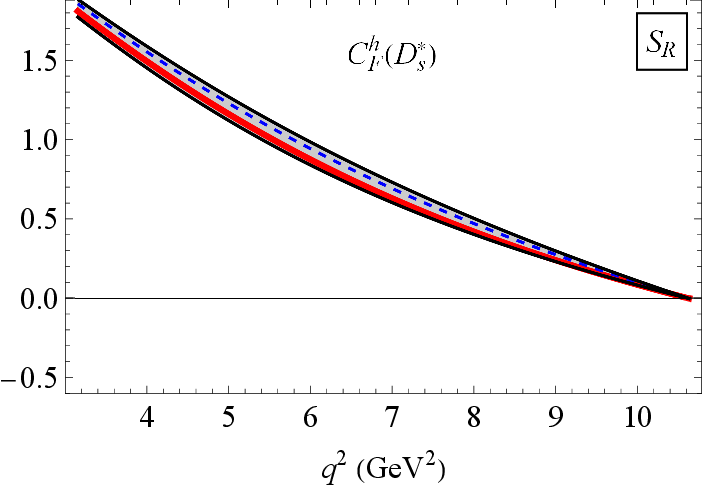}\\
		\includegraphics[width=0.33\textwidth]{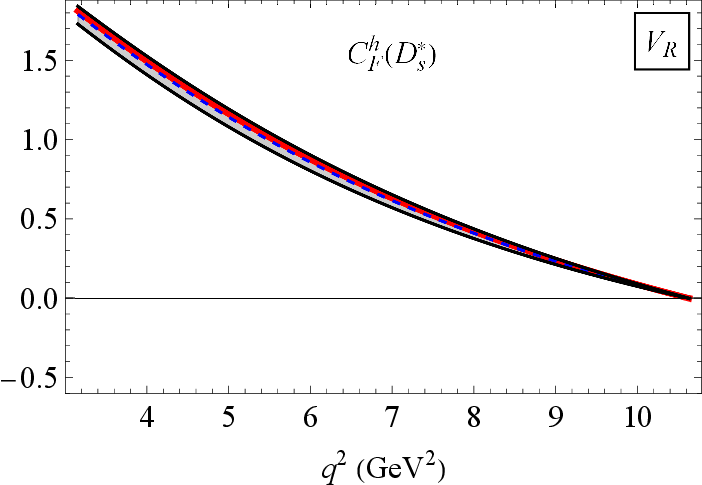}&
		\includegraphics[width=0.33\textwidth]{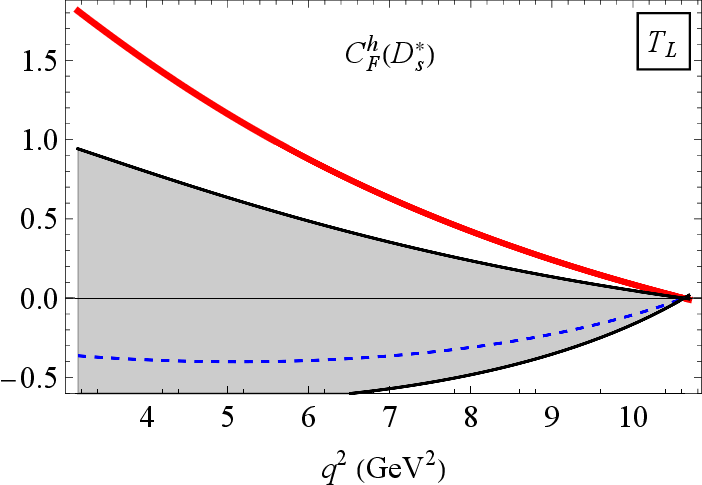}
	\end{tabular}
	\caption{Hadron-side convexity parameter $C_F^h(q^2)$. Notations are the same as in Fig.~\ref{fig:R}.}
	\label{fig:CFH}
\end{figure}

The sensitivity of the hadron-side convexity to NP is illustrated in Fig.~\ref{fig:CFH}. The various NP operators exert distinct influences on the $D_s^*$ polarization. The vector operator $\mathcal{O}_{V_R}$ has a negligible effect on $C_F^h(q^2)$, with the distribution remaining nearly identical to the SM prediction. The inclusion of scalar NP operators $\mathcal{O}_{S_L}$ and $\mathcal{O}_{S_R}$ consistently increases the convexity parameter across almost the entire $q^2$ spectrum. However, the effects are small. The tensor operator $\mathcal{O}_{T_L}$ provides the most dramatic signature. It significantly suppresses the convexity parameter, especially in the low-$q^2$ region. Crucially, $\mathcal{O}_{T_L}$ can push $C_F^h(q^2)$ into negative values -- a phenomenon that is forbidden within the SM. Finally, in Table~\ref{tab:hadron-side-avg} we present the $q^2$ averages of the hadron-side convexity parameter and the longitudinal polarization fraction.

\begin{table}[htbp] 
	\caption{The $q^{2}$ averages of the hadron-side convexity parameter and the longitudinal polarization fraction of $D^*_s$.}
	\centering
	\begin{tabular}{lcccc}
		\hline\hline	
		&  $\left\langle C_{F}^h\right\rangle_\tau$ 			  
		&  $\left\langle C_{F}^h\right\rangle_{\mu/e}$		
		&  $\left\langle F_L\right\rangle_\tau$
		&  $\left\langle F_L\right\rangle_{\mu/e}$ 		
		\\			\hline
		SM \quad & $0.56(6)$ & $0.91(9)$ & $0.46(5)$ & $0.54(6)$\\
		$\mathcal{O}_{S_L}$\quad & $(0.54,0.63)$ && $(0.45,0.47)$ &\\
		$\mathcal{O}_{S_R}$\quad & $(0.54,0.65)$ && $(0.45,0.48)$ &\\
		$\mathcal{O}_{V_R}$\quad & $(0.52,0.58)$ && $(0.45,0.46)$ &\\
		$\mathcal{O}_{T_L}$\quad & $(-0.52,0.32)$ && $(0.22,0.41)$ &\\
		\hline\hline
	\end{tabular}
	\label{tab:hadron-side-avg}
\end{table}

\subsection{The \boldmath{$\chi$} distribution and the trigonometric moments}
By integrating the master decay rate over both polar angles $\cos\theta$ and $\cos\theta^\ast$, we obtain the distribution in the azimuthal angle $\chi$. The normalized form of this distribution is expressed as
\be
\hat{J}^{(I)}(\chi)=\frac{1}{2\pi}\Big[1+A_C^{(1)}(q^2)\cos 2\chi+A_T^{(1)}(q^2)\sin 2\chi\Big].
\en 

To further isolate specific angular coefficients, we can define alternative distributions $J^{(II)}(\chi)$ and $J^{(III)}(\chi)$ by applying asymmetric integration ranges over the polar angles. Distribution $J^{(II)}(\chi)$ involves an asymmetric integration over the hadron-side angle $\cos\theta^\ast$, while distribution $J^{(III)}(\chi)$ involves asymmetric integrations over both $\cos\theta$ and $\cos\theta^\ast$. The two distributions are defined as follows~\cite{Duraisamy:2013pia}:
\bea
J^{(II)}(\chi) &=&\Big[\int_0^1-\int_{-1}^0\Big]d\cos \theta^\ast\int_{-1}^1d\cos \theta\frac{d^4\Gamma}
{dq^2 d\cos\theta d\chi d\cos\theta^\ast},\nn
J^{(III)}(\chi) &=& \Big[\int_0^1-\int_{-1}^0\Big]d\cos \theta^\ast\Big[\int_0^1-\int_{-1}^0\Big]d\cos \theta\frac{d^4\Gamma}
{dq^2 d\cos\theta d\chi d\cos\theta^\ast}.
\ena
The normalized forms of these distributions read
\bea
\hat{J}^{(II)}(\chi) &=&\frac14\Big[A_C^{(2)}(q^2)\cos\chi+A_T^{(2)}(q^2)\sin\chi\Big],
\nn
\hat{J}^{(III)}(\chi) &=& \frac{2}{3\pi}\Big[A_C^{(3)}(q^2)\cos\chi+A_T^{(3)}(q^2)\sin\chi\Big].
\ena
The resulting $A_{C,T}^{(i)}$ coefficients directly map to the angular functions $J_4$, $J_5$, $J_7$, and $J_8$, providing a clear experimental path to determining these values.

An alternative and robust method for extracting the $J_i$ coefficients (specifically for $i=3,4,5,7,8,9$) is the use of trigonometric moments. These moments, denoted as $W_X$, are the expectation values of specific trigonometric kernels $M_X$ weighted by the normalized decay distribution $\hat{J}(\theta^\ast,\theta,\chi)$~\cite{Ivanov:2015tru}:
\be
W_{X}(q^2) = \int d\cos\theta d\cos\theta^\ast d\chi
M_{X}(\theta^\ast,\theta,\chi)\hat{J}(\theta^\ast,\theta,\chi) 
\equiv  \left\langle M_{X}(\theta^\ast,\theta,\chi) \right\rangle.
\en
The mapping between these moments and the angular coefficients is summarized in Table~\ref{tab:W}.
\begin{table}[htbp]
	\caption{Relations between trigonometric moments $W_X(q^2)$ and angular coefficients $A_{C,T}^{(i)}(q^2)$.}
	\renewcommand{\arraystretch}{1.2}
	\centering
	\begin{tabular}{cccc}
		\hline\hline
Moment & Kernel $M_X$  & Relation to $J_i$  &  Relation to $A_{C,T}^{(i)}$\\\hline
$W_T$ & $\cos 2\chi$  & ${2J_3}/{J_{\rm tot}}$  & 	$\frac12 A_C^{(1)}$\\
$W_{IT}$ & $\sin 2\chi$ & ${2J_9}/{J_{\rm tot}}$ & $\frac12 A_T^{(1)}$\\
$W_A$ & $\sin\theta\cos\theta^{\ast}\cos \chi$ & $\frac{3\pi}{8} {J_5}/{J_{\rm tot}}$ & $\frac{\pi}{8} A_C^{(2)}$\\
$W_{IA}$ & $\sin\theta\cos\theta^{\ast}\sin \chi$ & $\frac{3\pi}{8} {J_7}/{J_{\rm tot}}$ & $\frac{\pi}{8} A_T^{(2)}$\\
$W_I$ & $\cos\theta\cos\theta^{\ast}\cos \chi$ & $\frac{9\pi^2}{128}{J_4}/{J_{\rm tot}}$ & $\frac{3\pi^2}{128} A_C^{(3)}$\\
$W_{II}$ & $\cos\theta\cos\theta^{\ast}\sin \chi$ & $\frac{9\pi^2}{128}{J_8}/{J_{\rm tot}}$ & $\frac{3\pi^2}{128} A_T^{(3)}$\\
		\hline\hline
	\end{tabular}
	\label{tab:W}
\end{table}
Beyond moments and full integrations, these coefficients can also be isolated by analyzing piecewise sums and differences of events in different sectors of the angular phase space. This method is often preferred in experimental environments where detector efficiency and acceptance across the full angular range must be carefully controlled~\cite{Korner:1989ve,Korner:1989qb,Becirevic:2016hea}.

\begin{figure}[htbp]
	\begin{tabular}{cccc}
		\includegraphics[width=0.25\textwidth]{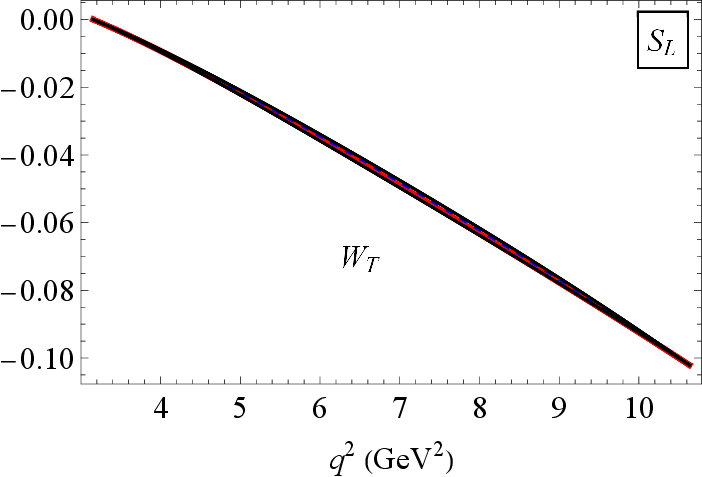}&
		\includegraphics[width=0.25\textwidth]{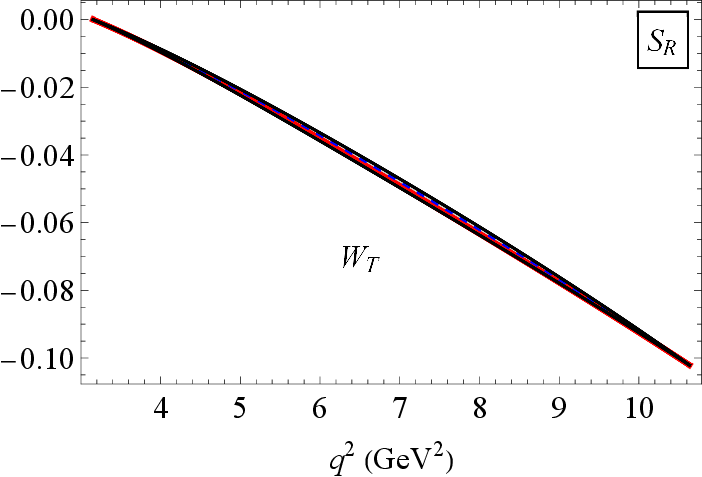}&
		\includegraphics[width=0.25\textwidth]{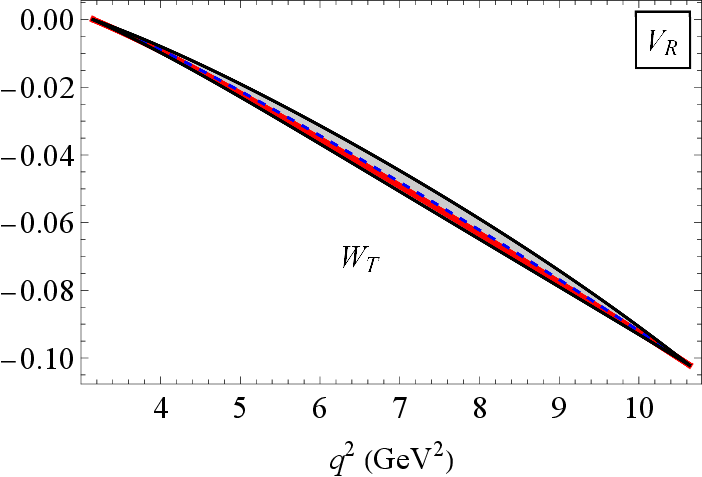}&
		\includegraphics[width=0.25\textwidth]{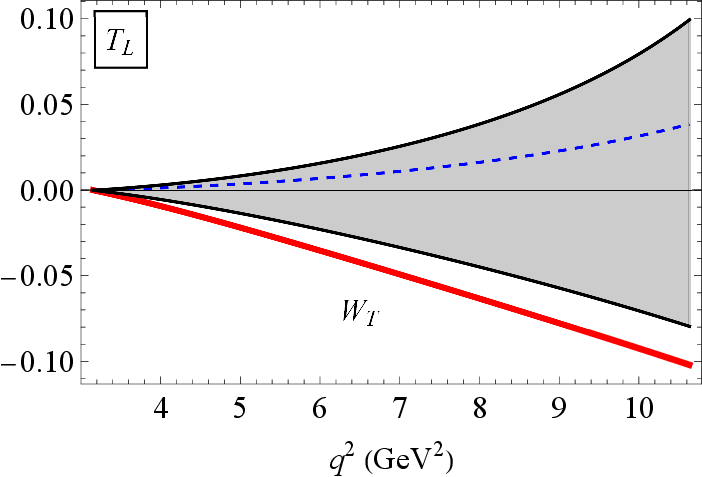}\\
		\includegraphics[width=0.25\textwidth]{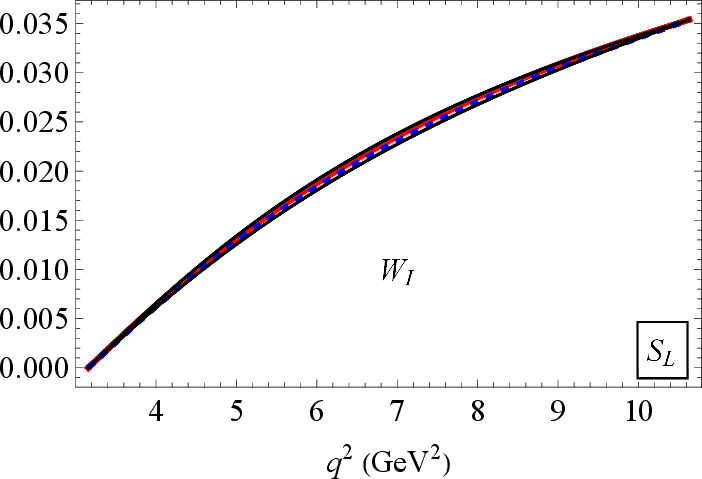}&
		\includegraphics[width=0.25\textwidth]{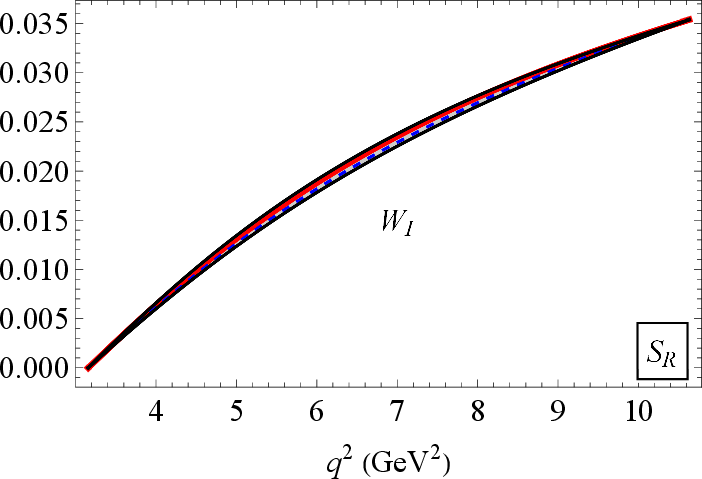}&
		\includegraphics[width=0.25\textwidth]{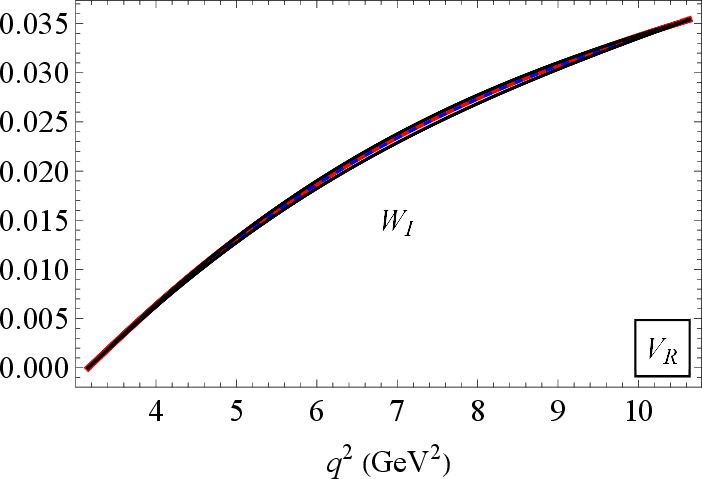}&
		\includegraphics[width=0.25\textwidth]{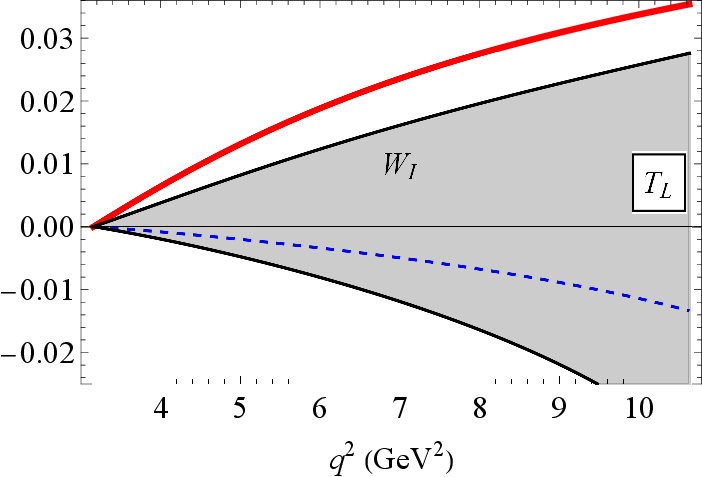}\\
		\includegraphics[width=0.25\textwidth]{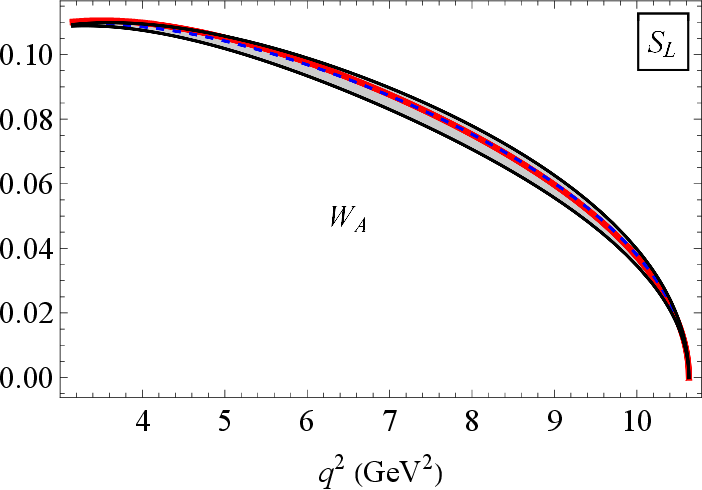}&
		\includegraphics[width=0.25\textwidth]{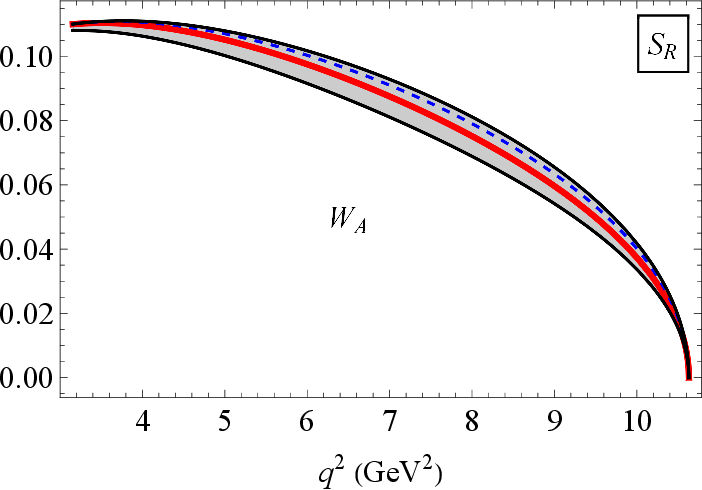}&
		\includegraphics[width=0.25\textwidth]{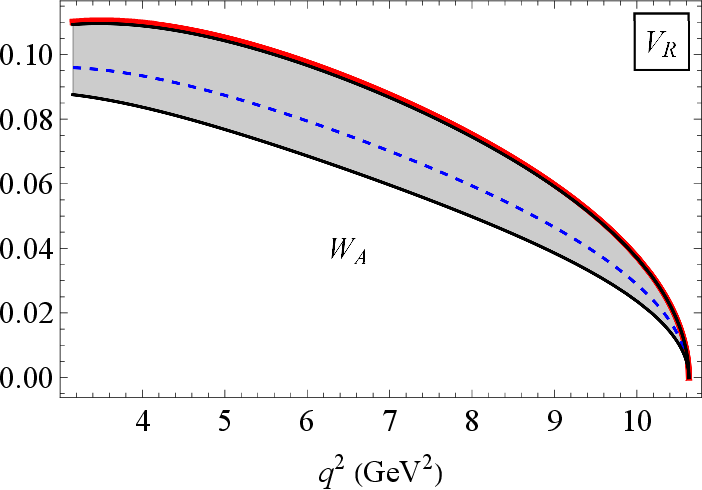}&	
		\includegraphics[width=0.25\textwidth]{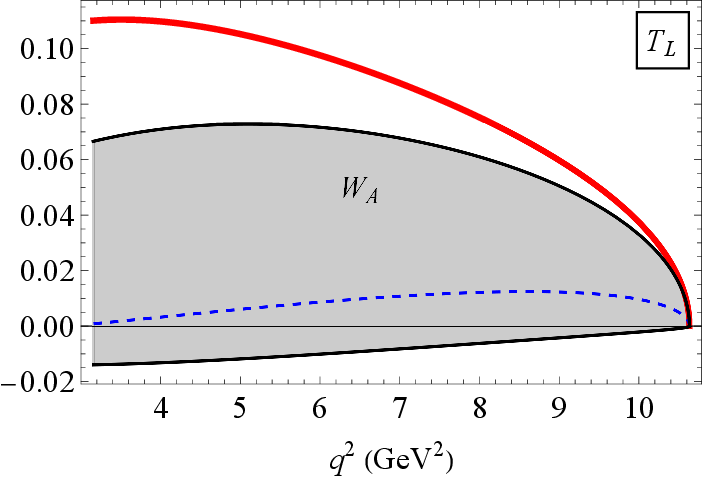}\\
		\includegraphics[width=0.25\textwidth]{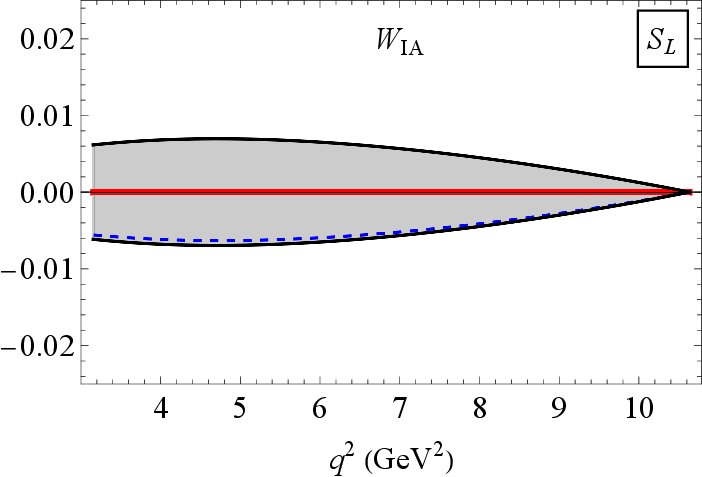}&
		\includegraphics[width=0.25\textwidth]{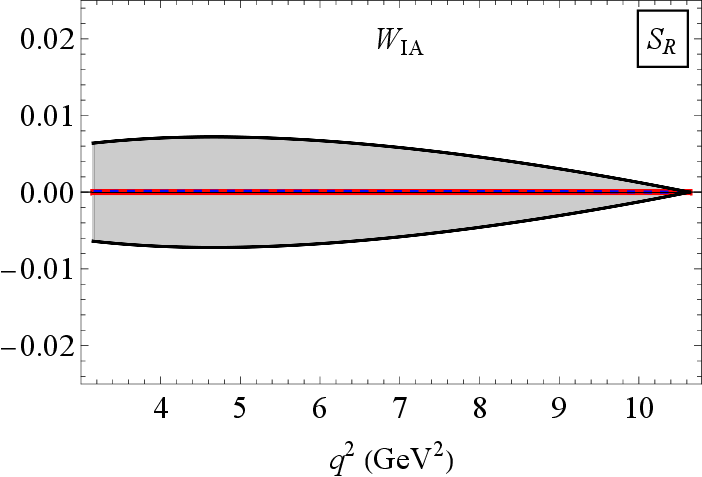}&
		\includegraphics[width=0.25\textwidth]{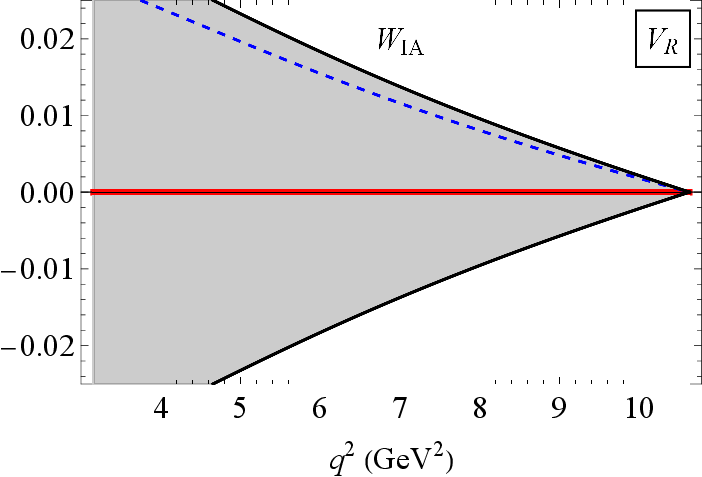}&
		\includegraphics[width=0.25\textwidth]{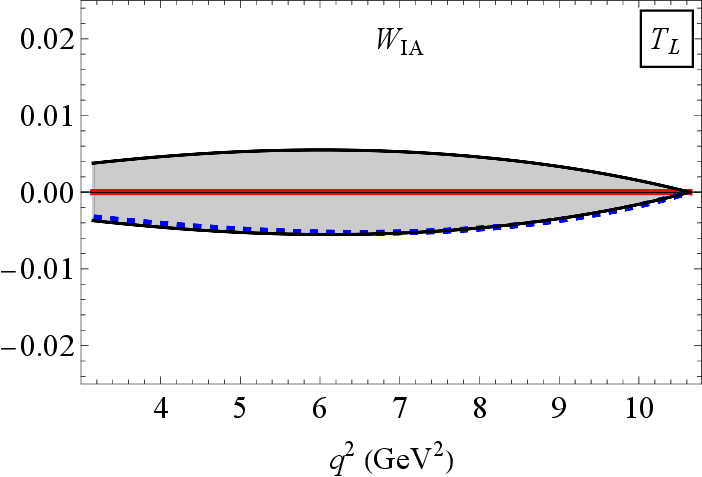}
	\end{tabular}
	\caption{Trigonometric moments $W_T(q^2)$, $W_I(q^2)$, $W_A(q^2)$, and $W_{IA}(q^2)$. Notations are the same as in Fig.~\ref{fig:R}.}
	\label{fig:W1}
\end{figure}
\begin{figure}[htbp]
	\begin{tabular}{lr}
		\includegraphics[width=0.33\textwidth]{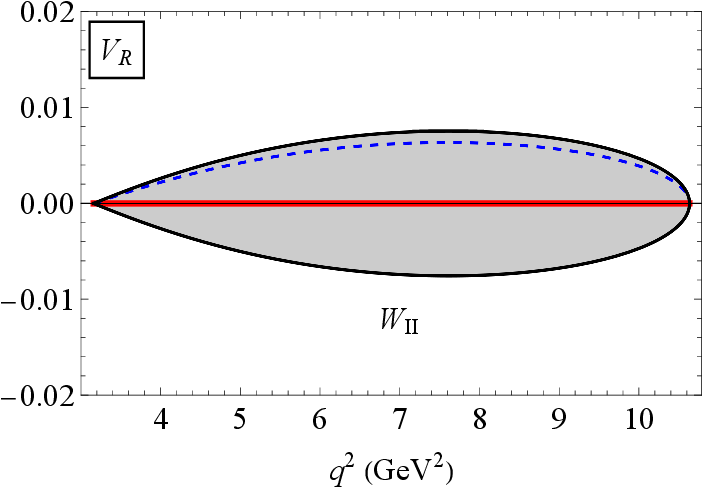}&
		\includegraphics[width=0.33\textwidth]{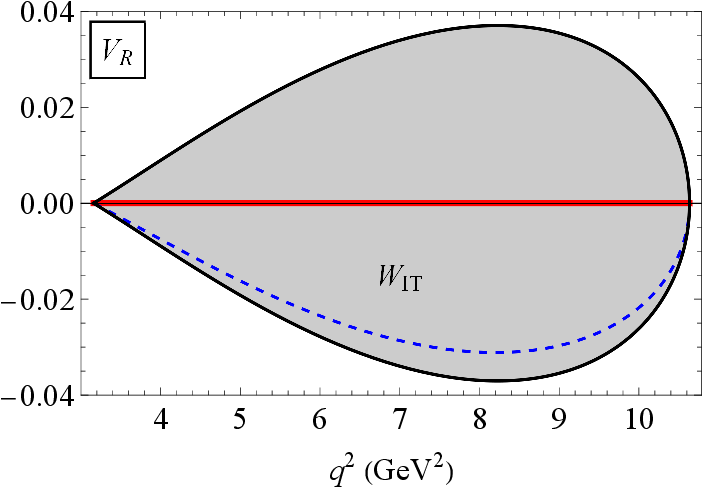}
	\end{tabular}
	\caption{Trigonometric moments $W_{II}(q^2)$ (left panel) and $W_{IT}(q^2)$ (right panel). Notations are the same as in Fig.~\ref{fig:R}.}
	\label{fig:W2}
\end{figure}

The $q^2$ dependence of the trigonometric moments is depicted in Figs.~\ref{fig:W1} and~\ref{fig:W2}. These observables provide distinct signatures for various NP scenarios. The moments $W_T(q^2)$ and  $W_I(q^2)$ remain largely unaffected by the scalar ($\mathcal{O}_{S_L}$, $\mathcal{O}_{S_R}$) and right-handed vector ($\mathcal{O}_{V_R}$) operators, but they exhibit profound sensitivity to the tensor operator $\mathcal{O}_{T_L}$. The presence of $\mathcal{O}_{T_L}$ typically leads to an enhancement in $W_T$ and a corresponding suppression in $W_I$.
The $W_A$ moment is significantly influenced by $\mathcal{O}_{V_R}$ and $\mathcal{O}_{T_L}$. Specifically, both operators act to reduce its value.
A critical feature of the tensor operator $\mathcal{O}_{T_L}$ is its ability to induce sign changes in $W_T$, $W_I$, and $W_A$, marking a clear departure from the SM predictions.
The moments $W_{IA}$, $W_{II}$, and $W_{IT}$ serve as critical null tests.
For the $W_{IA}$ moment, the NP operators can cause deviations in either direction, with the most pronounced sensitivity observed for the $\mathcal{O}_{V_R}$ coupling. For $W_{II}$ and $W_{IT}$, only $\mathcal{O}_{V_R}$ can modify the observables. 
Because the SM predicts these values to be effectively zero, any experimental observation of a nonzero signal would point directly to a right-handed vector current. Finally, the $q^2$ averages of the trigonometric moments are listed in Table~\ref{tab:azimuthal-avg} [see also~\cite{Zhang:2020dla}].

\begin{table}[htbp] 
	\caption{The $q^{2}$ averages of the trigonometric moments.}
	\centering
	\begin{tabular}{cccccc}
		\hline\hline	
		{} & SM & $\mathcal{O}_{S_L}$ & $\mathcal{O}_{S_R}$ & $\mathcal{O}_{V_R}$ & $\mathcal{O}_{T_L}$\\
		\hline
		$\left\langle W_T\right\rangle_\tau$ & $-0.057(6)$ & $(-0.057,-0.055)$ & $(-0.057,-0.055)$ & $(-0.058,-0.053)$ & $(-0.039,0.031)$ \\
		$\left\langle W_T\right\rangle_{\mu/e}$ & $-0.094(9)$ &\\
		$\left\langle W_{IT}\right\rangle_\tau$ & 0 & 0 & 0 & $(-0.030,0.030)$ & 0 \\	
		$\left\langle W_A\right\rangle_\tau$ & $0.077(8)$ & $(0.073,0.079)$ & $(0.071,0.082)$ & $(0.053,0.076)$ & $(-0.008,0.060)$\\
		$\left\langle W_A\right\rangle_{\mu/e}$ & $0.062(6)$ & \\		
		$\left\langle W_{IA}\right\rangle_\tau$ & 0 & $(-0.005,0.005)$ & $(-0.005,0.005)$ & $(-0.012,0.012)$ & $(-0.005,0.005)$\\		
		$\left\langle W_I\right\rangle_\tau$ & $0.025(3)$ & $(0.024,0.025)$ & $(0.024,0.025)$ & $(0.024,0.025)$ & $(-0.013,0.017)$\\
		$\left\langle W_I\right\rangle_{\mu/e}$ & $0.054(6)$ &\\		
		$\left\langle W_{II}\right\rangle_\tau$ & 0 & 0 & 0 & $(-0.006,0.006)$ & 0\\		
		\hline\hline
	\end{tabular}
	\label{tab:azimuthal-avg}
\end{table}

\subsection{Final lepton polarization}
\label{subsec:lep-pol}

Similar to what has been discussed in Refs.~\cite{Chen:2017eby, Tanaka:2010se, Ivanov:2017mrj}, one can use the polarization of the $\tau$ in the semileptonic decays $B_s\to D_s^{(*)}\tau\nu$ to probe for NP. The longitudinal ($L$), transverse ($T$), and normal ($N$) polarization components of the $\tau$ are defined as
\be
\label{eq:poldef}
P_i(q^2) = \frac{d\Gamma(s^\mu_i)/dq^2-d\Gamma(-s^\mu_i)/dq^2}{d\Gamma(s^\mu_i)/dq^2+d\Gamma(-s^\mu_i)/dq^2},\qquad i=L, N, T,
\en
where $s^\mu_i$ are the polarization four-vectors of the $\tau$ in the $W^-$ rest frame. One has
\be
s^\mu_L=\Big(\frac{|\vec{p}_\tau|}{m_\tau},\frac{E_\tau}{m_\tau}\frac{\vec{p}_\tau}{|\vec{p}_\tau|}\Big),\qquad
s^\mu_N=\Big(0,\frac{\vec{p}_\tau \times\vec{p}_2}{|\vec{p}_\tau \times\vec{p}_2|}\Big),\qquad 
s^\mu_T=\Big(0,\frac{\vec{p}_\tau \times\vec{p}_2}{|\vec{p}_\tau \times\vec{p}_2|}\times
\frac{\vec{p}_\tau}{|\vec{p}_\tau|}\Big).
\en
Here, $\vec{p}_\tau$ and $\vec{p}_2$ are the three-momenta of $\tau$ and the final meson ($D_s^*$ or $D_s$), respectively, in the $W^-$ rest frame. While the reconstruction of these polarization states via subsequent $\tau$ decay products is a rich field of study, our present analysis is restricted to identifying the characteristic shifts induced by NP operators on these observables. A detailed analysis of the tau polarization with the help of its subsequent decays can be found in Refs.~\cite{Ivanov:2017mrj, Alonso:2016gym, Alonso:2017ktd}. 

\begin{figure}[htbp]
	\centering
	\begin{tabular}{ccc}
		\includegraphics[width=0.33\textwidth]{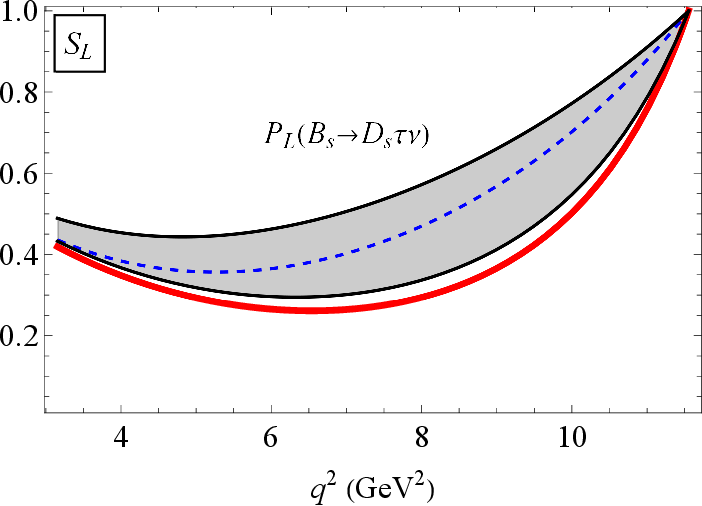}& 
		\includegraphics[width=0.33\textwidth]{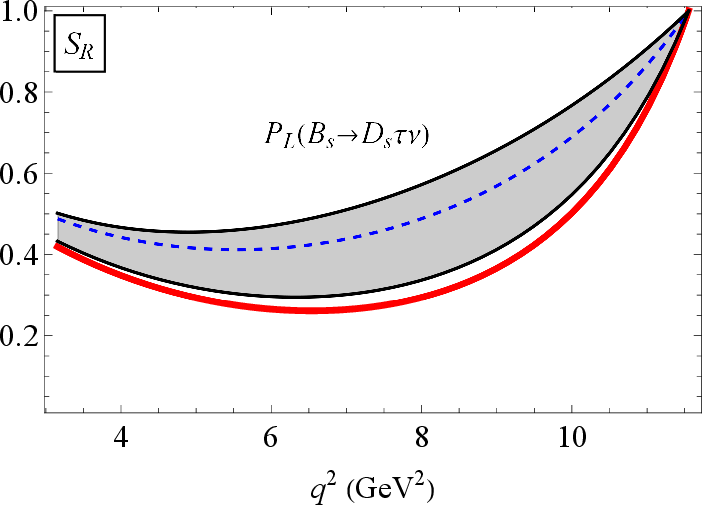}& 
		\includegraphics[width=0.33\textwidth]{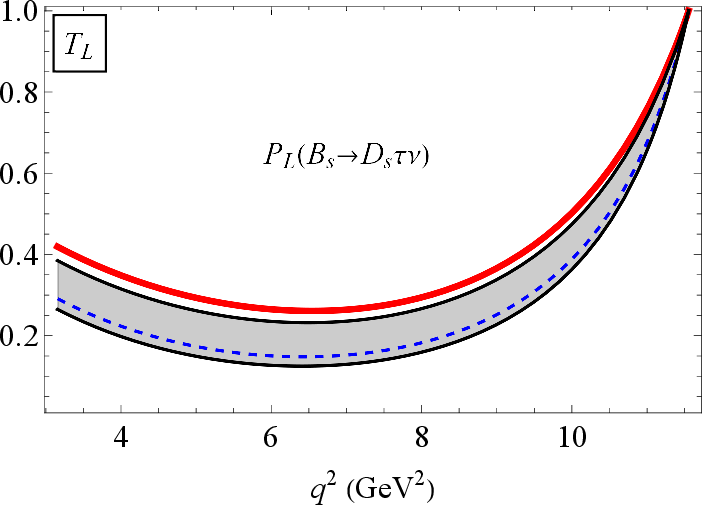}\\
		\includegraphics[width=0.33\textwidth]{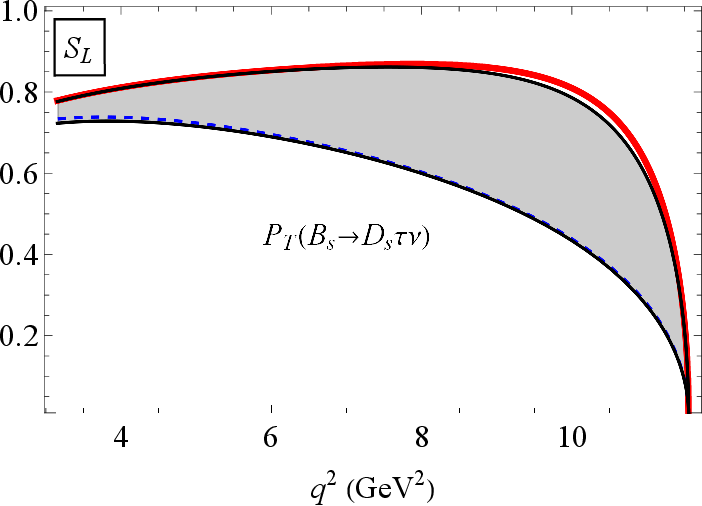}& \includegraphics[width=0.33\textwidth]{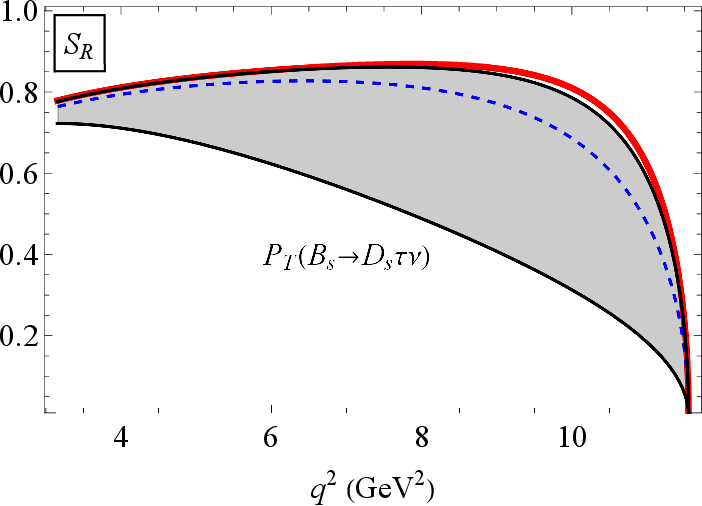}&
		\includegraphics[width=0.33\textwidth]{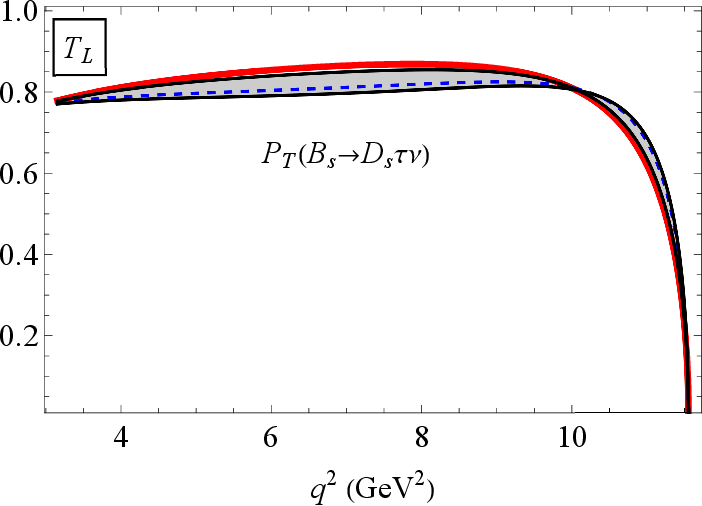}\\
		\includegraphics[width=0.33\textwidth]{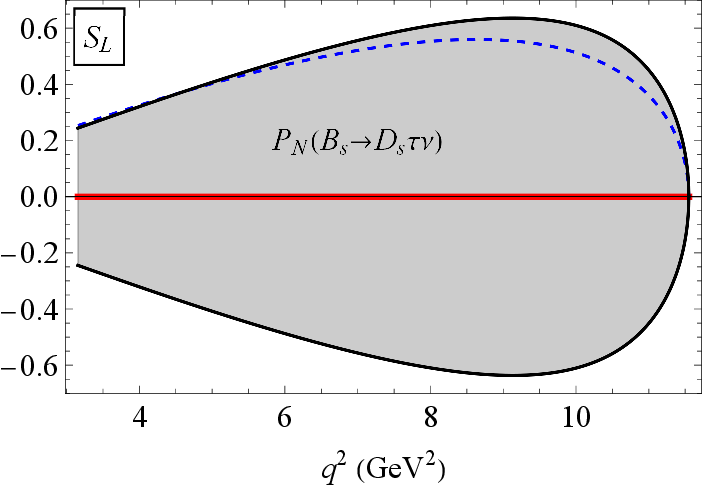}& 
		\includegraphics[width=0.33\textwidth]{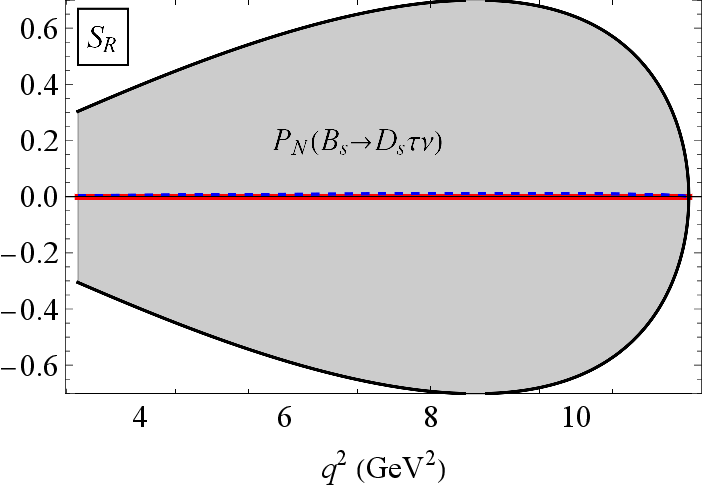}& 
		\includegraphics[width=0.33\textwidth]{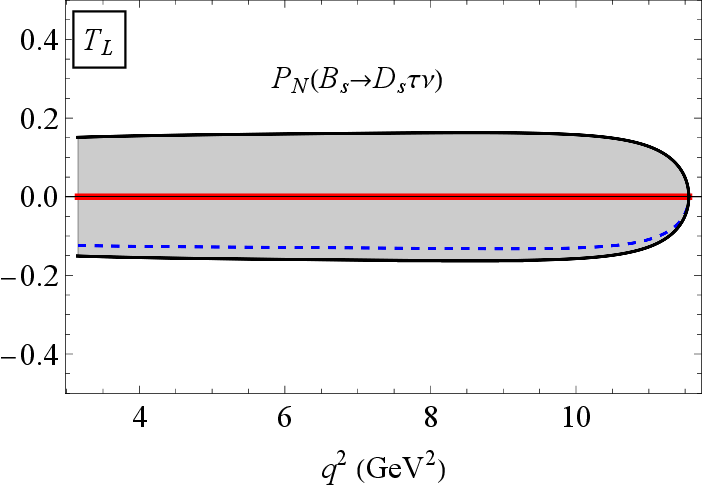}
	\end{tabular}
	\caption{Longitudinal (upper panels), transverse (middle panels), and normal (lower panels) polarization of the $\tau$ in the decays $B_s \to D_s\tau\nu$. Notations are the same as in Fig.~\ref{fig:R}. In this case, $\mathcal{O}_{V_R}$ does not affect these observables.}
	\label{fig:pol-Ds}
\end{figure}
\begin{figure}[htbp]
\centering
\begin{tabular}{ccc}
\includegraphics[width=0.33\textwidth]{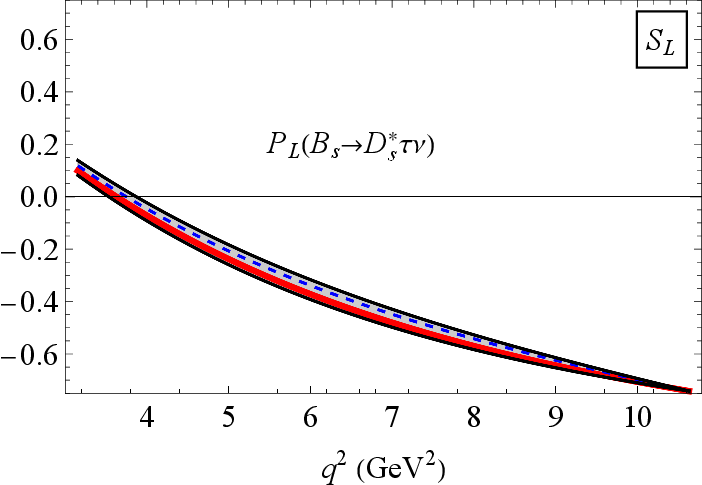}& 
\includegraphics[width=0.33\textwidth]{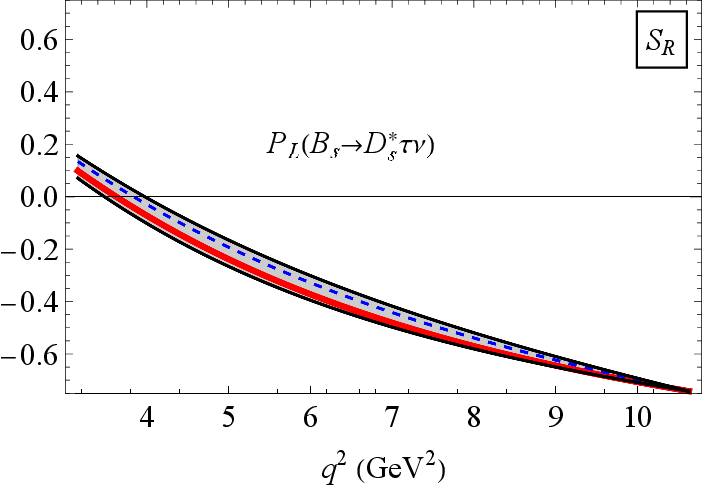}&
\includegraphics[width=0.33\textwidth]{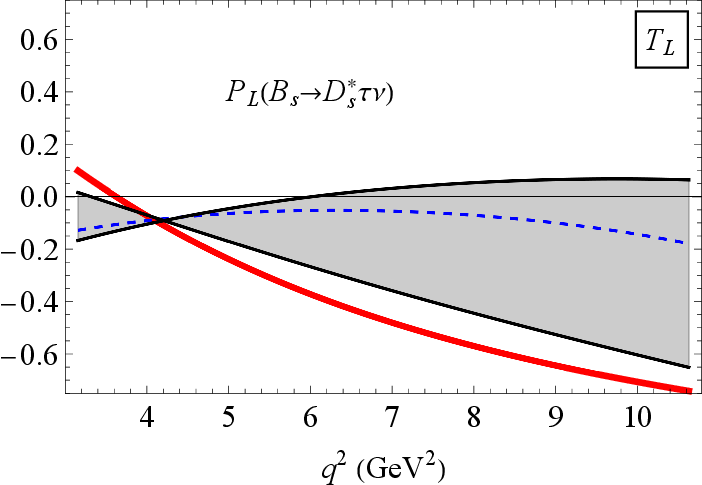}\\
\includegraphics[width=0.33\textwidth]{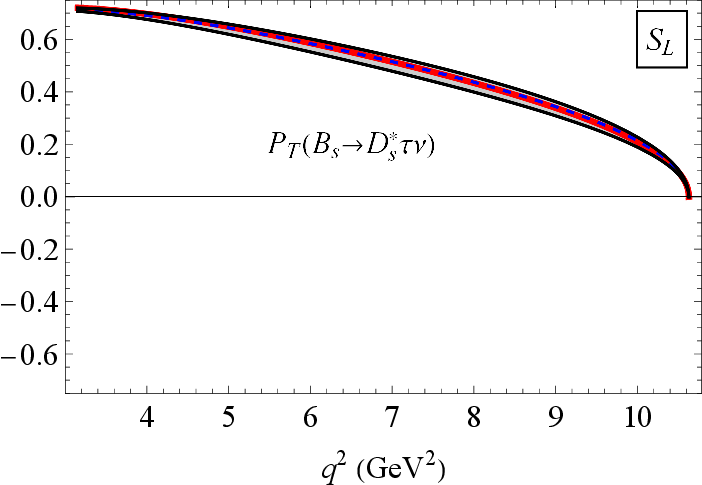}& 
\includegraphics[width=0.33\textwidth]{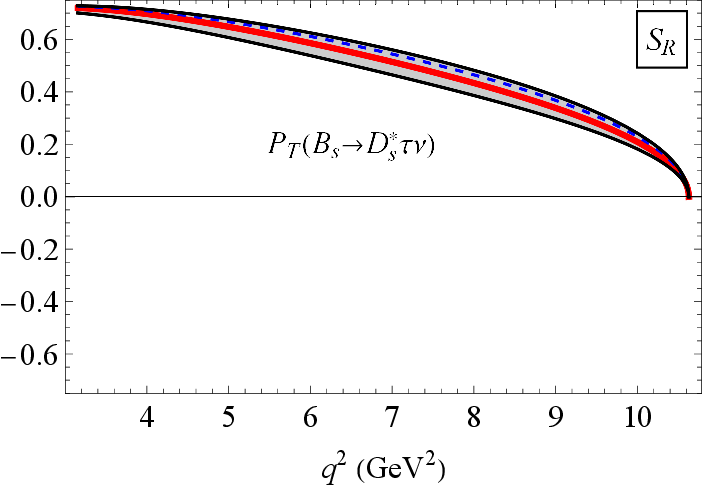}&
\includegraphics[width=0.33\textwidth]{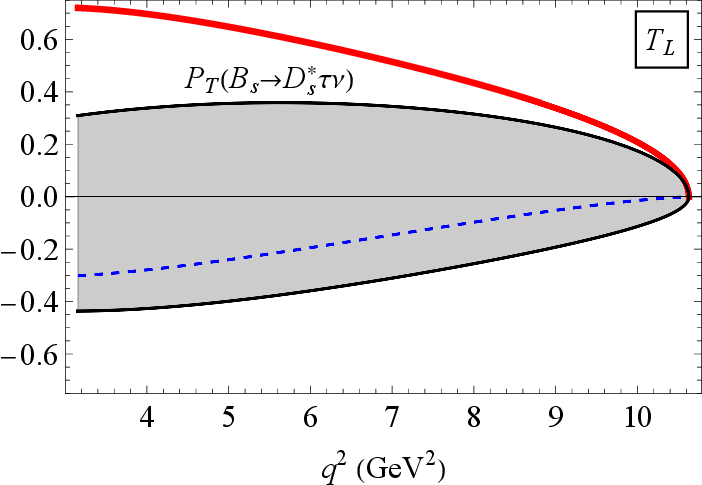}\\
\includegraphics[width=0.33\textwidth]{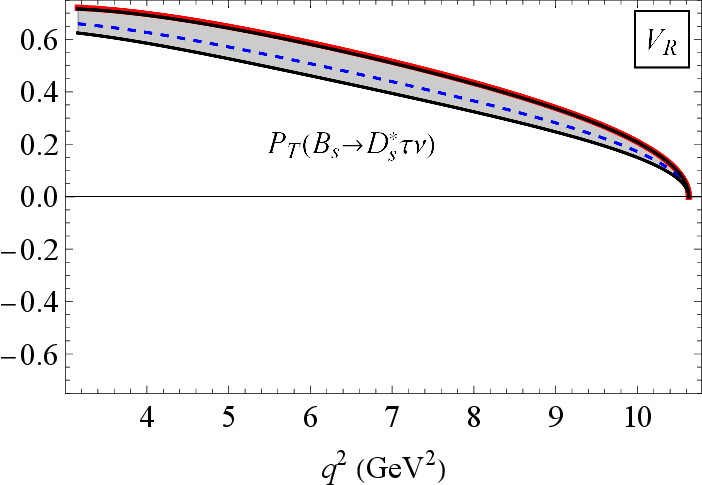}&&\\
\includegraphics[width=0.33\textwidth]{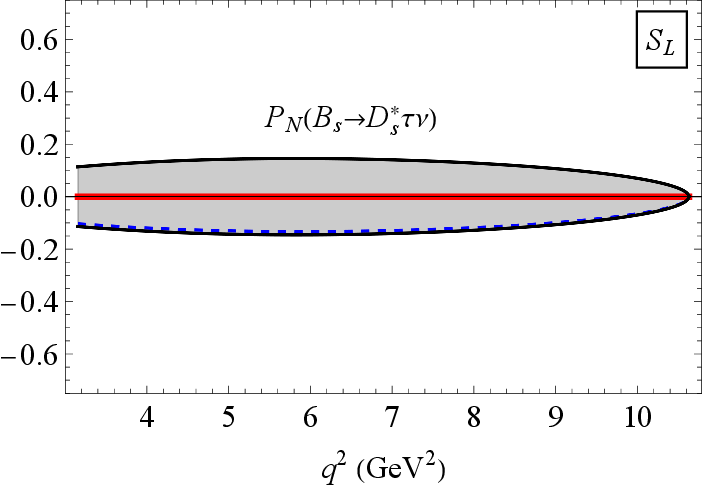}&
\includegraphics[width=0.33\textwidth]{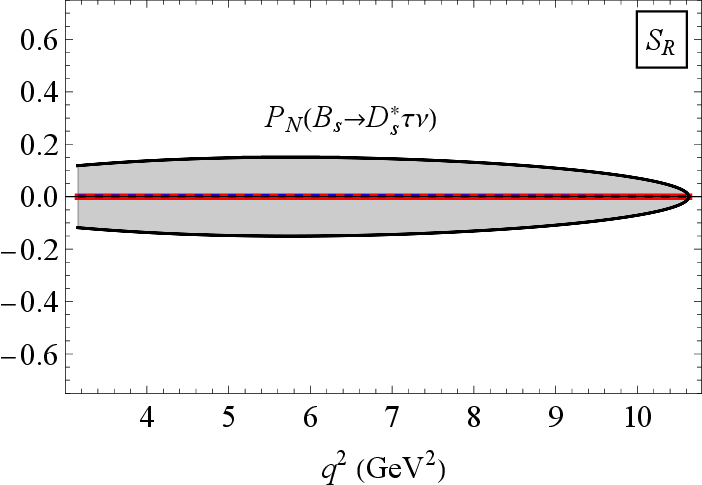}&  
\includegraphics[width=0.33\textwidth]{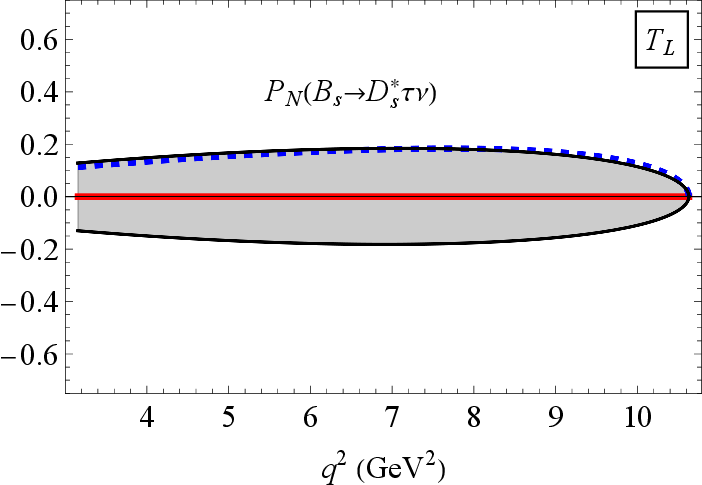}
\end{tabular}
\caption{Longitudinal (upper panels), transverse (middle panels), and normal (lower panels) polarization of the $\tau$ in the decays $B_s \to D_s^*\tau\nu$. Notations are the same as in Fig.~\ref{fig:R}.}
\label{fig:pol-Dvs}
\end{figure}

The $q^2$ dependence of the tau polarizations are presented in Figs.~\ref{fig:pol-Ds} and~\ref{fig:pol-Dvs}. First, we discuss the longitudinal ($P_L$) and transverse ($P_T$) polarizations. In the pseudoscalar channel $B_s \to D_s\tau\nu$, the scalar operators $\mathcal{O}_{S_L}$ and $\mathcal{O}_{S_R}$ significantly enhance $P_L$ while simultaneously suppressing $P_T$ relative to the SM. The vector channel $B_s \to D^*_s\tau\nu$ is remarkably sensitive to the tensor operator $\mathcal{O}_{T_L}$. In $P_L$, the tensor contribution induces a dramatic sign reversal at low $q^2$ values, shifting the polarization from positive (SM) to negative (NP). Similarly, $T_L$ causes a massive suppression of the transverse polarization $P_T$, potentially leading to values that are entirely excluded by the SM prediction. The $P_L$ and $P_T$ components in the $D_s^*$ channel remain relatively stable under $\mathcal{O}_{S_L}$, $\mathcal{O}_{S_R}$, and $\mathcal{O}_{V_R}$ operators, making these observables specific smoking guns for tensor-type interactions. 
The normal polarization $P_N$ is a particularly clean observable because the SM predicts it to be effectively zero. Our results show that all NP scenarios ($S_{L,R}$, $V_R$, and $T_L$) allow for nonzero values of $P_N$. Especially, in the $D_s$ channel, scalar operators produce wide allowed bands for $P_N$, indicating strong sensitivity to these couplings. The $q^2$ averages of the final lepton polarization are given in Table~\ref{tab:lepton-pol-avg}.
\begin{table}[htbp] 
	\caption{The $q^{2}$ averages of the final lepton polarization.}
	\centering
	\begin{tabular}{lccccccc}
		\hline\hline
		\multicolumn{8}{c}{ $B_s\to D_s$ }\\
		\hline
	{}	&  $\left\langle P^\tau_{L}\right\rangle$ 			  
	&  $\left\langle P^\mu_{L}\right\rangle$
	&  $\left\langle P^e_{L}\right\rangle$ 
	&  $\left\langle P^\tau_{T}\right\rangle$
	&  $\left\langle P^\mu_{T}\right\rangle$ 			 
	&  $\left\langle P^e_{T}\right\rangle$
	&  $\left\langle P^\tau_{N}\right\rangle$ 
		\\		\hline
		SM \quad & $0.34(4)$  & $-0.96(10)$ & $-1.0(1)$ & $0.84(9)$ & $0.21(2)$ & $1.1(1)\times 10^{-3}$ & 0\\ 
		$\mathcal{O}_{S_L}$\quad & $(0.37,0.59)$ &&& $(0.58,0.83)$ & & & $(-0.53,0.53)$ \\
		$\mathcal{O}_{S_R}$\quad & $(0.37,0.59)$ &&& $(0.50,0.83)$ & & & $(-0.61,0.61)$\\
		$\mathcal{O}_{T_L}$\quad & $(0.20,0.31)$ &&& $(0.79,0.83)$ & & & $(-0.16,0.16)$\\
		\hline\hline
		\multicolumn{8}{c}{ $B_s\to D_s^*$ }\\ 	\hline
		&  $\left\langle P^\tau_{L}\right\rangle$ 			  
		&  $\left\langle P^\mu_{L}\right\rangle$
		&  $\left\langle P^e_{L}\right\rangle$ 
		&  $\left\langle P^\tau_{T}\right\rangle$
		&  $\left\langle P^\mu_{T}\right\rangle$ 			 
		&  $\left\langle P^e_{T}\right\rangle$
		&  $\left\langle P^\tau_{N}\right\rangle$ 
		\\			\hline
		SM \quad & $-0.50(5)$ & $-0.98(10)$ & $-1.0(1)$ & $0.45(5)$ & $0.089(9)$ & $4.7(5)\times 10^{-4}$ & 0	\\
		$\mathcal{O}_{S_L}$\quad & $(-0.52,-0.46)$ &&& $(0.42,0.47)$ & & & $(-0.12,0.12)$\\
		$\mathcal{O}_{S_R}$\quad & $(-0.52,-0.45)$ &&& $(0.41,0.50)$ & & & $(-0.13,0.13)$\\
		$\mathcal{O}_{V_R}$\quad & $(-0.51,-0.50)$ &&& $(0.35,0.45)$ & & & 0\\
		$\mathcal{O}_{T_L}$\quad & $(-0.39,0.019)$ &&& $(-0.29,0.31)$ & & & $(-0.17,0.17)$\\
		\hline\hline
	\end{tabular}
	\label{tab:lepton-pol-avg}
\end{table}

\subsection{Summary of NP signatures in \boldmath{$B_s \to D_s^{(*)} \tau \nu$}}
\label{subsec:strategy}
To summarize how every observable we have discussed responds to the various NP operators we create a comprehensive table as a master map (Table~\ref{tab:sum}). It highlights the specific signatures that can be used to distinguish the origin of NP. A primary highlight of this analysis is the diagnostic power of the hadron-side convexity parameter $C_F^h(D_s^*)$. In the SM, $C_F^h(D_s^*)$ remains strictly positive across the entire $q^2$ range. The tensor operator $\mathcal{O}_{T_L}$ is the only one capable of driving $C_F^h(D_s^*)$ to negative values. An experimental measurement of negative hadron-side convexity would provide unambiguous evidence of tensor interactions. The lepton-side convexity parameter $C_F^\tau(D_s^*)$ is another powerful tool to probe for the tensor current since scalar  and vector operators largely preserve the SM trajectory, while $\mathcal{O}_{T_L}$ massively changes the values of the parameter. Beyond tensor physics, other operators leave distinct fingerprints in the angular and polarization distributions. Scalar operators ($\mathcal{O}_{S_L},\,\mathcal{O}_{S_R}$) are most effectively probed via the forward-backward asymmetry in the $D_s$ channel, where they induce a dramatic magnitude suppression. They also significantly suppress the transverse $\tau$ polarization $P_T(D_s)$. The right-handed vector operator $\mathcal{O}_{V_R}$ is uniquely identified by the azimuthal moments $W_{II}$ and $W_{IT}$. Since the SM predicts these to be zero, any nonzero measurement serves as a direct null-test violation indicating right-handed currents.
\begin{table}[htbp]
	\caption{Master summary of NP signatures in $B_s \to D_s^{(*)} \tau \nu$.}
	\centering
	\begin{tabularx}{\textwidth}{cccX}
		\hline\hline
		Category & Observable  & Primary sensitivity   &  Key analyzing feature\\\hline
		Global rates & $R(D_s)$  & All NP  & 	Significant enhancement in scalar scenarios\\
		{} & $R(D_s^*)$ & $\mathcal{O}_{V_L},\,\mathcal{O}_{V_R},\,\mathcal{O}_{T_L}$ & $\mathcal{O}_{T_L}$ uniquely alters $R(D_s^{*})$ functional form \\
		Lepton polar & $\mathcal{A}_{FB}(D_s^*)$ & $\mathcal{O}_{V_R},\, \mathcal{O}_{T_L}$ & $\mathcal{O}_{V_R}$ significantly shifts the zero-crossing point toward high $q^2$\\
		{} & $\mathcal{A}_{FB}(D_s)$ & $\mathcal{O}_{S_L},\,\mathcal{O}_{S_R}$ & Dramatic magnitude decrease.\\
		{} & $C_F^\tau(D_s^*)$ & $\mathcal{O}_{T_L}$ & Massive shift toward more negative values\\
		{} & $C_F^\tau(D_s)$ & $\mathcal{O}_{S_L},\,\mathcal{O}_{S_R},\,\mathcal{O}_{T_L}$ & Significant shift toward less negative values\\
		Hadron polar & $C_F^h(D_s^*)$ & $\mathcal{O}_{T_L}$ & Extreme sensitivity; $\mathcal{O}_{T_L}$ can push $C_F^h(D_s^*)$ into negative values (impossible in the SM)\\
		Azimuthal & $W_T,\, W_I$ & $\mathcal{O}_{T_L}$ & High sensitivity; $\mathcal{O}_{T_L}$ is the only operator reversing the SM trend.\\
		{} & $W_A$ & $\mathcal{O}_{V_R},\, \mathcal{O}_{T_L}$ & Significant value decrease \\
		{} & $W_{IA},\, W_{IT},\, W_{II}$ & $\mathcal{O}_{V_R}$ & The ultimate null test: any nonzero signal indicates right-handed vector currents\\
		$\tau$ polarization & $P_L(D_s^*)$ & $\mathcal{O}_{T_L}$ & Unique sign flip at low $q^2$ (SM positive $\to$ NP negative)\\
		{} & $P_T(D_s^*)$ & $\mathcal{O}_{T_L}$ & Extreme suppression and potential for broad sign reversal across spectrum\\
		{} & $P_T(D_s)$ & $\mathcal{O}_{S_L},\,\mathcal{O}_{S_R}$ & Massive and consistent suppression from the SM baseline\\
		{} & $P_N$ (Any) & General NP & SM null test; allows for non-zero values identifying CP-violating phases\\
		\hline\hline
	\end{tabularx}
	\label{tab:sum}
\end{table}

Before ending this subsection, we discuss the difference in the behavior of the $q^2$ distribution of the various NP signatures presented in Table~\ref{tab:sum}. In particular, we focus on why the SM baseline (represented by the red lines) is excluded from the $2\sigma$ NP bands (the gray bands) for some observables, whereas it is cleanly enclosed for others. We take the lepton-side convexity parameter $C_F^\tau(D_s^*)$ and the normal polarization $P_N(D_s^{(*)})$ under the tensor operator scenario as two examples.

On the one hand, for the normal polarizations $P_N(D_s^{(*)})$, the red SM lines remain perfectly and symmetrically in the center of the gray band (see Figs.~\ref{fig:pol-Ds} and~\ref{fig:pol-Dvs}). This containment is the result of the underlying Lorentz and helicity structure of $P_N(D_s^{(*)})$. The analytical expressions of the normal polarizations under the effect of the tensor operator are given by~\cite{Ivanov:2017mrj} 
\bea
\label{eq:PN}
P_N^D(q^2)&=&\frac{6\pi m_\tau^2}{{\cal H}_{tot}^D} {\rm Im}T_L H_t H_T,\nn
P_N^{D^\ast}(q^2)&=&-\frac{3\pi}{2{\cal H}_{tot}^{D^\ast}}
{\rm Im}T_L
\big[H_{+}H_T^+-H_{-}H_T^- - \frac{m_\tau^2}{q^2}\big(H_{+}H_T^+-H_{-}H_T^- -2 H_{t}H_T^0\big) \big]
.
\ena
Since $P_N$ is strictly proportional to the imaginary part of the tensor coupling, the SM baseline vanishes. At the same time, the allowed region of $T_L$ is symmetric with respect to the real axis (see Fig.~\ref{fig:2sigma}). When our scanning procedure samples this symmetric allowed region to establish the extreme upper and lower boundaries of the envelope, it selects mirror-image values of $T_L$, which forces the zero-valued SM baseline to act as the exact geometric mirror axis for the gray band. 
\begin{figure}[htbp]
	\centering
	\begin{tabular}{cc}
		\includegraphics[width=0.5\textwidth]{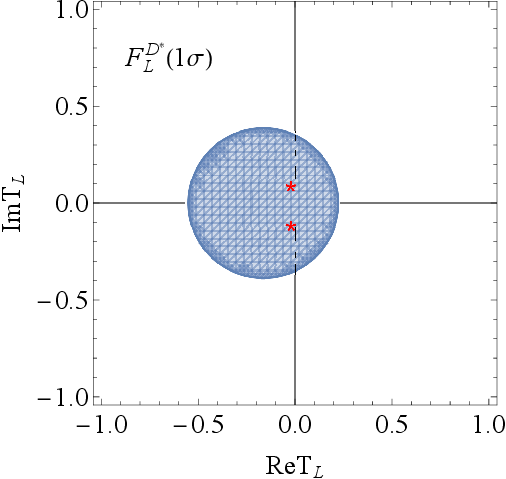}&
		\includegraphics[width=0.5\textwidth]{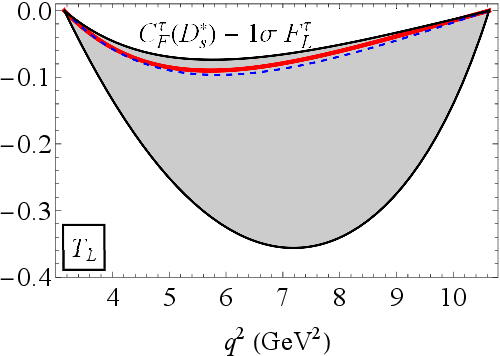}
	\end{tabular}
	\caption{The $1\sigma$ constraint on $T_L$ from experimental data of $F_L^{D^*}$ (left) and the corresponding $q^2$ behavior of $C_F^\tau(D_s^*)$ (right). Best-fit values are indicated by the red asterisks and the blue dashed line. The red line represents the SM prediction.}
	\label{fig:TL1sigmaFL}
\end{figure}

On the other hand, as seen in the last plot of Fig.~\ref{fig:CFL}, the gray band for $C_F^\tau(D_s^*)$ under $\mathcal{O}_{T_L}$ floats entirely clear of the red SM line across the full kinematic range. This tension is a direct consequence of the global fit framework, which captures both the internal tensions within the experimental dataset and the underlying model-dependence of the hadronic form factors. Because the global fit is performed using CCQM form factors across all channels, any numerical variation between the baseline CCQM SM predictions and the experimental data causes the $\chi^2$ optimizer to favor relatively large, nonzero Wilson coefficients to bridge the gap. However, this source of tension has a small effect since we have added a conservative 10\% theoretical uncertainty to the CCQM predictions used in the fit. The fit can therefore tolerate a large amount of distance between the SM and the data.

The main reason of this behavior is the irreducible experimental excess [such as the large central values of $R(J/\psi)$ and $R(D^{(*)})$] and possible hidden tensions among various experimental constraints used in the fit. As a result, the minimizer is forced to shift the entire $2\sigma$ parameter space into a region where the Wilson coefficients are definitively nonzero.
When these highly constrained parameter sets are mapped onto the $B_s \rightarrow D_s^{(*)}$ observables, the resulting envelopes are structurally translated away from the $C_i = 0$ coordinate, and the gray bands naturally float cleanly away from the red SM lines.
To demonstrate this point, we fit $T_L$ using only one constraint:  $F_L^{D^*}({B} \to D^{\ast} \tau\bar{\nu}_\tau)=0.43\pm 0.06\pm 0.03$~\cite{LHCb:2023ssl}. The $1\sigma$ allowed region for $T_L$ and the corresponding $q^2$ behavior of $C_F^\tau(D_s^*)$ are depicted in Fig.~\ref{fig:TL1sigmaFL}. The resulting $1\sigma$ allowed parameter space for $T_L$ forms a disk that fully encloses the origin $(0,0)$. Consequently, when this isolated parameter space is mapped onto $C_F^\tau(D_s^*)$, the SM baseline is fully enclosed within the gray envelope. We emphasize that, in this test, our conclusions regarding the sensitivity and overall behavior of the tensor operator effect remain completely intact and unchanged.
We note that a detailed study using a more rigorous statistical approach to precisely isolate and quantify the exact tensions among  various experimental constraints and individual observables is highly needed [see, e.g.,~Refs.~\cite{Fedele:2023ewe, Martinelli:2024bov, Bordone:2025jur}]. However, such an extensive analysis is beyond the scope of the present paper and will be addressed in our future study. 

\subsection{Binwise exploration of coefficient functions}
\label{subsec:Jbin}
Recently, the Belle Collaboration reported the first measurement of all angular coefficients of the decay $\bar{B}\to D^*\ell\bar{\nu}_\ell$~\cite{Belle:2023xgj}. More importantly, the results were obtained for four bins that almost cover the entire region of momentum transfer squared. This allows tests of LFU based on the $q^2$ dependence of the angular coefficients, rather than just the average values over the whole $q^2$ range as often seen earlier. Inspired by this new achievement, we decided to adopt a similar approach in defining and predicting the physical observables in the decay $B_s\to D_s^{*}\ell\nu$. To be more specific, we directly consider the angular coefficients as observables instead of combining them into traditional ones. A recent study using the measured angular coefficients by Belle~\cite{Belle:2023xgj} to constrain the NP couplings in $\bar{B}\to D^*\ell\bar{\nu}_\ell$ was provided by Colangelo \textit{et al.} in~\cite{Colangelo:2024mxe}.

Note that the definition of the angular coefficients by the Belle Collaboration in Ref.~\cite{Belle:2023xgj} is not the same as the one we use in this study [Eq.~(\ref{eq:J})]. To be more specific, the Belle Collaboration absorbed the $q^2$-dependent factor $|\mathbf{p_2}|q^2$ into the coefficients. Besides, they used the variable $w\equiv (m_B^2-m_{D^*}^2-q^2)/(2m_Bm_{D^*})$ instead of $q^2$. Moreover, the Belle Collaboration determined the angular coefficients in bins of $w$, $\bar{J}_i = \int_{\Delta w} J_i(w) dw$, and quoted the normalized angular coefficients as $\hat{J}_i = \bar{J}_i/\mathcal{N}$, where $\mathcal{N} = \frac89 \pi \sum_{k=1}^{4}\left( 3\bar{J}^k_{1c} + 6\bar{J}^k_{1s}-\bar{J}^k_{2c}-2\bar{J}^k_{2s} \right)$. For easy comparison with the Belle Collaboration's result and future experiments, we define $J_i^{\rm Belle}\equiv |\mathbf{p_2}|q^2 J_i^{\rm our}$. The corresponding normalized values are calculated according to $\hat{J}^{\rm Belle}_i = J^{\rm Belle}_i/J^{\rm Belle}_{\rm tot}$, with $J^{\rm Belle}_{\rm tot} = \frac89 \pi \left( 3J^{\rm Belle}_{1c}+6J^{\rm Belle}_{1s}-J^{\rm Belle}_{2c}-2J^{\rm Belle}_{2s}\right)$.
We then study the $w$-dependence of these normalized coefficients and calculate their values in several $w$ bins.
\begin{figure}[htbp]
	\centering
	\begin{tabular}{ccc}
		\includegraphics[width=0.33\textwidth]{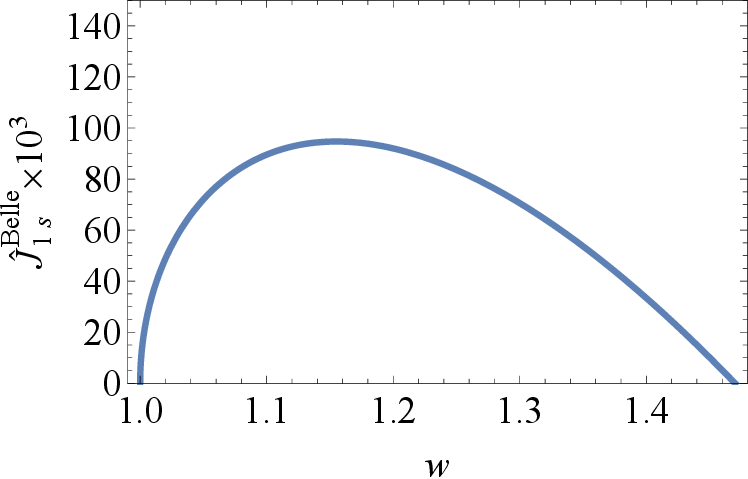}&
		\includegraphics[width=0.33\textwidth]{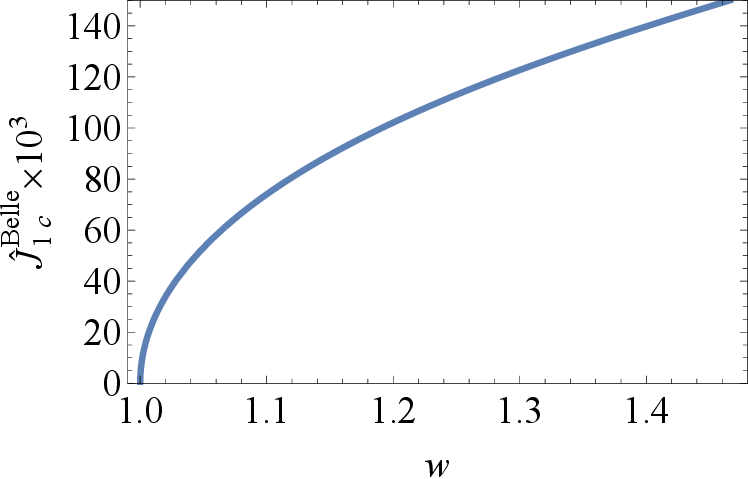}&
		\includegraphics[width=0.33\textwidth]{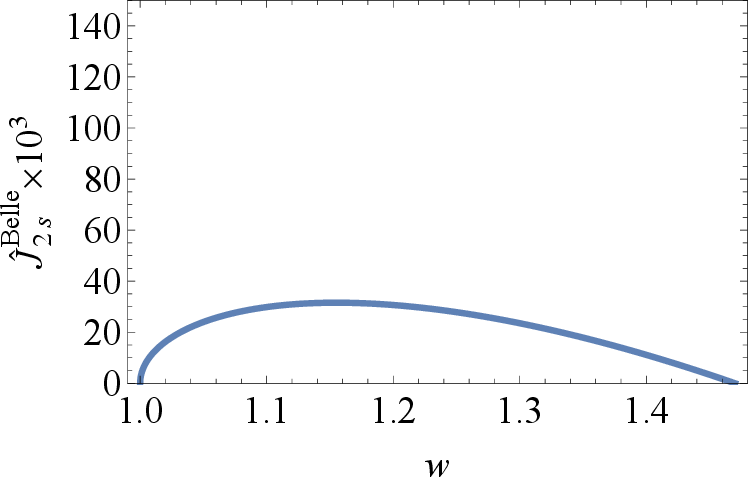}\\
		\includegraphics[width=0.33\textwidth]{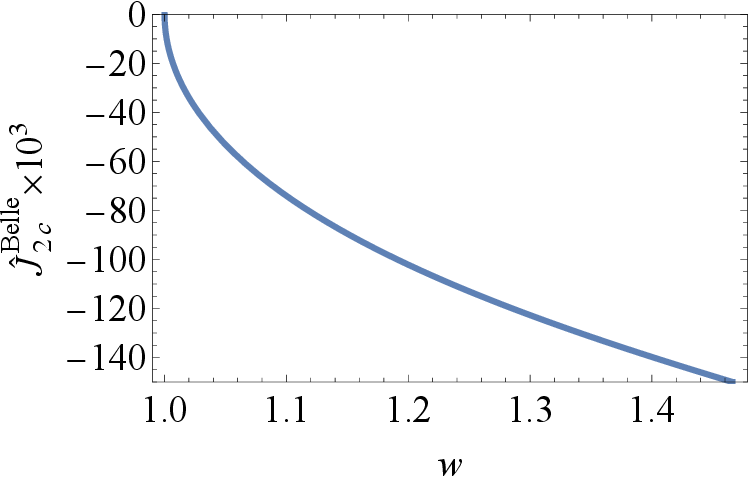}&
		\includegraphics[width=0.33\textwidth]{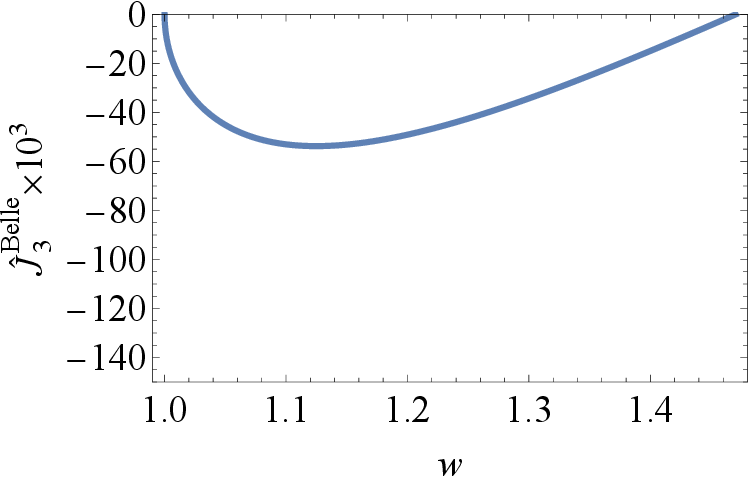}&	
		\includegraphics[width=0.33\textwidth]{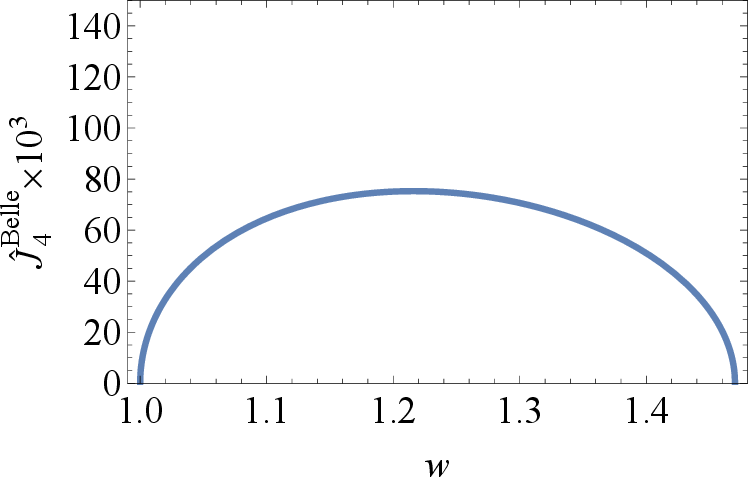}\\
		\includegraphics[width=0.33\textwidth]{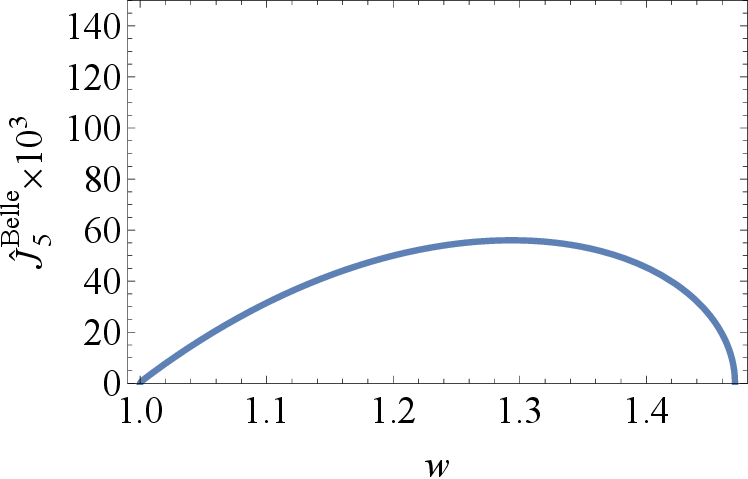}&
		\includegraphics[width=0.33\textwidth]{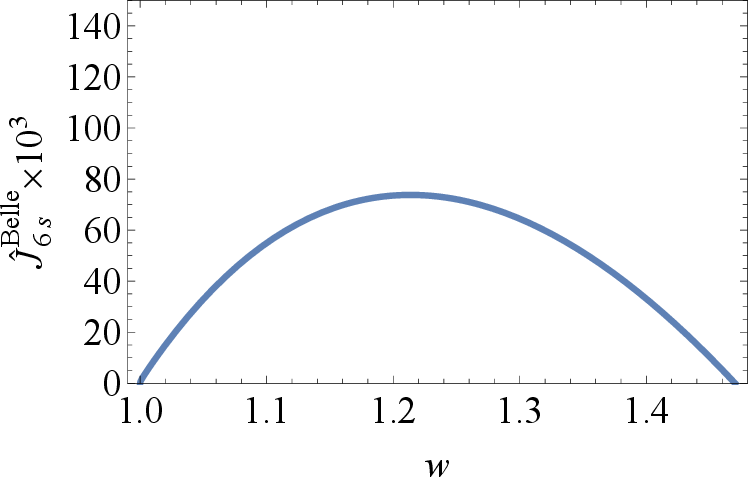}&	
		\includegraphics[width=0.33\textwidth]{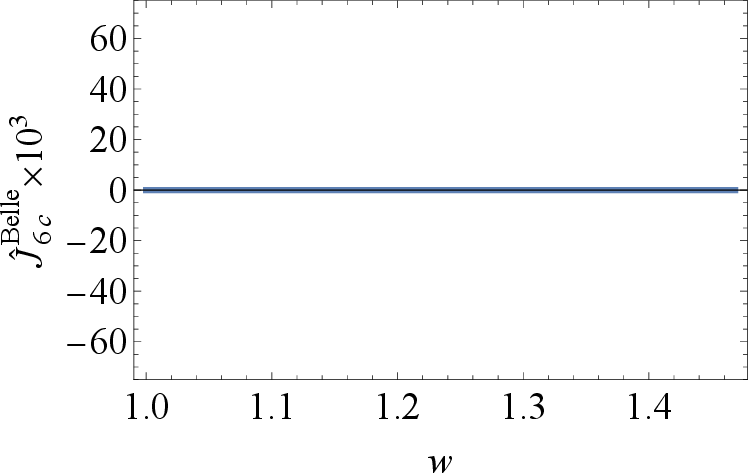}\\
	\end{tabular}
	\caption{$w$-Dependence of the normalized coefficient functions $\hat{J}^{\rm Belle}_i$ for the decay $B_s\to D^*_s e \nu_e$ in the SM.}
	\label{fig:Ji-ell-SM}
\end{figure}

First, we show in Fig.~\ref{fig:Ji-ell-SM} the $w$-dependence of the angular coefficients $\hat{J}^{\rm Belle}_i$ for the electron mode $B_s\to D^*_s e \nu_e$. The results for the muon mode do not differ significantly from the electron one, and are not displayed. Note again that NP operators are assumed to affect the tau mode only. Within the SM, the coefficients $J_7$, $J_8$, and $J_9$ are equal zero. In the zero-mass limit of the lepton, one also has $J_{6c} = 0$. The Belle Collaboration also used their measured coefficients to calculate several observables in four $w$ bins. These observables include the forward-backward asymmetry $\mathcal{A}_{FB}$, longitudinal polarization fraction $F_L(D^*)$, and two nonzero observables $S_{3,5}$ directly proportional to $J_{3,5}$:
\begin{eqnarray}
	S_3 &=& \frac{1}{\pi}\frac{4J_3}{3J_{1c}-J_{2c}+2(3J_{1s}-J_{2s})},\nn
	S_5 &=& \frac{3J_5}{3J_{1c}-J_{2c}+2(3J_{1s}-J_{2s})}.
\end{eqnarray}  
In Table~\ref{tab:ell-bins}, we provide our predictions for these observables in the decay $B_s\to D^*_s e \nu_e$. As a crosscheck, we also list the corresponding values in the case of $B\to D^* e \nu_e$ and compare with the Belle Collaboration's results~\cite{Belle:2023xgj}. Our predictions for $B\to D^* e \nu_e$ agree well with those from the Belle Collaboration. Note that a recent detailed analysis of the Belle Collaboration dataset for the angular functions $J_i$~\cite{Belle:2023xgj} was performed by Martinelli~\emph{et al.} in Ref.~\cite{Martinelli:2024vde}. The authors also found a good agreement between the Belle Collaboration results~\cite{Belle:2023xgj} and the SM predictions based on LQCD form factors.
\begin{table}[htbp] 
	\caption{Observables for electron mode in four $w$ bins. The Belle Collaboration's results~\cite{Belle:2023xgj} are given in bold text.}
	\centering
	\begin{tabular}{ccccc}
		\hline\hline
		\multicolumn{5}{c}{ $B\to D^* e\nu$ }\\
		\hline
		$w$ bin & $\mathcal{A}_{FB}$ & $F_L$ & $S_3$ & $S_5$
		\\		\hline
		$1.00<w<1.15$ & $0.19(2)$ & $0.37(4)$ & $-0.09(1)$ & $0.11(1)$\\
		{} & \bm{$0.23(3)$} & \bm{$0.29(4)$} & \bm{$-0.11(2)$} & \bm{$0.07(5)$}
		\\
		$1.15<w<1.25$ & $0.24(3)$ & $0.44(5)$ & $-0.070(7)$ & $0.16(2)$\\
		{} & \bm{$0.30(3)$} & \bm{$0.45(4)$} & \bm{$-0.06(2)$} & \bm{$0.23(4)$}\\
		$1.25<w<1.35$ & $0.23(2)$ & $0.54(6)$ & $-0.050(5)$ & $0.19(2)$ \\
		{} & \bm{$0.29(3)$} & \bm{$0.47(3)$} & \bm{$-0.03(2)$} & \bm{$0.26(4)$}\\
		$1.35<w<1.50$ & $0.13(1)$ & $0.75(8)$ & $-0.030(3)$ & $0.17(2)$ \\
		{} & \bm{$0.16(3)$} & \bm{$0.70(3)$} & \bm{$-0.01(2)$} & \bm{$0.20(3)$}\\
		\hline\hline
		\multicolumn{5}{c}{ $B_s\to D_s^* e\nu$ }\\ 	\hline
		$w$ bin & $\mathcal{A}_{FB}$ & $F_L$ & $S_3$ & $S_5$ 
		\\			\hline
		$1.00<w<1.15$ & 0.19(2) & 0.38(4) & $-0.09(1)$ & 0.11(1)
		\\
		$1.15<w<1.25$ & 0.24(3) & 0.46(5) & $-0.070(7)$ & 0.17(2) \\
		$1.25<w<1.35$ & 0.22(2) & 0.57(6) & $-0.050(5)$ & 0.19(2) \\
		$1.35<w<1.47$ & 0.12(1) & 0.79(8) & $-0.020(2)$ & 0.17(2) \\
		\hline\hline
	\end{tabular}
	\label{tab:ell-bins}
\end{table}

\begin{figure}[htbp]
	\begin{tabular}{lr}
		\includegraphics[scale=0.8]{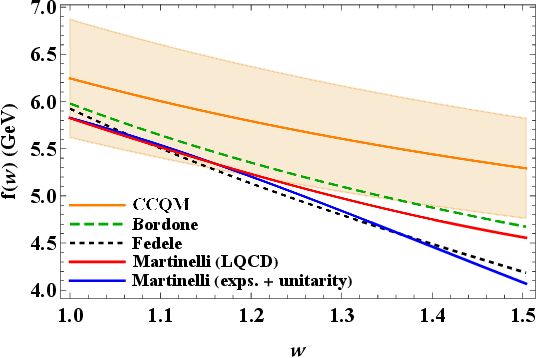}
		& 
		\includegraphics[scale=0.8]{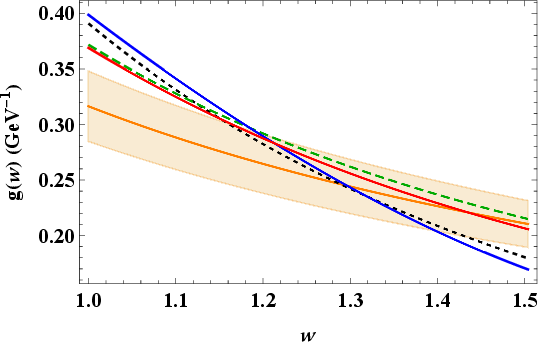}\\
		\includegraphics[scale=0.8]{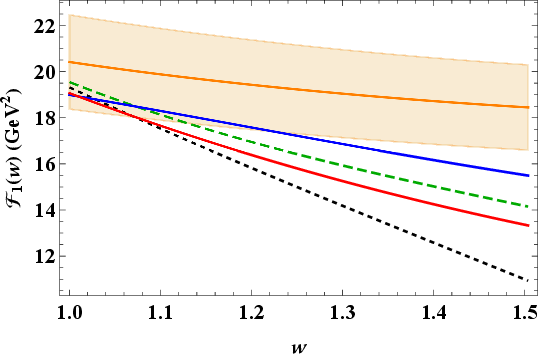}
		& 
		\includegraphics[scale=0.8]{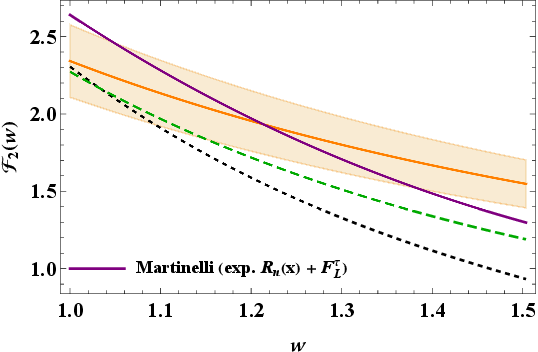}
	\end{tabular}
	\caption{$B\to D^*$ form factors in the CCQM and in other approaches~\cite{Bordone:2025jur, Fedele:2023ewe, Martinelli:2024bov}.}
	\label{fig:fwCompare}
\end{figure}
Recently, an important debate is going on in the community regarding the shapes of the FFs in $B\to D^* \ell \nu_\ell$. The question is, which shape of the FFs is preferred: by LQCD only, by experimental data only, or by a combination of the two (same question for QCDSR)? Several groups have discussed this issue in great details, including Fedele~\emph{et al.}~\cite{Fedele:2023ewe}, Martinelli~\emph{et al.}~\cite{Martinelli:2024bov}, and Bordone~\emph{et al.}~\cite{Bordone:2025jur}. Given that our predictions for the angular observables in the decay $B\to D^* \ell \nu_\ell$ agree well with the Belle Collaboration's results~\cite{Belle:2023xgj}, it is instructive to compare explicitly our SM $B\to D^*$ FFs with those obtained in the papers mentioned above. Note that our $B\to D^{(*)}$ FFs were calculated in our previous paper~\cite{Ivanov:2015tru}. 

The comparison is given in Fig.~\ref{fig:fwCompare}. For easy comparison, we have converted all FFs to the standard ones, namely, $f(w)$, $g(w)$, $\mathcal{F}_1(w)$, and $\mathcal{F}_2(w)$~\cite{Boyd:1997kz}.  We show our FFs with an error band; while for the FFs given by other groups, we take only their central lines for simplicity. The green dashed lines are related to Fig.~2 of Ref.~\cite{Bordone:2025jur} by Bordone~\emph{et al.} [their nominal results in the $3/2/1^*$ model with $\textrm{SU(3)}_F$ symmetry]. The black dashed lines are taken from Fig.~3 of Ref.~\cite{Fedele:2023ewe} by Fedele~\emph{et al.} [obtained in the dispersive matrix (DM) approach when performing predictions]. The red solid and blue solid lines are taken from Fig.~3 of Ref.~\cite{Martinelli:2024bov} by Martinelli~\emph{et al.} (the red line is obtained in the DM approach using LQCD data, while the blue line is extracted from experimental data of $B\to D^*\ell\nu$). The purple solid line [appearing only in $\mathcal{F}_2(w)$)] is taken from Fig.~6 of Ref.~\cite{Martinelli:2024bov}, where several results for $\mathcal{F}_2(w)$ (depending on which additional $\tau$ observable is used in the fit) are presented. From these results, we choose the central line of the band that is obtained by combining experimental data for $B\to D^*\ell\nu$ and $F_L^\tau$~\cite{Martinelli:2024bov}. This is a reasonable choice for our comparison since the corresponding band overlaps well with other bands, and the central line is close to the LQCD points [see~\cite{Martinelli:2024bov} for more details].  

We make the following observations:
\begin{enumerate}[label=\roman*.]
\item 
Our FFs exhibit a generally less steep behavior compared to those in other approaches. This is similar to the case of $B_s\to D_s^{(*)}$ discussed in Sec.~\ref{sec:FF}. The slopes of our FFs agree best with the Bordone ones, and least with the Fedele ones.
\item 
Comparing with the Martinelli FFs, our $f(w)$ and $g(w)$ prefer the LQCD results over the experimental fit; while for $\mathcal{F}_1(w)$, the tendency is reversed.
\item 
Regarding $\mathcal{F}_2(w)$, our prediction for the zero-recoil point $(w=1)$ is close to the Bordone and Fedele ones. Meanwhile, over the whole physical range, our $\mathcal{F}_2(w)$ agree well with Martinelli (the purple line), which also means a good agreement with LQCD since this line is close to the LQCD points.
\end{enumerate}
%
\begin{figure}[htbp]
	\centering
	\begin{tabular}{cccc}
		\includegraphics[width=0.2\textwidth]{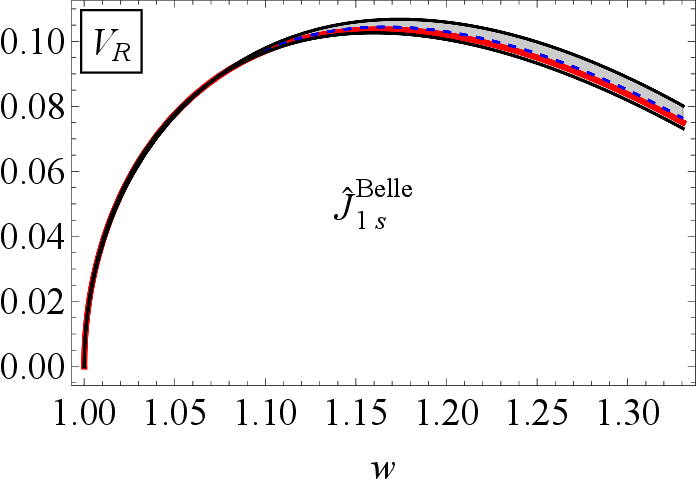}&
		\includegraphics[width=0.2\textwidth]{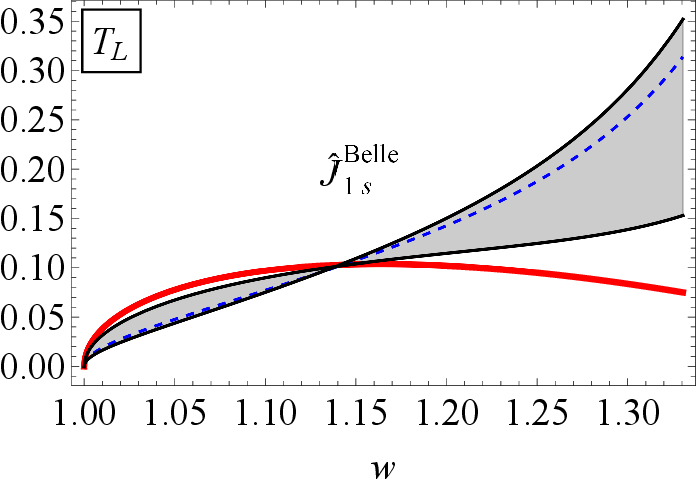}&
		\includegraphics[width=0.2\textwidth]{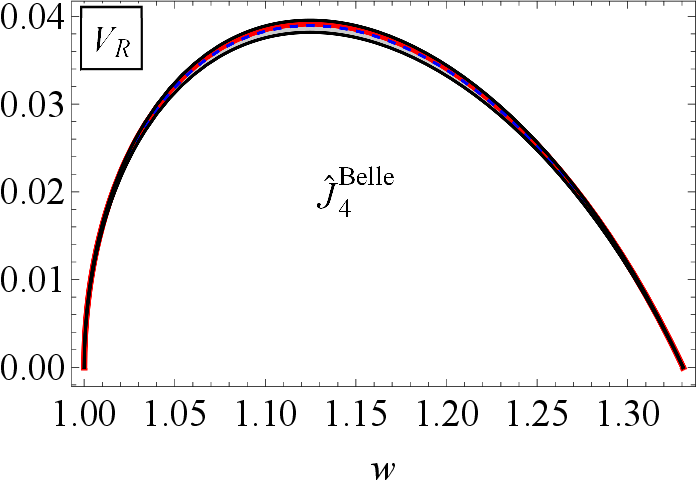}&
		\includegraphics[width=0.2\textwidth]{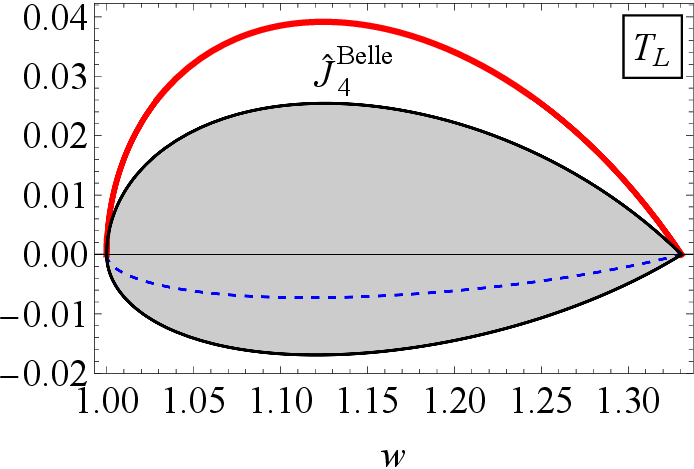}\\
		\includegraphics[width=0.2\textwidth]{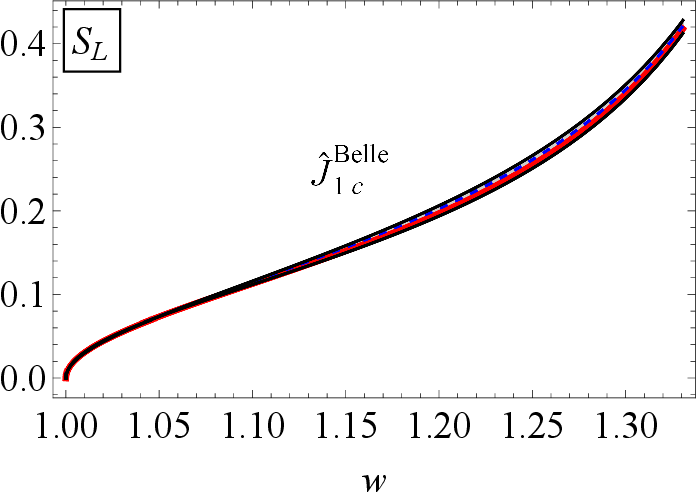}&
		\includegraphics[width=0.2\textwidth]{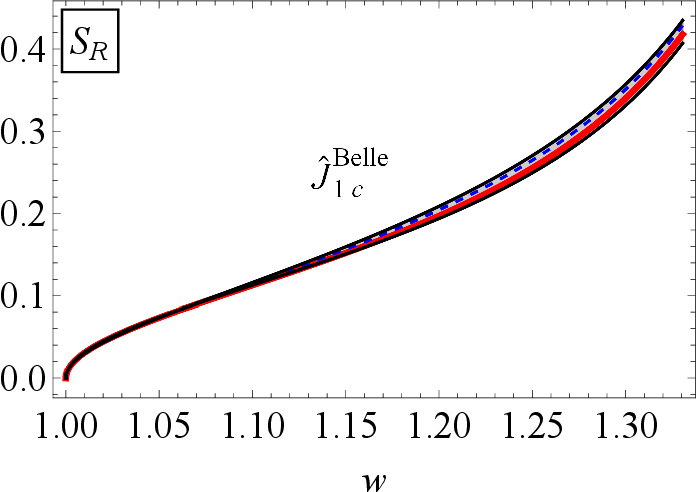}&	
		\includegraphics[width=0.2\textwidth]{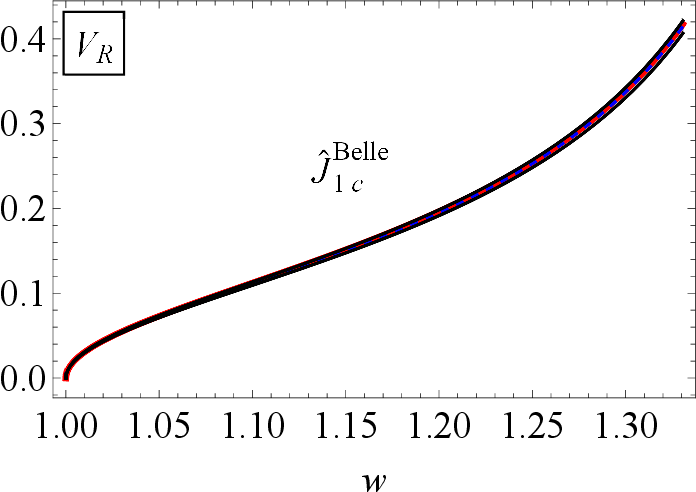}&
		\includegraphics[width=0.2\textwidth]{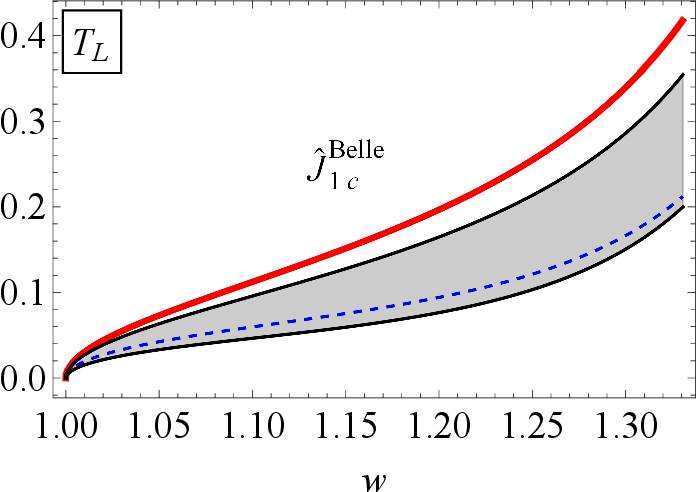}\\
		\includegraphics[width=0.2\textwidth]{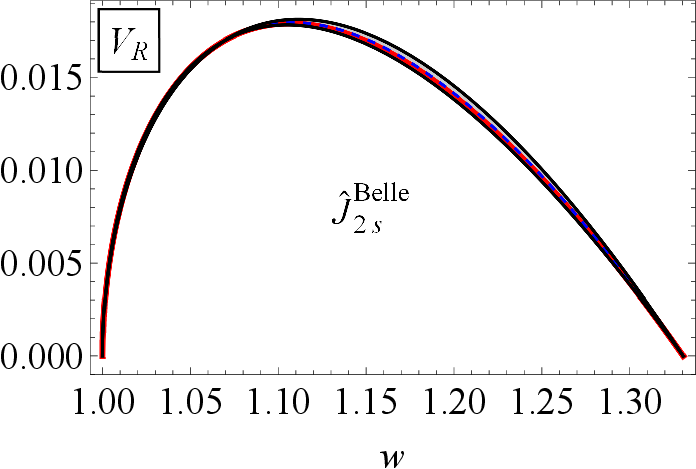}&
		\includegraphics[width=0.2\textwidth]{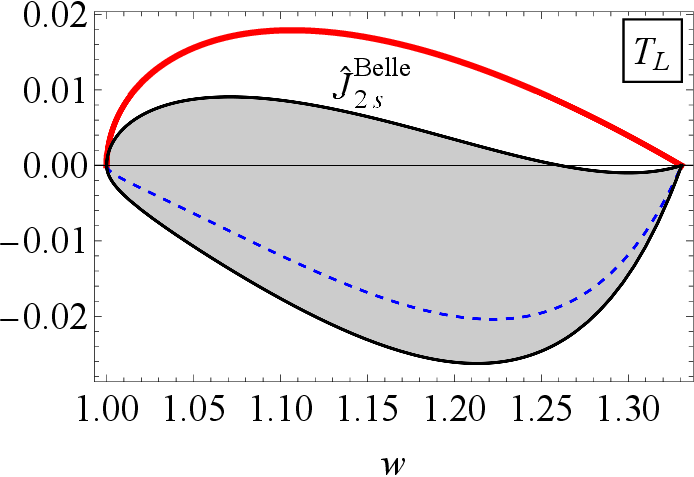}&	
		\includegraphics[width=0.2\textwidth]{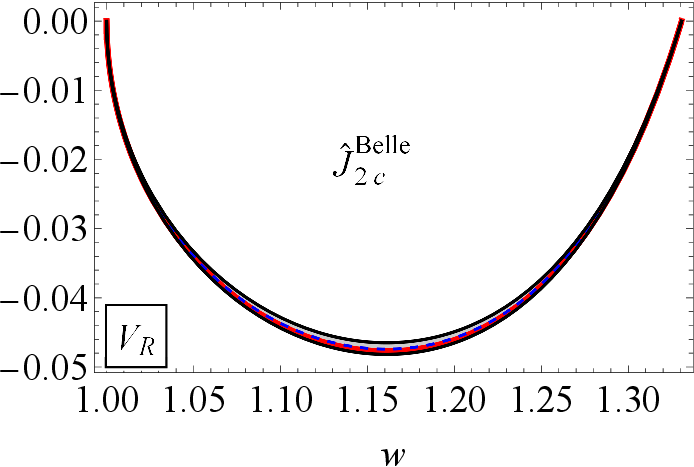}&
		\includegraphics[width=0.2\textwidth]{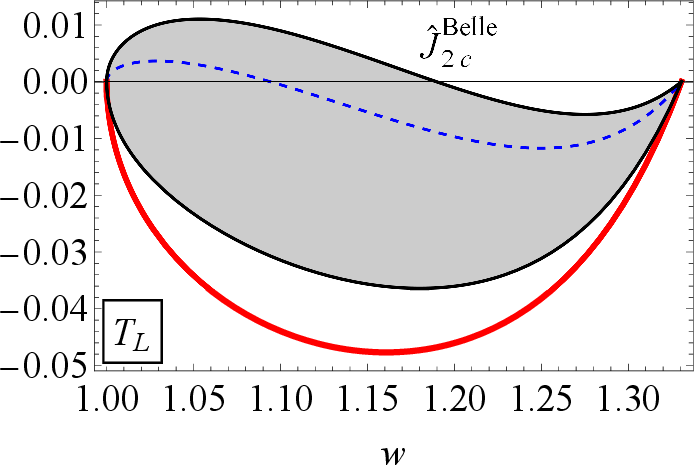}\\
		\includegraphics[width=0.2\textwidth]{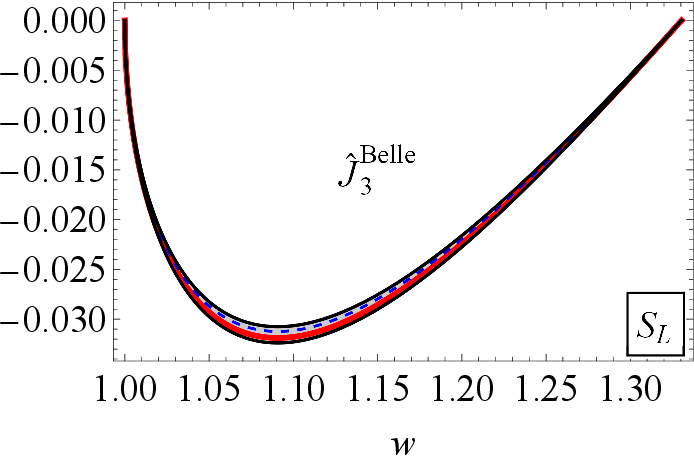}&
		\includegraphics[width=0.2\textwidth]{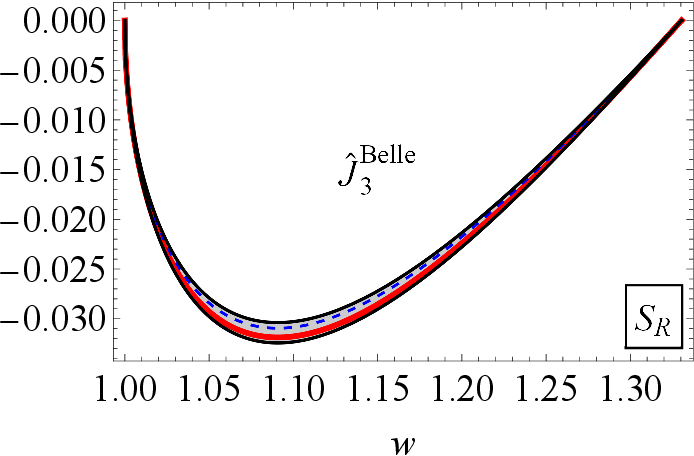}&	
		\includegraphics[width=0.2\textwidth]{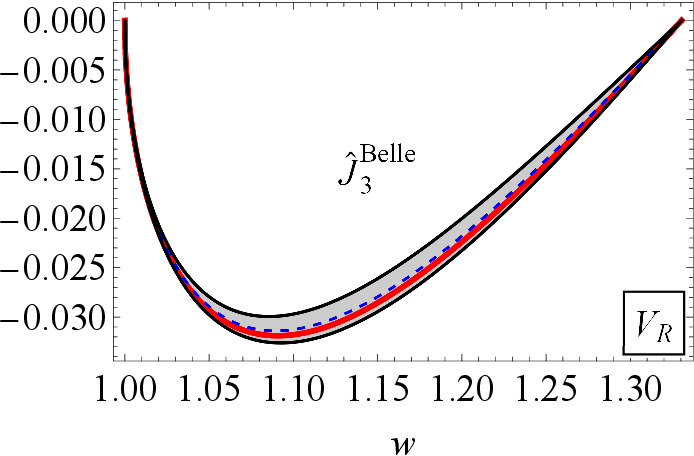}&
		\includegraphics[width=0.2\textwidth]{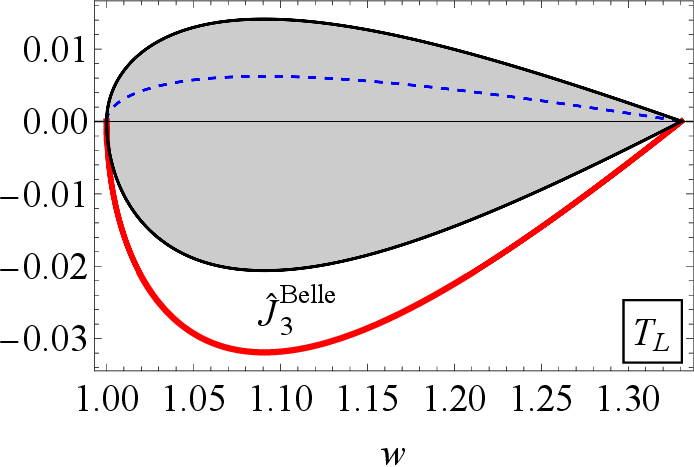}\\	
		\includegraphics[width=0.2\textwidth]{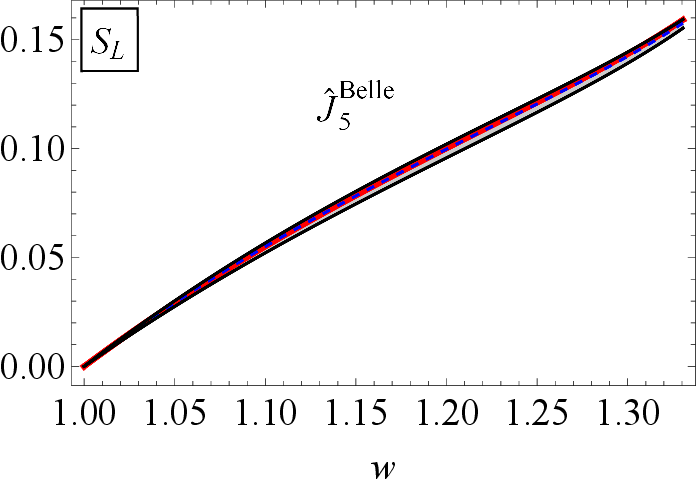}&
		\includegraphics[width=0.2\textwidth]{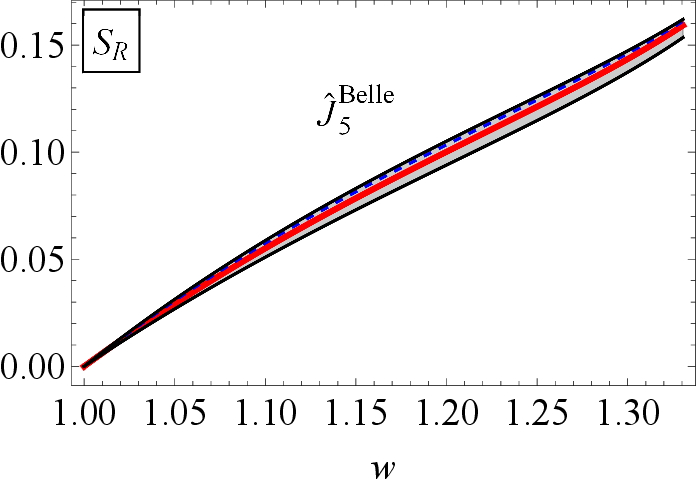}&	
		\includegraphics[width=0.2\textwidth]{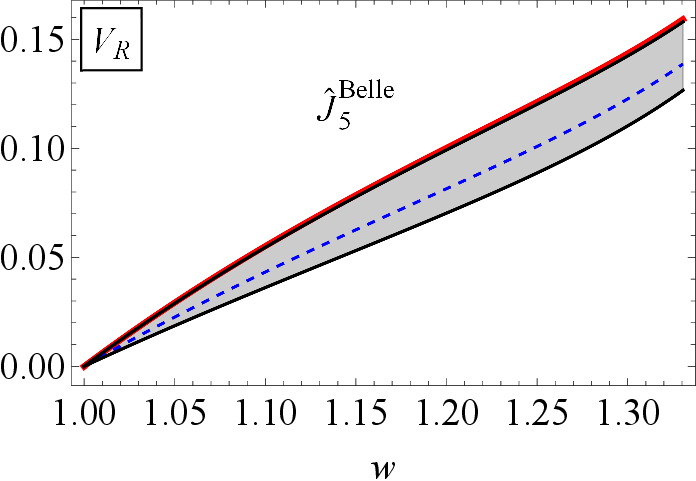}&
		\includegraphics[width=0.2\textwidth]{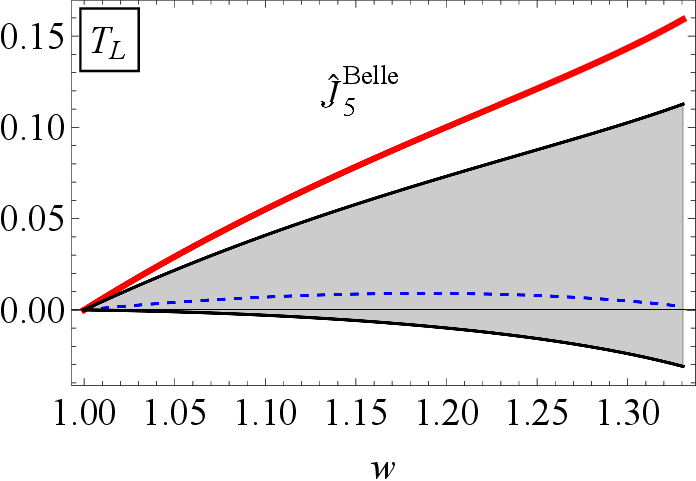}\\	
		\includegraphics[width=0.2\textwidth]{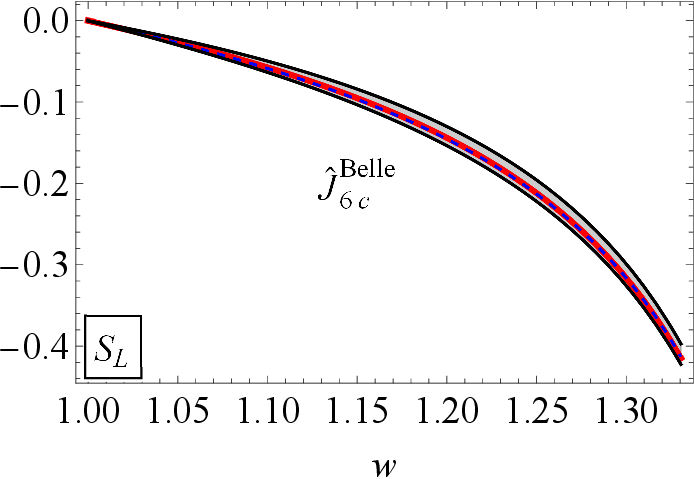}&
		\includegraphics[width=0.2\textwidth]{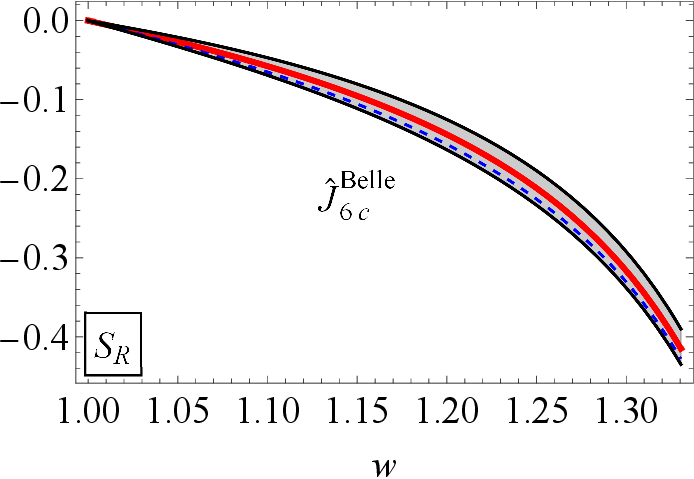}&	
		\includegraphics[width=0.2\textwidth]{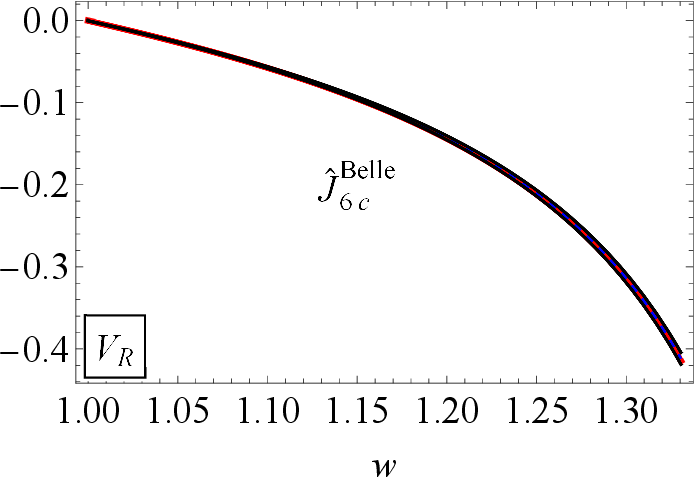}&
		\includegraphics[width=0.2\textwidth]{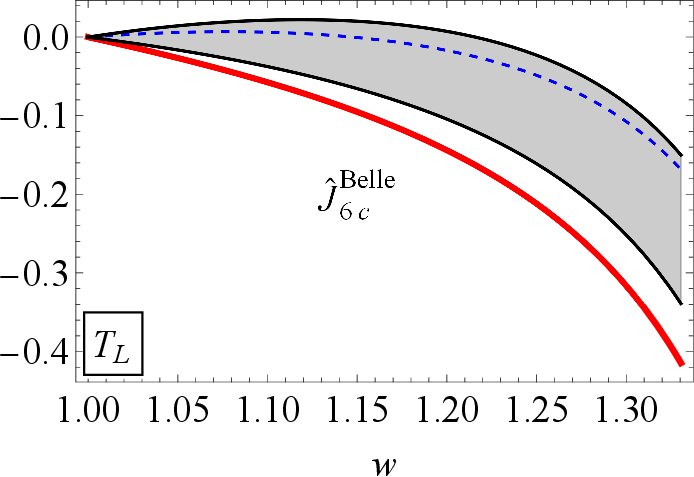}\\	
		\includegraphics[width=0.2\textwidth]{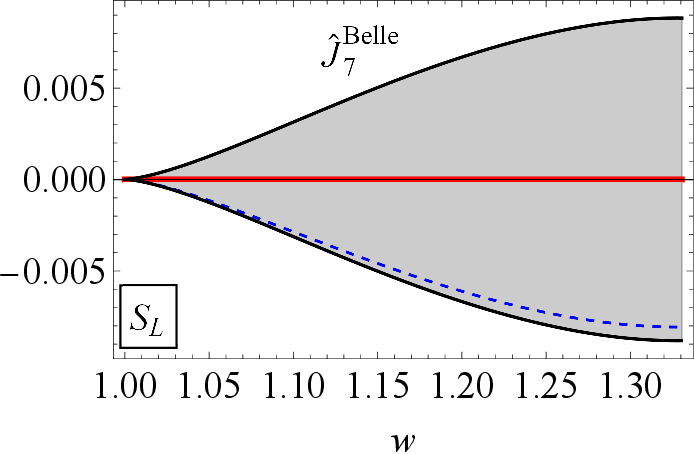}&
		\includegraphics[width=0.2\textwidth]{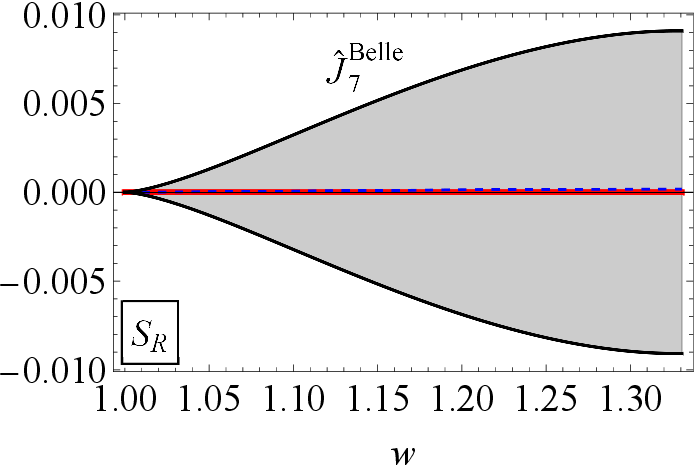}&	
		\includegraphics[width=0.2\textwidth]{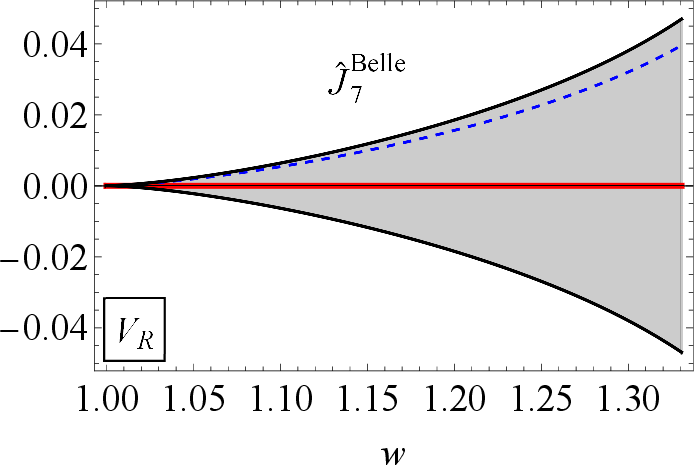}&
		\includegraphics[width=0.2\textwidth]{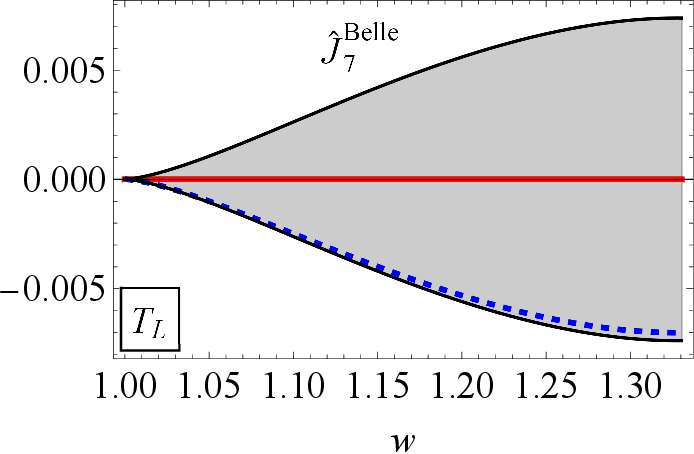}\\
		\includegraphics[width=0.2\textwidth]{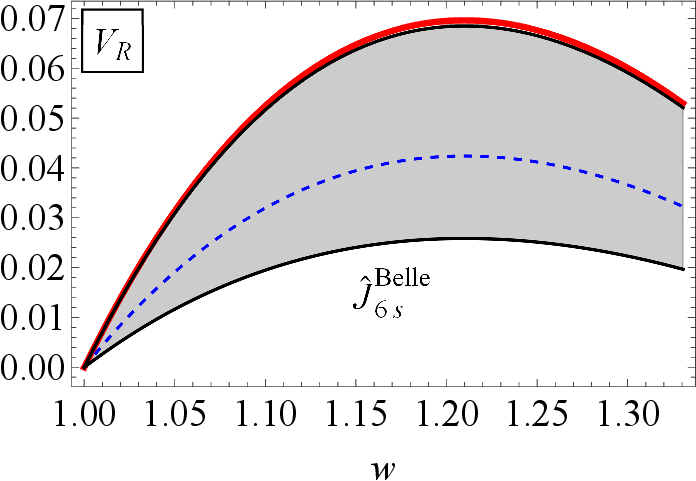}&
		\includegraphics[width=0.2\textwidth]{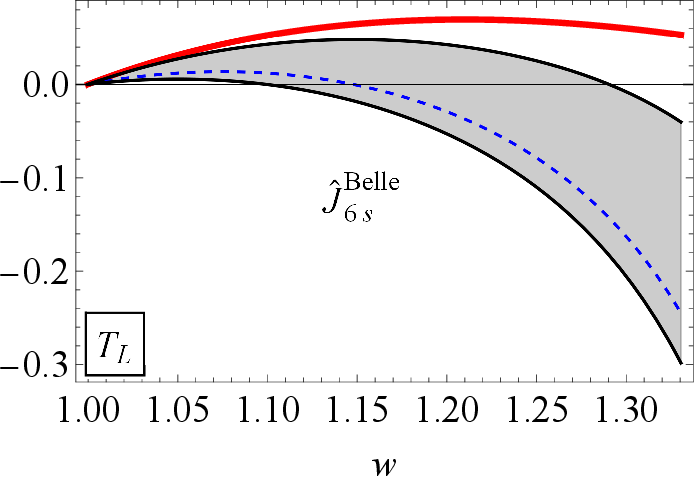}&	
		\includegraphics[width=0.2\textwidth]{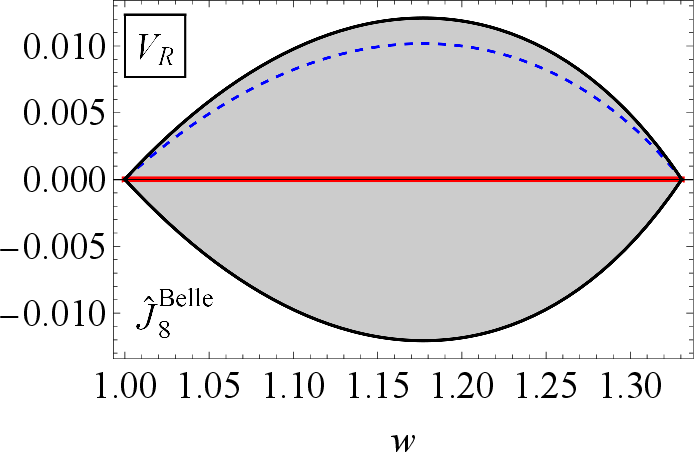}&
		\includegraphics[width=0.2\textwidth]{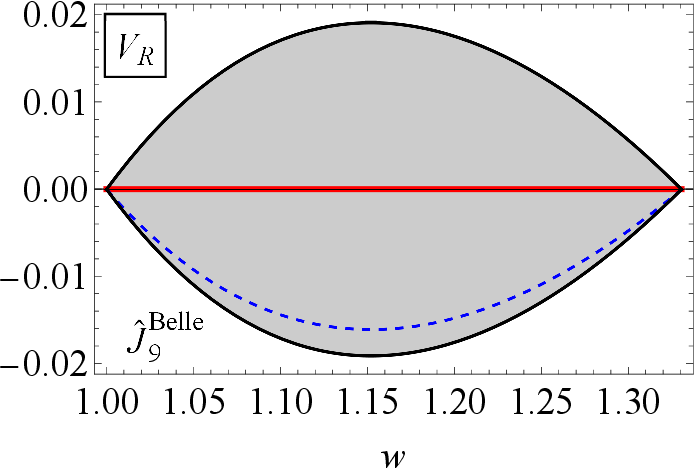}	
	\end{tabular}
	\caption{NP effects on the coefficient functions $\hat{J}^{\rm Belle}_i$. Notations are the same as in Fig.~\ref{fig:R}.}
	\label{fig:Ji}
\end{figure}

Next, we consider the effects of NP operators on the angular coefficients $\hat{J}^{\rm Belle}_i$ in the decay $B_s\to D_s^{*}\tau\nu$. In Fig.~\ref{fig:Ji}, we summarize all possible NP effects on the $w$ dependence of these coefficients. Based on these results, we propose the following strategy to isolate NP operators.  
\begin{enumerate}
	\item Step 1 ($\mathcal{O}_{V_R}$ extraction). We first focus on $\hat{J}_8$ and $\hat{J}_9$ since they are the most valuable null tests. They are immune to $\mathcal{O}_{S_{L,R}}$, $\mathcal{O}_{V_L}$, and $\mathcal{O}_{T_L}$. Any deviation from the SM line must be attributed to $\mathcal{O}_{V_R}$. If future experimental results for $\hat{J}_8$ and $\hat{J}_9$ match the SM, one can effectively set $V_R \approx 0$ for all other angular coefficients.
	\item Step 2 ($\mathcal{O}_{T_L}$ isolation). In this step, we focus on $\hat{J}_4$ and $\hat{J}_{6s}$. These are sensitive only to $\mathcal{O}_{V_R}$ and $\mathcal{O}_{T_L}$. However, the sensitivity of $\hat{J}_4$ to $\mathcal{O}_{V_R}$ is only minimal compared to $\mathcal{O}_{T_L}$. Therefore, any notable discrepancy in both $\hat{J}_4$ and $\hat{J}_{6s}$ is a direct confirmation of the tensor coupling.   
	\item Step 3 (Scalar degeneracy break). Now we focus on $\hat{J}_3$, $\hat{J}_5$, and $\hat{J}_7$. These coefficients respond to all operators (except for $\mathcal{O}_{V_L}$). We can use the $V_R$ and $T_L$ values from steps 1 and 2 to subtract their known effects. The leftover deviation, if any, is caused by the scalar operators $\mathcal{O}_{S_{L,R}}$. In order to confirm $\mathcal{O}_{S_L}$ and $\mathcal{O}_{S_R}$ we can consider in addition the lepton-side convexity parameter $C_F^\tau$ or the forward-backward asymmetry $\mathcal{A}_{FB}$ in both decays $B_s\to D_s\tau\nu$ and $B_s\to D^*_s\tau\nu$. These observables are extremely sensitive to $\mathcal{O}_{S_{L,R}}$ in the case of $B_s\to D_s\tau\nu$, while showing minimal to no sensitivity to the scalar operators in the case of $B_s\to D^*_s\tau\nu$. However, it would be difficult to distinguish between $\mathcal{O}_{S_L}$ and $\mathcal{O}_{S_R}$ because they affect the same set of observables with nearly the same sensitivity for any observable.
\end{enumerate}

It is worth noting that, in theory, $\hat{J}_8$ and $\hat{J}_9$ are golden observables because they are zero in the SM and only sensitive to $\mathcal{O}_{V_R}$. However, in practice, they are difficult to measure, especially for the $\tau$ mode, due to statistical limitations and the missing neutrino problem. We hope that our strategy will provide a useful roadmap for future experiments, such as Belle~II when it reaches its full luminosity. Finally, in Table~\ref{tab:Ji-avg}, we provide our predictions for the integrated values of all the coefficients functions in various bins of $w$, both in the SM and in the presence of NP. These values will be useful for future measurements of angular coefficients in the $\tau$ mode.

\begin{table}[htbp] 
	\caption{Integrated values of the coefficient functions $\hat{J}^{\rm Belle}_i$ in various bins of $w$ (in $10^{-3}$).}
	\centering
	\def\arraystretch{0.65}
	\begin{tabular}{cccccc}
		\hline\hline
		$\hat{J}^{\rm Belle}_i$ & Scenario & $\left[1.0,1.1\right]$ & $\left[1.1,1.2\right]$ & $\left[1.2,w_{\rm max}\right]$ & Total \\
		\hline	
\multirow{3}{*}{$\hat{J}_{1s}$} 
& SM    & 7.15(72) & 10.2(1.0) & 11.8(1.2) & 29.2(2.9)  \\
& $\mathcal{O}_{V_R}$ & $(7.16,7.12)$ & $(10.1,10.5)$ & $(11.6,12.5)$ & $(28.9,30.0)$  \\
& $\mathcal{O}_{T_L}$ & $(6.27,4.30)$ & $(10.3,11.0)$ & $(16.9,30.3)$ & $(33.5,45.6)$  \\
\hline
\multirow{5}{*}{$\hat{J}_{1c}$} 
& SM    & 7.04(0.70) & 15.3(1.5) & 37.4(3.8) & 59.7(6.0)  \\
& $\mathcal{O}_{S_L}$ & $(7.04,7.11)$ & $(15.1,15.9)$ & $(36.8,38.8)$ & $(58.9,61.8)$  \\
& $\mathcal{O}_{S_R}$ & $(7.10,7.08)$ & $(15.1,16.0)$ & $(36.5,39.4)$ & $(58.8,62.5)$  \\
& $\mathcal{O}_{V_R}$ & $(6.86,7.11)$ & $(14.9,15.4)$ & $(36.5,37.8)$ & $(58.2,60.3)$  \\
& $\mathcal{O}_{T_L}$ & $(3.15,6.09)$ & $(6.03,12.9)$ & $(16.0,31.5)$ & $(25.1,50.4)$  \\
\hline
\multirow{3}{*}{$\hat{J}_{2s}$} 
& SM    & 1.41(0.14) & 1.66(0.17) & 1.00(0.10) & 4.07(0.41)  \\
& $\mathcal{O}_{V_R}$ & $(1.41,1.40)$ & $(1.65,1.70)$ & $(0.98,1.05)$ & $(4.04,4.15)$  \\
& $\mathcal{O}_{T_L}$ & $(-1.04,0.78)$ & $(-2.27,0.63)$ & $(-2.56,0.05)$ & $(-5.87,1.47)$  \\
\hline
\multirow{3}{*}{$\hat{J}_{2c}$} 
& SM    & $-3.16(0.32)$ & $-4.67(0.47)$ & $-3.98(0.40)$ & $-11.8(1.2)$  \\
& $\mathcal{O}_{V_R}$ & $(-3.19,-3.08)$ & $(-4.72,-4.55)$ & $(-4.02,-3.88)$ & $(-11.9,-11.5)$  \\
& $\mathcal{O}_{T_L}$ & $(-2.18,0.95)$ & $(-3.50,0.41)$ & $(-3.25,-0.53)$ & $(-8.93,0.83)$  \\
\hline
\multirow{5}{*}{$\hat{J}_3$} 
& SM    & $-2.64(0.27)$ & $-2.81(0.28)$ & $-1.53(0.15)$ & $-6.97(0.70)$  \\
& $\mathcal{O}_{S_L}$ & $(-2.68,-2.54)$ & $(-2.85,-2.71)$ & $(-1.55,-1.47)$ & $(-7.09,-6.73)$  \\
& $\mathcal{O}_{S_R}$ & $(-2.68,-2.51)$ & $(-2.86,-2.68)$ & $(-1.55,-1.46)$ & $(-7.10,-6.65)$  \\
& $\mathcal{O}_{V_R}$ & $(-2.69,-2.51)$ & $(-2.90,2.57)$ & $(-1.60,-1.34)$ & $(-7.18,-6.42)$  \\
& $\mathcal{O}_{T_L}$ & $(-1.71,1.17)$ & $(-1.82,1.24)$ & $(-0.99,0.68)$ & $(-4.51,3.09)$  \\
\hline
\multirow{3}{*}{$\hat{J}_4$} 
& SM    & 2.94(0.30) & 3.78(0.38) & 2.64(0.27) & 9.36(0.94)  \\
& $\mathcal{O}_{V_R}$ & $(2.86,2.97)$ & $(3.68,3.82)$ & $(2.57,2.66)$ & $(9.12,9.45)$  \\
& $\mathcal{O}_{T_L}$ & $(-1.29,1.90)$ & $(-1.62,2.46)$ & $(-1.11,1.72)$ & $(-4.02,6.09)$\\
\hline
\multirow{5}{*}{$\hat{J}_5$} 
& SM    & 2.86(0.29) & 7.82(0.78) & 16.7(1.7) & 27.4(2.8)  \\
& $\mathcal{O}_{S_L}$ & $(2.70,2.95)$ & $(7.44,8.00)$ & $(16.2,16.9)$ & $(26.3,27.9)$  \\
& $\mathcal{O}_{S_R}$ & $(2.64,3.07)$ & $(7.28,8.27)$ & $(15.9,17.3)$ & $(25.8,28.6)$  \\
& $\mathcal{O}_{V_R}$ & $(1.84,2.83)$ & $(5.31,7.74)$ & $(12.5,16.6)$ & $(19.7,27.2)$  \\
& $\mathcal{O}_{T_L}$ & $(-0.12,2.13)$ & $(-0.60,5.76)$ & $(-2.44,12.1)$ & $(-3.17,19.9)$  \\
\hline
\multirow{3}{*}{$\hat{J}_{6s}$} 
& SM    & 2.97(0.30) & 6.35(0.64) & 8.39(0.84) & 17.7(1.8)  \\
& $\mathcal{O}_{V_R}$ & $(1.10,2.92)$ & $(2.36,6.25)$ & $(3.12,8.26)$ & $(6.58,17.4)$  \\
& $\mathcal{O}_{T_L}$ & $(0.37,2.54)$ & $(-2.13,4.64)$ & $(-19.2,1.56)$ & $(-21.0,8.73)$  \\
\hline
\multirow{5}{*}{$\hat{J}_{6c}$} 
& SM    & $-2.72(0.27)$ & $-9.68(0.97)$ & $-32.8(3.3)$ & $-45.2(4.5)$  \\
& $\mathcal{O}_{S_L}$ & $(-3.04,-2.31)$ & $(-10.5,-8.58)$ & $(-34.0,-30.7)$ & $(-47.6,-41.5)$  \\
& $\mathcal{O}_{S_R}$ & $(-3.33,-2.17)$ & $(-11.3,-8.19)$ & $(-35.5,-29.8)$ & $(-50.1,-40.2)$  \\
& $\mathcal{O}_{V_R}$ & $(-2.75,-2.65)$ & $(-9.77,-9.43)$ & $(-33.1,-31.9)$ & $(-45.6,-44.0)$  \\
& $\mathcal{O}_{T_L}$ & $(-1.71,-1.32)$ & $(-6.72,-1.83)$ & $(-25.5,-6.28)$ & $(-34.0,-3.13)$  \\
\hline
\multirow{5}{*}{$\hat{J}_{7}$} 
& SM    & 0 & 0 & 0 & 0  \\
& $\mathcal{O}_{S_L}$ & $(-0.14,0.14)$ & $(-0.50,0.50)$ & $(-1.05,1.05)$ & $(-1.69,1.69)$  \\
& $\mathcal{O}_{S_R}$ & $(-0.14,0.14)$ & $(-0.51,0.51)$ & $(-1.09,1.09)$ & $(-1.74,1.74)$  \\
& $\mathcal{O}_{V_R}$ & $(-0.26,0.26)$ & $(-1.20,1.20)$ & $(-4.03,4.03)$ & $(-5.48,5.48)$  \\
& $\mathcal{O}_{T_L}$ & $(-0.11,0.11)$ & $(-0.42,0.42)$ & $(-0.88,0.88)$ & $(-1.41,1.41)$  \\
\hline
\multirow{2}{*}{$\hat{J}_{8}$} 
& SM    & 0 & 0 & 0 & 0  \\
& $\mathcal{O}_{V_R}$ & $(-0.55,0.55)$ & $(-1.15,1.15)$ & $(-1.01,1.01)$ & $(-2.70,2.70)$  \\
\hline
\multirow{2}{*}{$\hat{J}_{9}$} 
& SM    & 0 & 0 & 0 & 0  \\
& $\mathcal{O}_{V_R}$ & $(-1.02,1.02)$ & $(-1.85,1.85)$ & $(-1.34,1.34)$ & $(-4.20,4.20)$  \\
		\hline\hline
	\end{tabular}
	\label{tab:Ji-avg}
\end{table}
\section{SUMMARY AND CONCLUSIONS}
\label{sec:summary}
In this study, we have performed a systematic and detailed analysis of the semileptonic decays $B_s \to D_s^{(*)} \tau \bar{\nu}$ as a powerful laboratory for searching for new physics in the $b\to c\tau\bar{\nu}$ transition. Using the standard model effective theory framework, we extended the standard model Hamiltonian to include a complete set of dimension-six four-fermion operators, characterizing the potential contributions from scalar, vector, and tensor NP. All relevant hadronic form factors were calculated directly within the covariant confined quark model with infrared confinement. By evaluating these form factors across the entire physical $q^2$ range without the need for extrapolations, we have ensured the reliability and self-consistency of our results independently of heavy quark effective theory.

Using the experimental data for $R_{D^{(\ast)}}$, $R_{J/\psi}$, and $F_L^{D^*}$ from the {\it BABAR}, Belle, and LHCb Collaborations, as well as the upper limit of $\mathcal{B}(B_c\to\tau\nu)$, we have obtained the constraints on the Wilson coefficients characterizing the NP contributions. It is worth mentioning that the constraints have been obtained under the assumption of one-operator dominance, where the interferences between different operators have been omitted. Finally, within the $2\sigma$ allowed regions of the corresponding Wilson coefficients, we have analyzed the effects of the NP operators on various physical observables. The sensitivity of these observables to NP operators has been examined in details. Some of the effects may help distinguish between NP operators. We have also discussed specific strategies to disentangle NP scenarios.

A major contribution of this work is the provision of a vast set of theoretical predictions for both traditional and novel observables. Our results include detailed predictions for the following: 
\begin{enumerate}[label=\roman*.]
\item Traditional observables: We have provided calculated values and $q^2$-distributions for the ratios of branching fractions $R(D_s^{(*)})$, forward-backward asymmetry, lepton-side and hadron-side convexity parameters, longitudinal polarization fraction of the final vector meson, longitudinal, transverse, and normal polarizations of the final tau lepton, and the full set of trigonometric moments $W_i$.
\item Angular coefficient functions: We analyzed the fourfold full angular distribution of the cascade decay $B_s \to D_s^*(\to D_s \pi) \tau \bar{\nu}$ and provided predictions for the complete set of 12 angular coefficient functions $J_i$, both in the SM and in various NP scenarios.
\item Binwise predictions: Recognizing the importance of direct comparison with experimental data, we have provided numerical predictions for the angular coefficient functions in specific kinematic bins of the recoil parameter $w$. This binwise approach is designed to provide ready-to-use templates for current and future measurements at Belle~II and LHCb Collaborations.
\end{enumerate} 


\begin{acknowledgments}
This  research  is  funded  by  Vietnam National Foundation for Science and Technology Development (NAFOSTED) 
under Grant~No.~103.01-2021.09. C.~T.~T. thanks Dang-Khoa N.~Nguyen for his help in the extraction of the form factors from other approaches.
\end{acknowledgments}
\section*{DATA AVAILABILITY}
The data that support the findings of this article are not publicly available. The data are available from the authors upon reasonable request.



\begin{thebibliography}{99}
\bibitem{BaBar:2012obs}
J.~P.~Lees \textit{et al.} (\emph{BABAR} Collaboration),
Phys. Rev. Lett. \textbf{109}, 101802 (2012)
[arXiv:1205.5442].
	
\bibitem{BaBar:2013mob}
J.~P.~Lees \textit{et al.} (\emph{BABAR} Collaboration),
Phys. Rev. D \textbf{88}, 072012 (2013)
[arXiv:1303.0571].

\bibitem{Belle:2015qfa}
M.~Huschle \textit{et al.} (Belle Collaboration),
Phys. Rev. D \textbf{92}, 072014 (2015)
[arXiv:1507.03233].

\bibitem{Belle:2016ure}
Y.~Sato \textit{et al.} (Belle Collaboration),
Phys. Rev. D \textbf{94}, 072007 (2016)
[arXiv:1607.07923].

\bibitem{Belle:2016dyj}
S.~Hirose \textit{et al.} (Belle Collaboration),
Phys. Rev. Lett. \textbf{118}, 211801 (2017)
[arXiv:1612.00529].

\bibitem{Belle:2017ilt}
S.~Hirose \textit{et al.} (Belle Collaboration),
Phys. Rev. D \textbf{97}, 012004 (2018)
[arXiv:1709.00129].

\bibitem{Belle:2019rba}
G.~Caria \textit{et al.} (Belle Collaboration),
Phys. Rev. Lett. \textbf{124}, 161803 (2020)
[arXiv:1910.05864].

\bibitem{LHCb:2023zxo}
R.~Aaij \textit{et al.} (LHCb Collaboration),
Phys. Rev. Lett. \textbf{131}, 111802 (2023)
[arXiv:2302.02886].

\bibitem{LHCb:2023uiv}
R.~Aaij \textit{et al.} (LHCb Collaboration),
Phys. Rev. D \textbf{108}, 012018 (2023); \textbf{109}, 119902(E) (2024)
[arXiv:2305.01463].

\bibitem{LHCb:2024jll}
R.~Aaij \textit{et al.} (LHCb Collaboration),
Phys. Rev. Lett. \textbf{134}, 061801 (2025)
[arXiv:2406.03387].

\bibitem{Belle-II:2024ami}
I.~Adachi \textit{et al.} (Belle-II Collaboration),
Phys. Rev. D \textbf{110}, 072020 (2024)
[arXiv:2401.02840].

\bibitem{Belle-II:2025yjp}
I.~Adachi \textit{et al.} (Belle-II Collaboration),
Phys. Rev. D \textbf{112}, 032010 (2025)
[arXiv:2504.11220].

\bibitem{HFLAV:2022esi}
Y.~S.~Amhis \textit{et al.} (Heavy Flavor Averaging Group),
Phys. Rev. D \textbf{107}, 052008 (2023)
[arXiv:2206.07501]. For update, see https://hflav-eos.web.cern.ch/hflav-eos/semi/ckm25/html/RDsDsstar/RDRDs.html.


\bibitem{Bigi:2016mdz}
D.~Bigi and P.~Gambino,
Phys. Rev. D \textbf{94}, 094008 (2016)
[arXiv:1606.08030].

\bibitem{Bordone:2019vic}
M.~Bordone, M.~Jung, and D.~van Dyk,
Eur. Phys. J. C \textbf{80}, 74 (2020)
[arXiv:1908.09398].

\bibitem{Martinelli:2021onb}
G.~Martinelli, S.~Simula, and L.~Vittorio,
Phys. Rev. D \textbf{105}, 034503 (2022)
[arXiv:2105.08674].

\bibitem{Bernlochner:2022ywh}
F.~U.~Bernlochner, Z.~Ligeti, M.~Papucci, M.~T.~Prim, D.~J.~Robinson, and C.~Xiong,
Phys. Rev. D \textbf{106}, 096015 (2022)
[arXiv:2206.11281].

\bibitem{Ray:2023xjn}
I.~Ray and S.~Nandi,
J. High Energy Phys. 01 (2024) 022 
[arXiv:2305.11855].

\bibitem{FlavourLatticeAveragingGroupFLAG:2024oxs}
Y.~Aoki \textit{et al.} (Flavour Lattice Averaging Group),
Phys. Rev. D \textbf{113}, 014508 (2026)
[arXiv:2411.04268].

\bibitem{BaBar:2019vpl}
J.~P.~Lees \textit{et al.} (\emph{BABAR} Collaboration),
Phys. Rev. Lett. \textbf{123}, 091801 (2019)
[arXiv:1903.10002].

\bibitem{Gambino:2019sif}
P.~Gambino, M.~Jung, and S.~Schacht,
Phys. Lett. B \textbf{795}, 386 (2019)
[arXiv:1905.08209].

\bibitem{Martinelli:2023fwm}
G.~Martinelli, S.~Simula, and L.~Vittorio,
Eur. Phys. J. C \textbf{84}, 400 (2024)
[arXiv:2310.03680].

\bibitem{MILC:2015uhg}
J.~A.~Bailey \textit{et al.} (MILC Collaboration),
Phys. Rev. D \textbf{92}, 034506 (2015)
[arXiv:1503.07237].

\bibitem{Na:2015kha}
H.~Na \textit{et al.} (HPQCD Collaboration),
Phys. Rev. D \textbf{92}, 054510 (2015); \textbf{93}, 119906(E) (2016)
[arXiv:1505.03925].

\bibitem{Fajfer:2012vx}
S.~Fajfer, J.~F.~Kamenik, and I.~Nisandzic,
Phys. Rev. D \textbf{85}, 094025 (2012)
[arXiv:1203.2654].

\bibitem{LHCb:2017vlu}
R.~Aaij \textit{et al.} (LHCb Collaboration),
Phys. Rev. Lett. \textbf{120}, 121801 (2018)
[arXiv:1711.05623].

\bibitem{CMS:2025jfx}
V.~Chekhovsky \textit{et al.} (CMS Collaboration),
Phys. Rev. D \textbf{113}, L111101 (2026)
[arXiv:2510.21559].

\bibitem{Iguro:2024hyk}
S.~Iguro, T.~Kitahara, and R.~Watanabe,
Phys. Rev. D \textbf{110}, 075005 (2024)
[arXiv:2405.06062].

\bibitem{Harrison:2020nrv}
J.~Harrison \textit{et al.} (HPQCD Collaboration),
Phys. Rev. Lett. \textbf{125}, 222003 (2020)
[arXiv:2007.06956].

\bibitem{Colangelo:2016ymy}
P.~Colangelo and F.~De Fazio,
Phys. Rev. D \textbf{95}, 011701 (2017)
[arXiv:1611.07387].

\bibitem{Bernlochner:2021vlv}
F.~U.~Bernlochner, M.~F.~Sevilla, D.~J.~Robinson, and G.~Wormser,
Rev. Mod. Phys. \textbf{94}, 015003 (2022)
[arXiv:2101.08326].

\bibitem{LHCb:2020cyw}
R.~Aaij \textit{et al.} (LHCb Collaboration),
Phys. Rev. D \textbf{101}, 072004 (2020)
[arXiv:2001.03225].

\bibitem{Goldberger:1999yh}
W.~D.~Goldberger,
arXiv:hep-ph/9902311.

\bibitem{Buchmuller:1985jz}
W.~Buchmuller and D.~Wyler,
Nucl. Phys. \textbf{B268}, 621 (1986)

\bibitem{Grzadkowski:2010es}
B.~Grzadkowski, M.~Iskrzynski, M.~Misiak, and J.~Rosiek,
J. High Energy Phys. 10, (2010) 085 
[arXiv:1008.4884].

\bibitem{Schacht:2020qot}
S.~Schacht and A.~Soni,
J. High Energy Phys. \textbf{10}, 163 (2020)
[arXiv:2007.06587].

\bibitem{Dutta:2018jxz}
R.~Dutta and N.~Rajeev,
Phys. Rev. D \textbf{97}, 095045 (2018)
[arXiv:1803.03038].

\bibitem{Monahan:2017uby}
C.~J.~Monahan, H.~Na, C.~M.~Bouchard, G.~P.~Lepage, and J.~Shigemitsu,
Phys. Rev. D \textbf{95}, 114506 (2017)
[arXiv:1703.09728].

\bibitem{Harrison:2021tol}
J.~Harrison \textit{et al.} (HPQCD Collaboration),
Phys. Rev. D \textbf{105}, 094506 (2022)
[arXiv:2105.11433].

\bibitem{Das:2019cpt}
N.~Das and R.~Dutta,
J. Phys. G \textbf{47}, 115001 (2020)
[arXiv:1912.06811].

\bibitem{Das:2021lws}
N.~Das and R.~Dutta,
Phys. Rev. D \textbf{105}, 055027 (2022)
[arXiv:2110.05526].

\bibitem{Harrison:2023dzh}
J.~Harrison \textit{et al.} (HPQCD Collaboration),
Phys. Rev. D \textbf{109}, 094515 (2024)
[arXiv:2304.03137].

\bibitem{McLean:2019qcx}
E.~McLean, C.~T.~H.~Davies, J.~Koponen, and A.~T.~Lytle,
Phys. Rev. D \textbf{101}, 074513 (2020)
[arXiv:1906.00701].

\bibitem{Penalva:2023snz}
N.~Penalva, J.~M.~Flynn, E.~Hern\'andez, and J.~Nieves,
J. High Energy Phys. 01 (2024) 163
[arXiv:2304.00250].

\bibitem{Yadav:2024zrn}
A.~K.~Yadav and S.~Sahoo,
Nucl. Phys. \textbf{B1020}, 117164 (2025)
[arXiv:2410.06100].

\bibitem{Crivellin:2012ye}
A.~Crivellin, C.~Greub, and A.~Kokulu,
Phys. Rev. D \textbf{86}, 054014 (2012)
[arXiv:1206.2634].

\bibitem{Wang:2021zfp}
S.~W.~Wang,
Int. J. Theor. Phys. \textbf{60}, 3225 (2021)

\bibitem{Sahoo:2020wnk}
S.~Sahoo and A.~Bhol,
[arXiv:2005.12630].

\bibitem{Wang:2020kov}
S.~W.~Wang,
Nucl. Phys. \textbf{B954}, 114997 (2020)

\bibitem{Crivellin:2023sig}
A.~Crivellin and S.~Iguro,
Phys. Rev. D \textbf{110}, 015014 (2024)
[arXiv:2311.03430].

\bibitem{Harrison:2017fmw}
J.~Harrison \textit{et al.} (HPQCD Collaboration),
Phys. Rev. D \textbf{97}, 054502 (2018)
[arXiv:1711.11013].

\bibitem{Bailey:2012rr}
J.~A.~Bailey, A.~Bazavov, C.~Bernard
 \textit{et al.}
Phys. Rev. D \textbf{85}, 114502 (2012); \textbf{86}, 039904(E) (2012)
[arXiv:1202.6346].

\bibitem{Blasi:1993fi}
P.~Blasi, P.~Colangelo, G.~Nardulli, and N.~Paver,
Phys. Rev. D \textbf{49}, 238 (1994)
[arXiv:hep-ph/9307290].

\bibitem{Azizi:2008tt}
K.~Azizi,
Nucl. Phys. \textbf{B801}, 70 (2008)
[arXiv:0805.2802].

\bibitem{Zhang:2021wnv}
Y.~Zhang, T.~Zhong, H.~B.~Fu, W.~Cheng, and X.~G.~Wu,
Phys. Rev. D \textbf{103}, 114024 (2021)
[arXiv:2104.00180].

\bibitem{Li:2009wq}
R.~H.~Li, C.~D.~L\"u, and Y.~M.~Wang,
Phys. Rev. D \textbf{80}, 014005 (2009)
[arXiv:0905.3259].

\bibitem{Bordone:2019guc}
M.~Bordone, N.~Gubernari, D.~van Dyk, and M.~Jung,
Eur. Phys. J. C \textbf{80}, 347 (2020)
[arXiv:1912.09335].

\bibitem{Cui:2023jiw}
B.~Y.~Cui, Y.~K.~Huang, Y.~M.~Wang, and X.~C.~Zhao,
Phys. Rev. D \textbf{108}, L071504 (2023)
[arXiv:2301.12391].

\bibitem{Zhang:2025tlr}
S.~H.~Zhang, T.~Zhong, H.~B.~Fu, Y.~X.~Wang, and W.~B.~Luo,
Chin. Phys. C\textbf{49}, 093104 (2025)
[arXiv:2503.06365].

\bibitem{Zhao:2006at}
S.~M.~Zhao, X.~Liu, and S.~J.~Li,
Eur. Phys. J. C \textbf{51}, 601 (2007)
[arXiv:hep-ph/0612008].

\bibitem{Cheng:2003sm}
H.~Y.~Cheng, C.~K.~Chua, and C.~W.~Hwang,
Phys. Rev. D \textbf{69}, 074025 (2004)
[arXiv:hep-ph/0310359].

\bibitem{Verma:2011yw}
R.~C.~Verma,
J. Phys. G \textbf{39}, 025005 (2012)
[arXiv:1103.2973].

\bibitem{S:2025uej}
T.~M.~S. and R.~Dhir,
Eur. Phys. J. C \textbf{85}, 1059 (2025)
[arXiv:2507.05104].

\bibitem{Chen:2011ut}
X.~J.~Chen, H.~F.~Fu, C.~S.~Kim, and G.~L.~Wang,
J. Phys. G \textbf{39}, 045002 (2012)
[arXiv:1106.3003].

\bibitem{Faustov:2012mt}
R.~N.~Faustov and V.~O.~Galkin,
Phys. Rev. D \textbf{87}, 034033 (2013)
[arXiv:1212.3167].

\bibitem{Patnaik:2025fry}
S.~Patnaik, L.~Nayak, and S.~K.~Swain,
Phys. Rev. D \textbf{112}, 033003 (2025)

\bibitem{Hu:2019bdf}
X.~Q.~Hu, S.~P.~Jin, and Z.~J.~Xiao,
Chin. Phys. C \textbf{44}, 053102 (2020)
[arXiv:1912.03981].

\bibitem{Soni:2021fky}
N.~R.~Soni, A.~Issadykov, A.~N.~Gadaria, Z.~Tyulemissov, J.~J.~Patel, and J.~N.~Pandya,
Eur. Phys. J. Plus \textbf{138}, 163 (2023)
[arXiv:2110.12740].

\bibitem{Pandya:2023ldv}
J.~N.~Pandya, P.~Santorelli, and N.~R.~Soni,
Eur. Phys. J. Special Topics \textbf{233}, 2075 (2024)
[arXiv:2307.14245].

\bibitem{Dubnicka:2025feg}
S.~Dubni{\v{c}}ka, A.~Z.~Dubni{\v{c}}kov{\'a}, M.~A.~Ivanov, and A.~Liptaj,
Eur. Phys. J. A \textbf{62}, 106 (2026)
[arXiv:2512.15321].

\bibitem{Efimov:1988yd}
G.~V.~Efimov and M.~A.~Ivanov,
Int.\ J.\ Mod.\ Phys.\ A {\bf 4}, 2031 (1989).

\bibitem{Efimov:zg}
G.~V.~Efimov and M.~A.~Ivanov,
The Quark Confinement Model Of Hadrons,
(CRC Press, London, 1993).  

\bibitem{Branz:2009cd} 
T.~Branz, A.~Faessler, T.~Gutsche, M.~A.~Ivanov, J.~G.~K\"{o}rner,
and V.~E.~Lyubovitskij,
Phys.\ Rev.\ D {\bf 81}, 034010 (2010)
[arXiv:0912.3710].

\bibitem{Ivanov:2011aa}
M.~A.~Ivanov, J.~G.~K\"{o}rner, S.~G.~Kovalenko, P.~Santorelli, 
and G.~G.~Saidullaeva,
Phys.\ Rev.\ D\ {\bf 85}, 034004 (2012),
[arXiv:1112.3536]. 

\bibitem{Gershtein:1976mv}
S.~S.~Gershtein and M.~Y.~Khlopov,
JETP Lett. \textbf{23}, 338 (1976).

\bibitem{Khlopov:1978id}
M.~Y.~Khlopov,
Sov. J. Nucl. Phys. \textbf{28}, 583 (1978)

\bibitem{Neubert:1993mb}
M.~Neubert,
Phys.\ Rep.\ {\bf 245}, 259 (1994),
[arXiv:hep-ph/9306320]. 


\bibitem{Grozin:2004yc}
A.~G.~Grozin,
Springer Tracts Mod. Phys. \textbf{201}, 1 (2004).

\bibitem{Belle:2023xgj}
M.~T.~Prim \textit{et al.} (Belle Collaboration),
Phys. Rev. Lett. \textbf{133}, 131801 (2024)
[arXiv:2310.20286].

\bibitem{Ivanov:2015tru}
M.~A.~Ivanov, J.~G.~K\"orner, and C.~T.~Tran,
Phys. Rev. D \textbf{92}, 114022 (2015)
[arXiv:1508.02678].

\bibitem{Ivanov:2016qtw}
M.~A.~Ivanov, J.~G.~K\"orner, and C.~T.~Tran,
Phys. Rev. D \textbf{94}, 094028 (2016)
[arXiv:1607.02932].

\bibitem{Tran:2018kuv}
C.~T.~Tran, M.~A.~Ivanov, J.~G.~K\"orner, and P.~Santorelli,
Phys. Rev. D \textbf{97}, 054014 (2018)
[arXiv:1801.06927].

\bibitem{Anikin:1995cf}
I.~V.~Anikin, M.~A.~Ivanov, N.~B.~Kulimanova, and V.~E.~Lyubovitskij,
Z. Phys. C \textbf{65}, 681 (1995).

\bibitem{Ivanov:1997ug}
M.~A.~Ivanov and V.~E.~Lyubovitskij,
Phys. Lett. B \textbf{408}, 435 (1997)
[arXiv:hep-ph/9705423].

\bibitem{Faessler:2003yf}
A.~Faessler, T.~Gutsche, M.~A.~Ivanov, V.~E.~Lyubovitskij, and P.~Wang,
Phys. Rev. D \textbf{68}, 014011 (2003)
[arXiv:hep-ph/0304031].

\bibitem{Salam:1962ap}
A.~Salam,
Nuovo Cimento \textbf{25}, 224 (1962).

\bibitem{Weinberg:1962hj}
S.~Weinberg,
Phys. Rev. \textbf{130}, 776 (1963).

\bibitem{Hayashi:1967bjx}
K.~Hayashi, M.~Hirayama, T.~Muta, N.~Seto, and T.~Shirafuji,
Fortschr. Phys. \textbf{15}, 625 (1967).

\bibitem{Tran:2023hrn}
C.~T.~Tran, M.~A.~Ivanov, P.~Santorelli, and Q.~C.~Vo,
Chin. Phys. C \textbf{48}, 023103 (2024)
[arXiv:2311.15248].

\bibitem{Ivanov:2017hun}
M.~A.~Ivanov, J.~G.~K\"orner, and C.~T.~Tran,
Phys. Part. Nucl. Lett. \textbf{14}, 669 (2017).

\bibitem{LHCb:2023ssl}
R.~Aaij \textit{et al.} (LHCb Collaboration),
Phys. Rev. D \textbf{110}, 092007 (2024)
[arXiv:2311.05224].

\bibitem{Alonso:2016oyd}
R.~Alonso, B.~Grinstein, and J.~Martin Camalich,
Phys. Rev. Lett. \textbf{118}, 081802 (2017)
[arXiv:1611.06676].

\bibitem{Akeroyd:2017mhr}
A.~G.~Akeroyd and C.~H.~Chen,
Phys. Rev. D \textbf{96}, 075011 (2017)
[arXiv:1708.04072].

\bibitem{Blanke:2018yud}
M.~Blanke, A.~Crivellin, S.~de Boer, T.~Kitahara, M.~Moscati, U.~Nierste, and I.~Ni{\v{s}}and{\v{z}}i{\'c},
Phys. Rev. D \textbf{99}, 075006 (2019)
[arXiv:1811.09603].

\bibitem{Martinelli:2024bov}
G.~Martinelli, S.~Simula, and L.~Vittorio,
Eur. Phys. J. C \textbf{85}, 242 (2025)
[arXiv:2410.17974].

\bibitem{Belle:2019ewo}
A.~Abdesselam \textit{et al.} (Belle Collaboration),
arXiv:1903.03102.

\bibitem{Ivanov:2020iad}
M.~A.~Ivanov, J.~G.~K{\"o}rner, P.~Santorelli, and C.~T.~Tran,
Particles \textbf{3}, 193 (2020)
[arXiv:2009.00306].

\bibitem{Crivellin:2017zlb}
A.~Crivellin, D.~M\"uller, and T.~Ota,
J. High Energy Phys. 09  (2017) 040
[arXiv:1703.09226].

\bibitem{Lee:2017kbi}
J.~P.~Lee,
Phys. Rev. D \textbf{96}, 055005 (2017)
[arXiv:1705.02465].

\bibitem{Iguro:2017ysu}
S.~Iguro and K.~Tobe,
Nucl. Phys. \textbf{B925}, 560 (2017)
[arXiv:1708.06176].

\bibitem{Huang:2018nnq}
Z.~R.~Huang, Y.~Li, C.~D.~L\"u, M.~A.~Paracha, and C.~Wang,
Phys. Rev. D \textbf{98}, 095018 (2018)
[arXiv:1808.03565].

\bibitem{Duraisamy:2013pia}
M.~Duraisamy and A.~Datta,
J. High Energy Phys. 09 (2013) 059 
[arXiv:1302.7031].

\bibitem{Korner:1989ve}
J.~G.~K\"orner and G.~A.~Schuler,
Phys.\ Lett.\ B {\bf 231}, 306 (1989).

\bibitem{Korner:1989qb}
J.~G.~K\"orner and G.~A.~Schuler,
Z.\ Phys.\ C {\bf 46}, 93 (1990).

\bibitem{Becirevic:2016hea}
D.~Becirevic, S.~Fajfer, I.~Ni{\v{s}}and{\v{z}}i{\'c}, and A.~Tayduganov,
Nucl. Phys. \textbf{B946}, 114707 (2019)
[arXiv:1602.03030].

\bibitem{Zhang:2020dla}
L.~Zhang, X.~W.~Kang, X.~H.~Guo, L.~Y.~Dai, T.~Luo, and C.~Wang,
J. High Energy Phys. 02  (2021) 179
[arXiv:2012.04417].

\bibitem{Chen:2017eby}
C.~H.~Chen and T.~Nomura,
Eur. Phys. J. C \textbf{77}, 631 (2017)
[arXiv:1703.03646].

\bibitem{Tanaka:2010se}
M.~Tanaka and R.~Watanabe,
Phys. Rev. D \textbf{82}, 034027 (2010)
[arXiv:1005.4306].

\bibitem{Ivanov:2017mrj}
M.~A.~Ivanov, J.~G.~K\"orner, and C.~T.~Tran,
Phys. Rev. D \textbf{95}, 036021 (2017)
[arXiv:1701.02937].

\bibitem{Alonso:2016gym}
R.~Alonso, A.~Kobach, and J.~Martin Camalich,
Phys. Rev. D \textbf{94}, 094021 (2016)
[arXiv:1602.07671].

\bibitem{Alonso:2017ktd}
R.~Alonso, J.~Martin Camalich, and S.~Westhoff,
Phys. Rev. D \textbf{95}, 093006 (2017)
[arXiv:1702.02773].

\bibitem{Fedele:2023ewe}
M.~Fedele, M.~Blanke, A.~Crivellin, S.~Iguro, U.~Nierste, S.~Simula, and L.~Vittorio,
Phys. Rev. D \textbf{108}, 5 (2023)
[arXiv:2305.15457].

\bibitem{Bordone:2025jur}
M.~Bordone, N.~Gubernari, M.~Jung, and D.~van Dyk,
J. High Energy Phys. 11  (2025) 051
[arXiv:2507.03569].

\bibitem{Colangelo:2024mxe}
P.~Colangelo, F.~De Fazio, F.~Loparco, and N.~Losacco,
Phys. Rev. D \textbf{109}, 075047 (2024)
[arXiv:2401.12304].

\bibitem{Martinelli:2024vde}
G.~Martinelli, S.~Simula, and L.~Vittorio,
Phys. Rev. D \textbf{111}, 013005 (2025)
[arXiv:2409.10492].

\bibitem{Boyd:1997kz}
C.~G.~Boyd, B.~Grinstein, and R.~F.~Lebed,
Phys. Rev. D \textbf{56}, 6895 (1997)
[arXiv:hep-ph/9705252].

\end{thebibliography}
\end{document}